\documentclass[a4paper,11pt]{article}
\pdfoutput=1 

\usepackage{jcappub} 
\usepackage{xcolor}
\usepackage{physics}
\usepackage{bigints}
\usepackage{relsize}
\usepackage[capitalise]{cleveref}
\crefname{section}{Sec.}{Secs.}
\usepackage{soul}
\usepackage{appendix}

\usepackage[normalem]{ulem}

\title{The Sun and core-collapse supernovae are leading probes of the neutrino lifetime }

\author[a]{Pablo Mart{\'\i}nez-Mirav{\'e},}
\author[a]{Irene Tamborra,}
\author[b]{Mariam T{\'o}rtola}

\affiliation[a]{Niels Bohr International Academy and DARK, Niels Bohr Institute, University of \mbox{Copenhagen}, Blegdamsvej 17, 2100, Copenhagen, Denmark}
\affiliation[b]{Instituto de F{\'\i}sica Corpuscular  (CSIC-Universitat de Val{\`e}ncia), Parc Cient{\'\i}fic UV,\\ \mbox{C/~Catedr{\'a}tico} Jos{\'e} Beltr{\'a}n, 2, 46980, Paterna, Spain}

\emailAdd{pablo.mirave@nbi.ku.dk}
\emailAdd{tamborra@nbi.ku.dk}
\emailAdd{mariam@ific.uv.es}

\abstract{
The large distances travelled by neutrinos emitted from the Sun and core-collapse supernovae together with the characteristic energy of such neutrinos provide ideal conditions to probe their lifetime,  when the decay products evade detection. We investigate the prospects of probing invisible neutrino decay capitalising on the detection of solar and supernova neutrinos as well as the diffuse supernova neutrino background (DSNB) in the next-generation neutrino observatories Hyper-Kamiokande, DUNE, JUNO, DARWIN, and \mbox{RES-NOVA}.
We find that future solar neutrino data will be sensitive to values of the lifetime-to-mass ratio $\tau_1/m_1$ and $\tau_2/m_2$ of $\mathcal{O}(10^{-1}$--$10^{-2})$~s/eV. From a core-collapse supernova explosion at $10$~kpc,  lifetime-to-mass ratios of the three mass eigenstates of  $\mathcal{O}(10^5)$~s/eV could be tested. After $20$~years of data taking, the DSNB would extend the sensitivity reach of $\tau_1/m_1$ to $10^{8}$~s/eV. These results promise an improvement of about $6$--$15$ orders of magnitude on the values of the decay parameters with respect to existing limits.}

\begin{document}
\maketitle
\flushbottom
\section{Introduction}
The observation of neutrino oscillations provides evidence that neutrinos are massive particles. 
As a direct implication of non-zero masses, one also finds that at least the two heaviest neutrino mass eigenstates may be unstable.~\footnote{Since there is a vast number of Standard Model extensions in which neutrinos become unstable, it is useful to introduce the following classification. We denote neutrino decays in which the final-state particles are experimentally accessible as \textit{visible} decays. Conversely, we refer to decays in which the final products escape detection as \textit{invisible} decays. Alternatively, decay modes can also be classified into two categories, $\textit{radiative}$ and $\textit{non-radiative}$ neutrino decays, depending on whether there is a photon in the final state.} Neutrino decay has not been observed yet, but in some extensions of the Standard Model, the predicted neutrino lifetime is larger than the age of the Universe, making neutrinos effectively stable~\cite{Giunti:2014ixa}. However, the size of the neutrino lifetime predicted varies significantly for different scenarios~\cite{Bahcall:1972my,Shrock:1974nd,Petcov:1976ff,Marciano:1977wx,Zatsepin:1978iy,Chikashige:1980qk,Gelmini:1980re,Pal:1981rm,Schechter:1981cv,Shrock:1982sc,Gelmini:1983ea,Bahcall:1986gq,Nussinov:1987pc,Frieman:1987as,Kim:1990km}. 

In this paper, we focus on non-radiative invisible decays of neutrinos (i.e.~those in which the decay products are not observable). We find it convenient to work in terms of the decay width of the mass eigenstate in its rest frame, $\Gamma_i$ (where the index $i$ indicates the mass eigenstate), instead of the usually adopted ratio between the neutrino lifetime ($\tau_i$) and the mass ($m_i$). The motivation for this choice is that the absolute neutrino decay width simply results from the sum of the decay width of each channel. 
Hence, the decay parameter is
\begin{align}
\alpha_i = m_i \Gamma^{\rm invisible}_i = \frac{m_i}{\tau_i} \, .
\label{eq:params}
\end{align}
One could think that neutrinos travelling longer distances would be especially suited to probe the neutrino lifetime. 
Note, however, that neutrino lifetimes are well-defined in their rest frame. Then, one expects to be sensitive to neutrino lifetimes $\tau_i \sim  t_{TOF}/\gamma_i$, where $\gamma_i$ is the Lorentz factor and $t_{TOF}$ is the neutrino time-of-flight.

Existing constraints on the neutrino lifetime have been derived relying on neutrinos emitted with different characteristic energy from artificial or natural  sources~\cite{Abdullahi:2020rge,Ackermann:2022rqc,Abrahao:2015rba, Gonzalez-Garcia:2008mgl,Gomes:2014yua,Choubey:2017dyu,Choubey_2018,deSalas:2018kri}.
For instance, reactor antineutrinos, which have energies of a few MeV and travel distances of $\mathcal{O}(1$--$100)$~km, can constrain neutrino invisible decay. With approximately $5$ years of data taking, the medium-baseline reactor experiment JUNO would be sensitive to $\tau_3/m_3 \, <\, 7.5 \times 10^{-11}$~s/eV at $95\%$ C.L.~\cite{Abrahao:2015rba}. 
For atmospheric and accelerator neutrinos, the neutrino time-of-flight is much larger than the characteristic one of reactors, but the neutrino energy is at least $2$--$3$ orders of magnitude larger. As a result, the related bounds are of the same order. For instance, a recent analysis of accelerator data from T2K, NOvA and MINOS/MINOS+ reports $\tau_3/m_3 > 2.4 \times 10^{-11}$~s/eV at 90\% C.L.~\cite{Ternes:2024qui}.
Similarly, a joint analysis of Super-Kamiokande atmospheric data~\cite{Super-Kamiokande:2006jvq}, MINOS~\cite{MINOS:2006foh} and K2K~\cite{K2K:2006yov} leads to 
$\tau_3/m_3\, > \, 2.9  \times 10^{-10}$~s/eV at 90\% C.L.~\cite{Gonzalez-Garcia:2008mgl}.
After 10 years of data, KM3NeT-ORCA  alone will constrain the invisible neutrino decay at the same level, with $\tau_3/m_3 \, > 2.5 \times 10^{-10}$ s/eV at $90\%$ C.L.~\cite{deSalas:2018kri}.

Invisible neutrino decay can also be constrained through cosmological observables. For instance, Big Bang Nucleosynthesis requires neutrino lifetimes larger than $10^{-3}$~s at $95\%$ C.L.~\cite{Escudero:2019gfk}. Additionally, the neutrino lifetime and its mass can  be explored exploiting  the fact that the data from the cosmic microwave background (CMB) agree well with neutrinos  free streaming at the temperature of photon decoupling~\cite{Basboll:2008fx,Archidiacono:2013dua,Hannestad:2004qu,Hannestad:2005ex,Escudero:2019gfk,Barenboim:2020vrr,Chen:2022idm}. For instance, for a neutrino mass eigenstate $\nu_i$ decaying into a lighter sterile neutrino $\nu_4$ and a massless scalar particle, $\phi$,one can constrain $\tau_{\nu_i \rightarrow \nu_4 + \phi} > (4 \times 10^5 \to 4 \times 10^6) \textrm{ s}\left({m_{\nu_i}}/{0.05 \, \text{eV}}\right)^5$, depending on the model~\cite{Barenboim:2020vrr}. 
Furthermore, one can study how the matter power spectrum and lensing of the CMB would be modified if neutrinos decayed invisibly when non-relativistic~\cite{Chacko:2019nej,FrancoAbellan:2021hdb}. In the near future, data from Euclid, when combined with other cosmological probes, would provide an independent measure of the sum of neutrino masses and  lifetime~\cite{Chacko:2020hmh}.

For what concerns neutrinos of astrophysical origin, one may consider setting limits on neutrino invisible decay from the high-energy neutrinos observed by the IceCube Neutrino Observatory~\cite{Valera:2023bud,Denton:2018aml,Song:2020nfh}. However, in this case, even though  neutrinos travel over cosmic distances, their energy is so large that the effect of neutrino decay on the detectable signal may be suppressed.
Better constraints come from a combination of solar datasets~\cite{Bandyopadhyay:2002qg, Picoreti:2015ika, Choubey:2000an}. 
Current limits from invisible neutrino decay take into account information from $^7$Be neutrinos at Borexino~\cite{Bellini:2011rx} and KamLAND~\cite{KamLAND:2014gul} and $^8$B neutrinos at SNO~\cite{SNO:2011hxd}, Super-Kamiokande~\cite{Super-Kamiokande:2016yck}, KamLAND~\cite{KamLAND:2011fld} and Borexino~\cite{Borexino:2017uhp}. They also include the analysis of data from the Homestake chlorine detector~\cite{Cleveland:1998nv} and from the neutrino capture rate in Gallium at GNO, GALLEX, and SAGE~\cite{SAGE:2009eeu}. At $2 \sigma$ C.L., the limits read~\cite{Berryman:2014qha,SNO:2018pvg}
\begin{align}
	\alpha_1 < 1.6 \times 10^{-13} \textrm{eV}^2\, \quad \textrm{and} \quad \alpha_2 < 4.5 \times 10^{-13} \textrm{eV}^2 \, ,
\label{eqn:current-solar-1}
\end{align}
or in terms of the ratio between neutrino lifetime and mass, 
\begin{align}
        \frac{\tau_1}{m_1} > 4.2 \times 10^{-3} \, \textrm{s}/\textrm{eV} \, \quad \textrm{and} \quad \frac{\tau_2}{m_2} > 1.5 \times 10^{-3} \, \textrm{s}/\textrm{eV} \, .
\end{align}

Despite the limited amount of available data, the detection of electron antineutrinos from SN 1987A also provides constraints on invisible neutrino decay.
Such limits are generally quoted as the electron-neutrino lifetime: $\tau_{\bar{\nu}_e} > 1.7 \times 10^{5} \, [m_{\bar{\nu}_e}/E_{\bar{\nu}_e}]$~\cite{Hirata:1988ad}, where $E_{\bar{\nu}_e}$ is the electron-neutrino energy and $m_{\bar{\nu}_e}$ is an effective mass, not properly defined. Current searches for the diffuse supernova neutrino background (DSNB) at Super-Kamiokande are already excluding the most optimistic  DSNB models~\cite{Super-Kamiokande:2021jaq}.  Nonetheless, since the signature of invisible neutrino decay is a reduction of the flux with respect to its theoretical prediction, the lack of data on the DSNB has not provided useful insights into this new physics scenario. This will change soon due to the ongoing enrichment of Super-Kamiokande with Gadolinium~\cite{Super-Kamiokande:2023xup}.
\begin{figure}
\centering
\includegraphics[width = 0.7\textwidth]{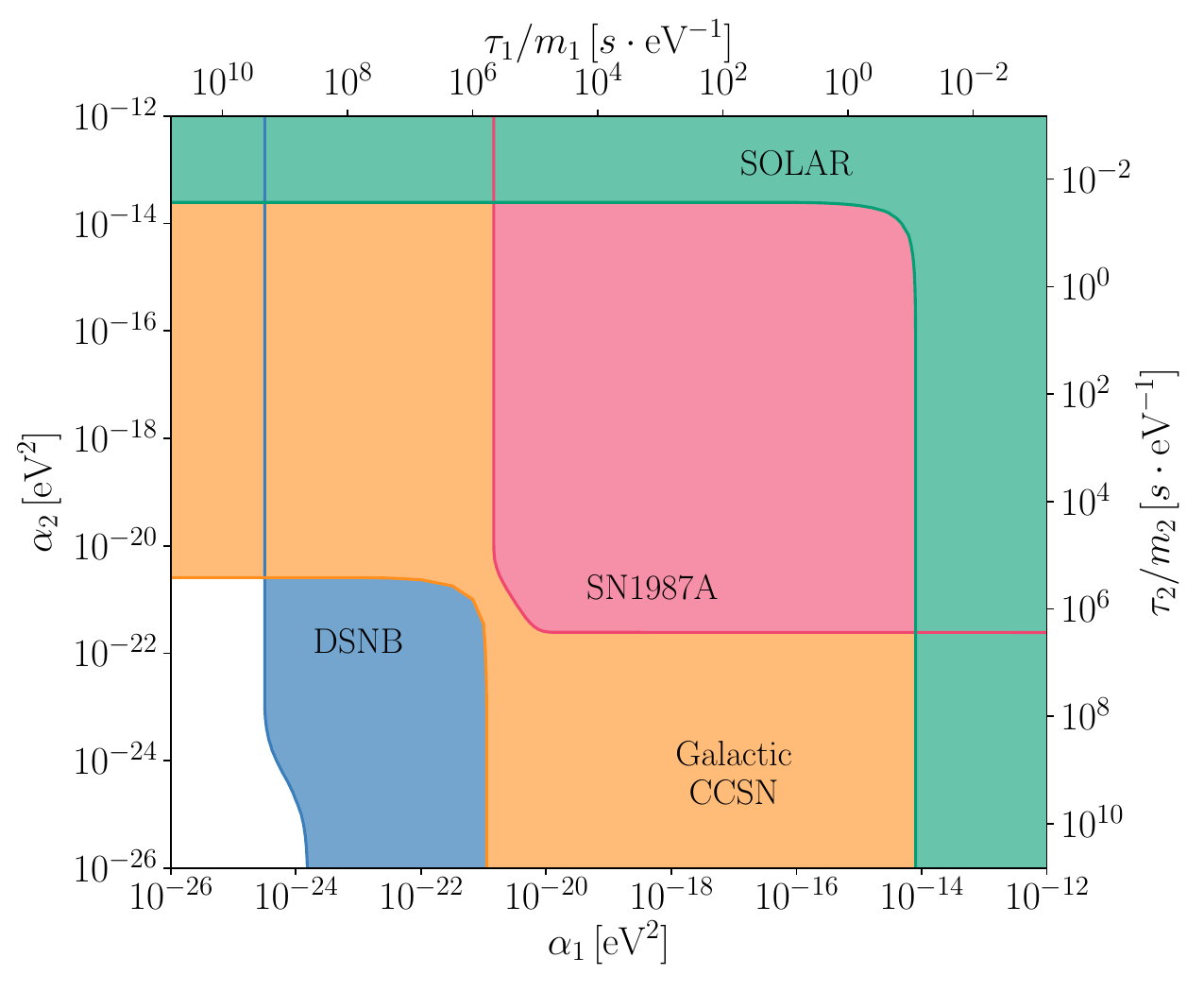}
\caption{Excluded regions at $90\%$~C.L.~in the plane spanned by the invisible neutrino decay parameters $\alpha_1$ and $\alpha_2$ obtained in this work from solar and CCSN neutrinos as well as the DSNB (shown in green, orange and blue, respectively).
We also derive limits on neutrino decay from the observation of neutrinos from SN 1987A in red. For details on the derivation of the projected limits, we refer the reader to~\cref{sec:Solar,sec:CCSN,sec:DSNB}. Solar data could set limits about $20$~times more stringent than the current ones, whereas from a CCSN at $10$~kpc, the sensitivity would improve up to $6$--$7$ orders of magnitude. After $20$ years of data taking, the DSNB could improve current bounds on $\alpha_1$ of about $10$ orders of magnitude. }
\label{fig:summary}
\end{figure}

In this work, we explore the prospects for constraining invisible neutrino decay through upcoming large-scale neutrino telescopes relying on neutrinos with energy $\mathcal{O}(0.1$--$100)$~MeV from the Sun and core-collapse supernovae (CCSNe), which seem to hold the optimal combination between neutrino energy and distance from the source.
Figure~\ref{fig:summary} summarises our forecast from solar neutrinos, neutrinos from a galactic CCSN, and the DSNB, together with our recast limits from SN 1987A. Our projected sensitivities take into account uncertainties on the sources for the first time. We can immediately see that the upcoming detection of neutrinos from these sources will provide the most stringent constraints on the invisible neutrino decay parameters $\alpha_1$ and $\alpha_2$ through direct neutrino detection. Depending on the value of neutrino masses, the bounds from cosmology could be more stringent than the ones obtained in this work,  but they would not be based on the direct detection of neutrinos.

This paper is organised as follows. In Sec.~\ref{sec:observatories}, we outline the main features of the next-generation neutrino observatories considered in this work (Hyper-Kamiokande, JUNO, DUNE, DARWIN, and RES-NOVA).  Section~\ref{sec:Solar} focuses on the impact of invisible neutrino decay on the signal expected from solar neutrinos, while  Sec.~\ref{sec:CCSN} presents the forecasted exclusion region of the invisible neutrino decay parameter space obtained relying on  CCSN  neutrinos, and Sec.~\ref{sec:DSNB} investigates the constraining power of the DSNB. Finally, Sec.~\ref{sec:summary} summarises our main findings. Details on the statistical analysis are provided in Appendix~\ref{sec:stats}.

\section{Next-generation neutrino observatories}
\label{sec:observatories}
In this section, we outline the main features of the upcoming neutrino observatories considered to explore the chances to constrain invisible neutrino decay: Hyper-Kamiokande, JUNO, DUNE, DARWIN, and RES-NOVA. We focus on the most relevant detection channels for each detector for the physics case under investigation and the energy range of relevance.

\subsection{Hyper-Kamiokande}
Hyper-Kamiokande~\cite{Hyper-Kamiokande:2018ofw} is the next-generation underground water Cherenkov detector, and successor of Super-Kamiokande. In its ideal configuration, it will consist of two tanks with a fiducial volume of $187$~kt each. The main detection channel for astrophysical neutrinos is inverse beta decay (IBD): $\bar{\nu}_e \, + \, p \longrightarrow e^+ \,  + \, n$, with an 
\begin{align}
    R_{{\rm IBD}} = N_t \, \varepsilon_{\rm IBD} \int {\rm d}E'_\nu \, \sigma_{\rm IBD} \int{\rm d} t' \, \Phi_{\bar{\nu}_e} (t')  \,\int \text{d}E_e \, K_{\rm IBD}(E'_e, E_{e})\, ,
    \label{eqn:rateIBD}
\end{align}
which depends on the number of targets ($N_t = 2.5 \times 10 ^{34}$ for each tank), and the cross-section, $\sigma_{\rm IBD}$~\cite{Strumia:2003zx}. The response function of the detector, $K_{\rm IBD}$,  relates the true and reconstructed energy of the positron (and, hence, of the neutrinos), which we denote by $E'_e$ and $E_e$ respectively. For  the detection channels of interest, we adopt the following parametrisation for the response function
\begin{align}
K (u,v) = \frac{1}{\sqrt{2\pi} \delta(v)} e^{-\left(\frac{u -v}{\sqrt{2}\delta(v)}\right)^2}\, ,
\end{align}
where $\delta(v)$ is the energy resolution. In particular, for the IBD detection channel in Hyper-Kamiokande, we take $\delta(E_e)/E_e = 0.1/\sqrt{E_e[\text{MeV}]}$.

Assuming a $0.1\%$ loading of both tanks with Gadolinium, the IBD detection efficiency is expected to be $\varepsilon_{\rm IBD} = 0.67$ \cite{Hyper-Kamiokande:2018ofw}, which includes the efficiency of neutron capture and the detection of $8$~MeV photons~\footnote{Note that, without Gadolinium loading, a lower detection efficiency is expected. The sensitivity reach would be degraded accordingly. For reference, Super-Kamiokande-IV reported $6\%$ signal efficiency in the search of electron antineutrinos of astrophysical origin with energies of $10$--$20$~MeV~\cite{Super-Kamiokande:2020frs}.}.

Another relevant channel for MeV neutrinos in Hyper-Kamiokande is the elastic scattering on electrons (ES), $\overset{_{(-)}}{\nu}_{\alpha} \, + \, e^-  \longrightarrow \overset{_{(-)}}{\nu}_{\alpha} \, + \, e^-$, with  $\alpha = e, \, x$, where $x$ denotes the non-electron flavours. Its importance relies on the fact that it provides a measurement of the total flux. 
The event rate, for a given neutrino flux $\Phi_{\bar{\nu}_e}$, is
\begin{align}
    R_{{\rm ES}}  =  N_t \, \varepsilon_{\rm ES} \sum_{\alpha}\int {\rm d}E'_\nu \int{\rm d} t'  \int {\rm d}T'_e  \, &\Bigg[ \sigma_{{\rm ES}, \nu_\alpha} ( T'_e) \, \Phi_{\nu_\alpha} (t')  +  \sigma_{{\rm ES}, \bar{\nu}_x} ( T'_e) \, \Phi_{\bar{\nu}_x} (t') \Bigg] \nonumber \\   \times \int \, \text{d}T_e  K_{\rm ES}(T'_e, T_e)\, ,
    \label{eqn:rateES}
\end{align}
which includes the contributions from neutrinos and antineutrinos of the three active flavours. Here $T'_e$ denotes the true electron recoil kinetic energy and $T_e$ is the reconstructed one. The cross-section $\sigma_{ES, \nu_x}$ depends on the flavour of the incoming neutrino or antineutrino~\cite{Strumia:2003zx}. The response function of the detector, $K_{\rm ES}$, with  the same resolution   of Super-Kamiokande-III~\cite{Super-Kamiokande:2010tar}, is
\begin{align}
    \frac{\delta (E_e)}{E_e} = 0.0349 + \frac{0.376}{\sqrt{E_e[\textrm{MeV}]}} -\frac{0.123}{E_e[{\rm MeV}]},
\end{align}
with $E_e = T_e + m_e$. We consider an efficiency $\varepsilon_{\rm ES} = 0.68$ arising from a cut in the angular distribution of the recoil electron~\cite{Laha:2013hva}. Note that the maximum electron recoil kinetic energy depends on the energy of the incoming neutrino, 
\begin{align}
    T'^{\rm max}_e = \frac{2E_\nu ^2}{m_e + 2E_\nu}\, ,
\end{align}
and the number of target electrons is $N_t = 1.25 \times 10^{35}$ per tank. 

Two other channels can contribute significantly to the number of events expected from a CCSN:  $ \nu_e \, + \, ^{16}{\rm O} \longrightarrow e^- \, + \, ^{16} {\rm F}^{*}$ and $\bar{\nu}_e \, + \,  ^{16}{\rm O} \longrightarrow e^{+} \, + \, ^{16}{\rm N}^{*}$~\cite{Kolbe:2002gk}, which however we neglect in this work. 
This choice is conservative since the inclusion of additional detection channels would increase the collected statistics.

\subsection{JUNO}
The Jiangmen Underground Neutrino Observatory (JUNO)~\cite{JUNO:2015zny} is a next-generation liquid scintillator detector suitable for the observation of electron antineutrinos from astrophysical sources via IBD. The expected event rate  is 
\begin{align}
    R_{{\rm IBD}}  = N_t \, \varepsilon_{\rm IBD} \int {\rm d}E'_\nu \, \sigma_{\rm IBD} \int{\rm d} t' \, \Phi_{\bar{\nu}_e} (t') \, \int \text{d} E_e K_{\rm IBD}(E'_e, E_e)\, .
    \label{eqn:rateIBD2}
\end{align}
For a fiducial volume of $17$~kt, the corresponding number of targets is $N_t = 1.2 \times 10^{33}$. The other channels of potential relevance for the study of low-energy astrophysical neutrinos, such as neutrino ES on protons, neutral current interactions on $^{12}$C and charged-current interactions of electron neutrinos and antineutrinos on $^{12}$C, are neglected in this work. 

\subsection{DUNE}
The upcoming Deep Underground Neutrino Experiment (DUNE)~\cite{DUNE:2020lwj} is also suitable for the study of astrophysical neutrinos with energies above $\mathcal{O}(10)$~MeV. 
The main detection channel is charged-current electron-neutrino scattering on argon ( $\nu_e \, +\, ^{40}\text{Ar} \, \rightarrow \, e^- \, + \,  ^{40} K^*$). In principle, other channels such as ES on electrons~\cite{Capozzi:2018dat,DUNE:2020ypp}, electron antineutrino charged-current interactions, and neutrino-argon neutral current interactions could be exploited~\cite{DUNE:2020zfm}.

The  event rate  from electron neutrino charged-current interactions as a function of the reconstructed neutrino energy is
\begin{align}
     R_{\rm DUNE} = \tau \, \varepsilon_{\rm DUNE}\, N_t \, \int \dd E_\nu \Phi_{\nu_e} \sigma_{\nu_e\text{CC}} \int \textrm{d}E_r K_{\rm DUNE}(E_\nu, E_r) \, ,
\end{align}
where  $\tau$ denotes the exposure time and $N_t = 6.02\times 10^{32}$ is the number of targets, corresponding to a fiducial volume of $40$~kt. The cross-section for the detection process ($\sigma_{\nu_e\text{CC}}$) is taken from \texttt{SNOWGLoBES}~\cite{snowglobes} and $\varepsilon_{ \rm DUNE}$ denotes the detection efficiency.
We also account for the response function of the detector, $K_{\rm DUNE}$, resulting from a Gaussian energy resolution~\cite{DUNE:2020ypp,Castiglioni:2020tsu,Barenboim:2023krl} with ${\delta(E_\nu)}/{E_\nu} = 0.2$.

\subsection{DARWIN}
Primarily conceived as dark-matter detectors, the physics program of ton-scale Xenon experiments extends to the investigation of astrophysical neutrinos~\cite{Aalbers:2022dzr}.
The capability of the DARk matter WImp search with liquid xenoN (DARWIN) ~\cite{DARWIN:2016hyl} to detect low-energy nuclear recoil energy opens the possibility of investigating low-energy astrophysical neutrinos interacting in the detector via Coherent Elastic Neutrino-Nucleus Scattering (CE$\nu$NS)~\cite{Lang:2016zhv}. The rate of neutrino-induced nuclear recoils is
\begin{align}
    R_{\text{CE}\nu\text{NS}} = N_t\, \varepsilon_{\text{CE}\nu\text{NS}} \int \text{d}t \int \text{d}E'_\nu \sum_{\alpha}\left[ \Phi_{\nu_\alpha} (t)+ \Phi_{\bar{\nu}_\alpha}(t) \right]\int \text{d}E'_R \,\sigma_{\text{CE}\nu\text{NS}} \nonumber \\ \times\int \text{d}E_R K_{\text{CE}\nu\text{NS}} (E_R, E'_R)\, ,
    \label{eqn:rateCEvNS}
\end{align}
where $N_t$ is the number of targets, $\varepsilon_{\text{CE}\nu\text{NS}}$ is the detection efficiency, $E'_R$ is the true nuclear recoil energy and $E_R$ is the reconstructed one. Notice that this is a flavour-insensitive channel that provides a measurement of the total neutrino flux. \mbox{DARWIN's} $30$~tons of Xe correspond to $N_t = 1.83 \times 10^{29}$ targets. We assume perfect detection efficiency and  energy resolution defined as~\cite{Schumann:2015cpa}
\begin{align}
    \frac{\delta (E_R)}{E_R} = 0.077 + \frac{0.232}{\sqrt{E_R[\text{keV}]}} + \frac{0.069}{E_R [\text{keV}]} \, . 
\end{align}

DARWIN will be sensitive to low-energy electron recoils, enabling the study of neutrino ES on electrons as well. The expected event rate for this channel is  the same as in Eq.~\eqref{eqn:rateES}, with an energy resolution given by
\begin{align}
    \frac{\delta(T_e)}{T_e} = \frac{0.3171}{\sqrt{T_e \, [{\rm keV}]}} + 0.0015\, ,
\end{align}
and an energy threshold of $1$~keV~\cite{DARWIN:2020bnc}. Notice that for ES on electrons, the number of target electrons is $N_t = 9.92\times10^{30}$.

\subsection{RES-NOVA}
Similar to DARWIN, astrophysical neutrinos can also be detected by the archaeological-lead-based detector  RES-NOVA~\cite{Pattavina:2020cqc}, relying on  CE$\nu$NS. 
We consider Phase-III of the planned detector, with a mass of $341$~tons of PbWO$_4$~\cite{RES-NOVAGroupofInterest:2022glt}. The interaction rate is the same as the one in Eq.~\ref{eqn:rateCEvNS}, but one needs to account for CE$\nu$NS on the three elements present in the crystals of PbWO$_4$. Besides that, for simplicity, we assume perfect detection efficiency and energy resolution, and a threshold of $1$~keV~\cite{RES-NOVA:2021gqp}.

\section{Solar neutrinos in the presence of invisible  neutrino decay}
\label{sec:Solar}
In this section, we introduce the signal expected from solar neutrinos and explore how such a signal is modified in the presence of invisible neutrino decay. We then present the bounds from solar neutrinos on invisible neutrino decay expected from the next-generation neutrino observatories Hyper-Kamiokande, DUNE, and DARWIN.

\subsection{Solar neutrino conversion probabilities}

Neutrinos are copiously produced in the nuclear reactions powering the Sun~\cite{Gann:2021ndb}. A flux of neutrinos with energy up to $\sim 18$~MeV results from the effective reaction $4p \, + \, 2e^- \, \longrightarrow \, ^4\textrm{He} \, + \, 2 \nu_e$, consisting of several  subprocesses, whose neutrinos are dubbed \textit{pp}, $^7$Be, \textit{pep}, $^8$B and \textit{hep} neutrinos, depending on the specific production channel. Additionally, neutrinos are also produced in the CNO fusion cycle, as confirmed by  Borexino~\cite{BOREXINO:2020aww, BOREXINO:2023ygs}. 

\begin{figure}
    \centering
    \includegraphics[width = \textwidth]{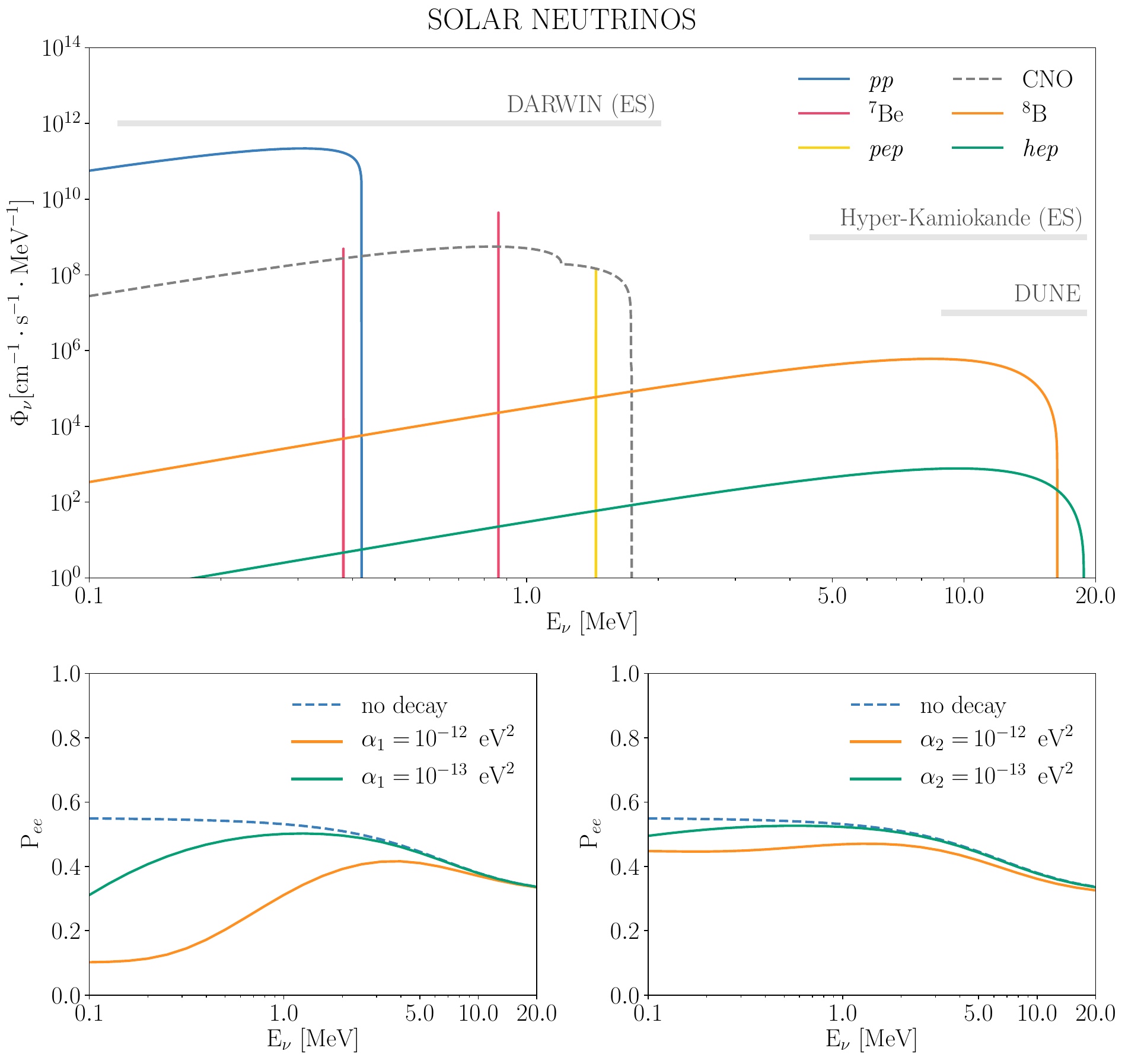}
    \caption{{\it Top panel:} Solar neutrino flux expected  at Earth for the different neutrino production channels (\textit{pp}, $^7$Be, \textit{pep}, CNO, $^8$B, and \textit{hep}), according to SSM GS98-16. The neutrino energy ranges explored by DARWIN (ES), Hyper-Kamiokande (ES) and DUNE are highlighted through grey bands for orientation. {\it Bottom panels:} Electron neutrino survival probability as a function of the neutrino energy for different choices of the decay parameters $\alpha_1$ (bottom left panel) and $\alpha_2$ (bottom right panel). For comparison, the survival probability in the absence of  invisible neutrino decay is displayed (blue dashed line). In all cases, the probability is computed for the SSM GS98-16 and weighted according to the contribution of each chain to the total solar neutrino flux for each neutrino energy. One can see that invisible neutrino decay induces an overall reduction of the survival  probability; this effect is  inversely proportional to the neutrino energy. }
    \label{fig:solar-prob}
\end{figure}

The predicted flux of solar neutrinos is given by the Standard Solar Model (SSM). In this work, we consider the high-metallicity SSM B16-GS98~\cite{Vinyoles:2016djt}, which is favoured by Borexino data~\cite{BOREXINO:2023ygs}. This choice of SSM also determines the uncertainty in the flux normalisation of each reaction chain. For illustrative purposes, we display the contributions to the total solar neutrino flux at Earth in the top panel of Fig.~\ref{fig:solar-prob}.

The Mikheyev-Smirnov-Wolfenstein (MSW) effect~\cite{Wolfenstein:1977ue,Mikheyev:1985zog} and the loss of coherence along propagation are responsible for the energy-dependent flavour composition of the solar neutrino flux detected on Earth~\cite{Maltoni:2015kca}. In the presence of neutrino decay, the picture is slightly modified. Assuming that neutrinos can only decay en route to Earth but not inside the Sun, the expected electron-neutrino oscillation probabilities to the flavour $\alpha$ are given by
\begin{align}
    P_{e\alpha} = \sum_i |U_{\alpha i}|^2 |\mathcal{A}(\nu_e \rightarrow \nu_i)|^2 e^{- i \alpha_i d_\odot/E_\nu} \, .
    \label{eq:solar-prob}
\end{align}
Here, $U_{\alpha i}$ denotes the elements of the lepton mixing matrix, $\mathcal{A}(\nu_e \rightarrow \nu_i)$ is the amplitude of conversion of $\nu_e$ into the mass eigenstate $\nu_i$ at the surface of the Sun for a given neutrino energy $E$, $d_\odot = 1.5 \times 10 ^8$~km is the distance of the Sun from Earth, and $\alpha_i$ are the decay parameters introduced in Eq.~\ref{eq:params}. The argument of the exponential in Eq.~\ref{eq:solar-prob} is
\begin{align}
\frac{\alpha_i d_\odot}{E_\nu}   \simeq 0.76\left(\frac{\alpha_i}{10^{-12}\, \text{eV}^2} \frac{1 \text{MeV}}{E_\nu}\right)\, ,
\label{eq:solar-alpha-exp}
\end{align}
which allows to estimate the order of magnitude of the decay parameters $\alpha_i$ that solar neutrinos can test.

We compute flavour conversions numerically and take into account the different production regions for each of the reactions responsible for the solar neutrino flux. For the oscillation parameters, we adopt the best-fit values from Ref.~\cite{deSalas:2020pgw}. 

For simplicity, we  neglect electron-neutrino regeneration as a consequence of matter effects on Earth, responsible for the so-called day-night asymmetry~\cite{Bouchez:1986kb,Cribier:1986ak,Akhmedov:2004rq}. Such matter effects become  more prominent as the  energy increases and can be neglected  below $3$~MeV. On the other hand,  the signature of neutrino decay is more evident as the neutrino energy decreases. Since the most stringent limits on neutrino decay are expected from the study of \textit{pp} neutrinos, with energy below $0.5$~MeV, it is safe to neglect matter effects on Earth.~\footnote{Note that, for the most energetic solar neutrinos, the day-night asymmetry due to  Earth matter effects is of the order of $2$--$3\%$~\cite{Super-Kamiokande:2023jbt}.}  

As a result of the reduction in the survival and flavour conversion probabilities, the main signature of  invisible neutrino decay in solar neutrino experiments is a reduction of the measured flux. As one can see from Eq.~\ref{eq:solar-prob}, the effect is inversely proportional to the neutrino energy. This can be seen in Fig.~\ref{fig:solar-prob}, where  the dependence of the $\nu_e$ survival probability on the decay parameters $\alpha_1$ and $\alpha_2$ is illustrated.
Note that the fraction of $\nu_3$ that arrives at Earth is approximately $\sin^2\theta_{13} \sim 0.02$ . Then, even if all $\nu_3$ decayed invisibly on their way to Earth, the variation in the total flux would be negligible given the uncertainties from the SSM predictions on the flux normalisation. Therefore, one does not expect any meaningful bound on $\tau_3/m_3$ from solar neutrino data for invisible neutrino decay.~\footnote{It is, however, possible to constrain the visible decay of $\nu_3$ using solar data, see Refs.~\cite{Funcke:2019grs, Picoreti:2015ika}.}

\subsection{Expected  solar neutrino event rate}
\begin{figure}
    \centering
    \includegraphics[width = \textwidth]{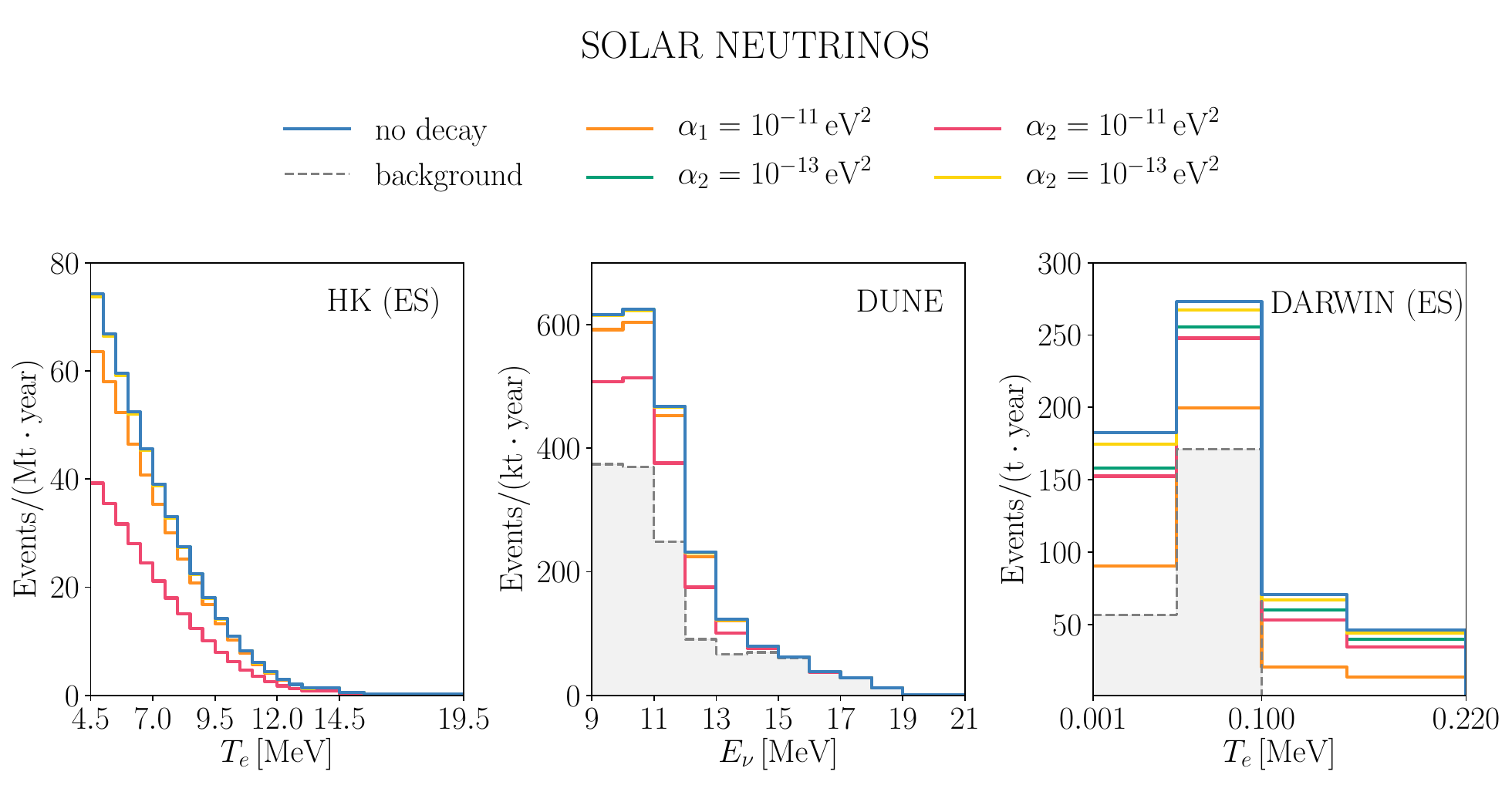}
    \caption{Number of  solar neutrino events and background events expected at Hyper-Kamiokande (left panel),  DUNE (middle panel) and DARWIN (right panel), respectively, as  a function of the electron recoil energy ($T_e$) or the reconstructed neutrino energy ($E_\nu$). The number of background events is also shown separately for comparison. }
    \label{fig:solar-events}
\end{figure}
Figure~\ref{fig:solar-events} shows the  prospects for solar neutrino detection at the next-generation observatories Hyper-Kamiokande, DUNE, and DARWIN. 
As its predecessor, Super-Kamiokande, Hyper-Kamiokande will be sensitive to $^8$B and \textit{hep} neutrinos from the Sun via ES on electrons~\cite{Yano:2021usb}. Our analysis considers the flux from both chains  and the corresponding electron-neutrino survival and non-electron neutrino appearance probabilities. 
We perform the analysis in terms of the ratio between the expected event rate and the event rate in the absence of flavour conversions and decays.~\footnote{Note that, following Refs.~\cite{Martinez-Mirave:2021cvh,Barenboim:2023krl}, working in terms of this ratio allows for the scaling of the statistical uncertainties with respect to those reported for Super-Kamiokande~\cite{Nakano:2016uws}.} We  account for energy-uncorrelated systematic uncertainties of the same order of magnitude as in Super-Kamiokande IV~\cite{Nakano:2016uws}, which encode the impact of backgrounds and other uncertainties in the analysis, and consider an energy threshold of $4.5$~MeV~\cite{Barenboim:2023krl}.
In the left panel of Fig.~\ref{fig:solar-events}, we display the expected number of events with and without  neutrino decay for Hyper-Kamiokande as a function of the electron recoil energy, assuming perfect detection efficiency. 

The low-energy physics program of  DUNE features the investigation of solar neutrinos from $^8$B and \textit{hep} nuclear chains~\cite{Capozzi:2018dat,DUNE:2020ypp,Barenboim:2023krl}. 
Our expected number of events is displayed in the middle panel of Fig.~\ref{fig:solar-events}.
We account for backgrounds from $^{222}$Rn and neutron capture with a free normalisation and $90\%$ detection efficiency across all energies~\cite{DUNE:2020ypp}. Moreover, we set a low-energy threshold of $9$~MeV for the reconstructed neutrino energy~\cite{Barenboim:2023krl}, since below those energies the backgrounds would completely overwhelm the signal. 

DARWIN stands out for its sensitivity to low-energy neutrinos from \textit{pp} thermonuclear reactions in the Sun via neutrino ES  on electrons~\cite{DARWIN:2020bnc,deGouvea:2021ymm}, see the right panel of Fig.~\ref{fig:solar-events}. 
Note that, in addition, DARWIN would  be sensitive to $^7$Be neutrinos via ES and $^8$B neutrinos via CE$\nu$NS; however, lower statistics is expected from these channels~\cite{DARWIN:2020bnc}.
We take into account the background from $^{124}$Xe decays via double electron capture, assuming an abundance of $0.1\%$ for this isotope~\cite{DARWIN:2020bnc}. Relevant background contributions are also expected from $^{136}$Xe, which undergoes double-beta decays~\cite{DARWIN:2020jme}. However, they could be removed by isotopic depletion and hence we neglect  them~\cite{DARWIN:2020bnc,deGouvea:2021ymm}.

In what follows, we do not  consider the number of events expected from solar neutrinos in  RES-NOVA. In fact, since the energies of solar neutrinos are relatively low, most of the nuclear recoil energies are lower than realistic energy thresholds for this experiment~\cite{Suliga:2020jfa}.
Regarding JUNO, it will be sensitive to solar neutrinos, but its sensitivity to $^8$B neutrinos will not be competitive with the one of Hyper-Kamiokande~\cite{JUNO:2022jkf}. JUNO is also expected to further reduce the uncertainties in the measurement of $^7$Be and $pep$ neutrinos~\cite{JUNO:2023zty}. However, as shown in the next section, the bounds on neutrino decay are  dominated by the measurement of the less energetic $pp$ neutrinos.

\subsection{Constraints on invisible neutrino decay from solar neutrinos}

\begin{figure}
    \centering
    \includegraphics[width = \textwidth]{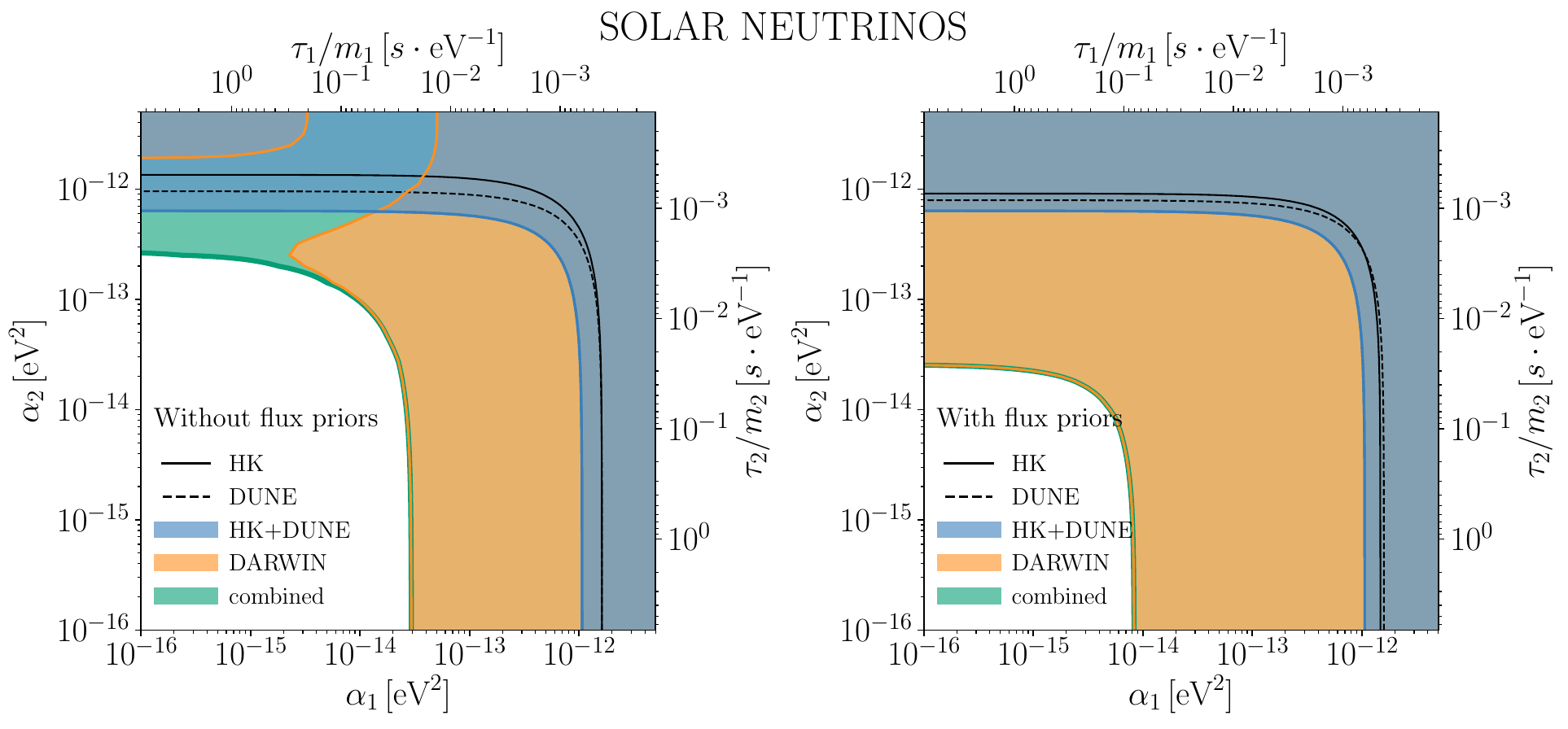}
    \caption{Projected sensitivity to the neutrino decay parameters $\alpha_1$ and $\alpha_2$ at $90\%$ C.L. The left and right panel correspond to the results without and with priors on the normalisation of the solar neutrino flux, respectively.  Solid and dashed black lines correspond to the  results for Hyper-Kamiokande (HK) and DUNE. Their joint constraints are represented by  the blue-shaded exclusion region. The orange-shaded region would be excluded by DARWIN, whereas the green region corresponds to the exclusion prospects from a combined analysis of all three neutrino observatories. }
    \label{fig:solar-decay}
\end{figure}

In our analysis, we assume an exposure of $300$~t$\cdot$year  for DARWIN and $400$~kt$\cdot$year for DUNE. For Hyper-Kamiokande, we consider an exposure of 3740 kt$\cdot$year. In all three cases, the exposures correspond to $10$~years of data taking with the full detector configuration.
We  follow two  approaches. In the first one, we allow for the absolute flux normalisation of each neutrino chain to vary freely. In the second one, we include priors on such flux normalisations, mainly $\sigma_{pp} = 0.5\%$, $\sigma_{^8\text{B}} = 4\%$ and $\sigma_{hep} = 30\%$, based on the uncertainties of the SSM adopted~\cite{Vinyoles:2016djt}. The sensitivity of each experiment and a combined fit is presented in Fig.~\ref{fig:solar-decay}, both with and without priors.
For further details of the statistical analysis, we refer the reader to Appendix~\ref{sec:stats}.

Note that we do not explore the  degeneracy between the oscillation and decay parameters~\cite{Beacom:2002cb}. In fact,  the medium-baseline reactor experiment JUNO will provide an accurate determination of the solar and reactor mixing angles, as well as the two mass splittings~\cite{JUNO:2015zny}. With an average baseline of $\sim 52$~km and given the current constraints in Eq.~\ref{eqn:current-solar-1},  the determination of the oscillation parameters by JUNO will be independent of neutrino decay. 

A joint analysis of Hyper-Kamiokande and DUNE solar data will provide an accurate determination of the $^8$B flux, since using two detection channels allows to break the degeneracy between the $^8$B flux normalisation and the solar mixing angle~\cite{Barenboim:2023krl,Martinez-Mirave:2023fyb}. 
The reason is that the ES signal in Hyper-Kamiokande provides complementary information on the flavour composition of the flux and its absolute normalisation, while DUNE is only sensitive to electron neutrinos.
Hence, one expects a joint fit to significantly improve the results with respect to the individual experimental capabilities and break the degeneracy with the solar mixing angle~\cite{Capozzi:2018dat}.

Concerning the sensitivity of DARWIN, there exists a large degeneracy between the decay parameters and the absolute $pp$-neutrino flux normalisation. As shown in the left panel of Fig.~\ref{fig:solar-decay}, without prior information on the $pp$ flux normalisation, DARWIN constraints on $\alpha_1$ and $\alpha_2$ are compatible with $\nu_2$ almost decaying  completely.
Notice, however, that a combined fit with Hyper-Kamiokande and DUNE would exclude that region of the parameter space. Conversely, if one includes a prior on the $pp$ flux normalisation, the bounds from DARWIN become significantly more stringent, as shown in the right panel of Fig.~\ref{fig:solar-decay}. In this case, a combined fit with $^8$B neutrino data would not be of such relevance as DARWIN on its own  dominates the constraints. 

The expected exclusion regions from the combined analysis at $90\%$ C.L.~for one non-zero decay width at a time are
\begin{align}
\alpha_1 < 6.4 \times 10^{-15} \, \text{eV}^2 \quad \text{ and } \quad \alpha_2 < 1.9 \times 10^{-14} \, \text{eV}^2\, ,
\end{align}
or analogously,
\begin{align}
        \frac{\tau_1}{m_1} > 0.10 \, \textrm{s}/\textrm{eV} \, \quad \textrm{and} \quad \frac{\tau_2}{m_2} > 3.5 \times 10^{-2} \, \textrm{s}/\textrm{eV} \, .
\end{align}
These limits are more than one order of magnitude more stringent than the current ones. This highlights the potential of DARWIN.  Note that existing sensitivity studies on neutrino decay in xenon-based experiments consider a smaller energy range for the study of the signal and lower exposures~\cite{Huang:2018nxj}, leading to weaker constraints than the ones presented in this work.

\section{Supernova neutrinos in the presence of invisible neutrino decay}
\label{sec:CCSN}
In this section, we introduce the signal expected from CCSN neutrinos  and explore how it is modified by invisible neutrino decay. We then present the bounds from CCSN neutrinos on invisible neutrino decay, investigating the impact of the uncertainties on the neutrino flavour conversion physics in the stellar envelope, as well as the uncertainties on the CCSN properties and its distance. We explore the sensitivity of Hyper-Kamiokande, DUNE, RES-NOVA and DARWIN, as well as the impact of a combined analysis. We also present bounds on invisible neutrino decay obtained relying on the SN 1987A neutrino observations.

\subsection{Supernova neutrino flux}
Despite the significant progress made in recent years on the theory underlying the collapse of massive stars, our understanding remains incomplete, especially for what concerns the expected neutrino signal~\cite{Mirizzi:2015eza,Tamborra:2020cul,Richers:2022zug, Mezzacappa:2022hmk,Mezzacappa:2020oyq,Janka:2017vlw}. In order to derive relevant bounds on invisible neutrino decay, we choose to focus on the  neutronisation burst, which is with good approximation insensitive to astrophysical uncertainties~\cite{OConnor:2018sti, Kachelriess:2004ds}. 

In order to take into account variations in the neutrino emission properties due to the supernova mass, we rely on one-dimensional spherically symmetric hydrodynamical simulations~\cite{Mirizzi:2015eza,garching} and  consider CCSN models with mass of $11.2 M_\odot$ and $27 M_\odot$. In addition, we use a black-hole forming collapse model with mass of $40 M_\odot$ (model s40s7b2)~\cite{Mirizzi:2015eza}. All models employ Lattimer and Swesty equation of state with a nuclear incompressibility modulus of $K = 220$~MeV~\cite{Lattimer:1991nc}.

The flavour composition of the observed neutrino signal depends on neutrino interactions with the medium and among themselves~\cite{Mirizzi:2015eza}. During the neutronisation burst, it is expected that the neutrino signal should be mostly affected by MSW conversion~\cite{Dighe:1999bi}. Hence, we have:
\begin{align}
    \Phi_{\nu_e} &= \texttt{p}\, \Phi^0_{\nu_e} + (1 - \texttt{p})\,\Phi^0_{\nu_x}\, , \\
    \Phi_{\bar{\nu}_e} &=  \bar{\texttt{p}}\,\Phi^0_{\bar{\nu}_e} + (1 - \bar{\texttt{p}})\,\Phi^0_{\bar{\nu}_x}\, , \\
    \Phi_{\nu_x} &= \frac{1}{2}\left[(1-\texttt{p})\,\Phi^0_{\nu_e} + (1+\texttt{p})\,\Phi^0_{\nu_x}\right]\, , \\
    \Phi_{\bar{\nu}_x} &= \frac{1}{2}\left[(1-\bar{\texttt{p}})\,\Phi^0_{\bar{\nu}_e} + (1+\bar{\texttt{p}})\,\Phi^0_{\bar{\nu}_x}\right]\, ,
\end{align}
where  $\Phi_{\nu_\alpha}$ and $\Phi_{\nu_\alpha}$ denote the final number of neutrinos and antineutrinos of a given flavour $\alpha$, i.e.~for electron and non-electron flavours (the latter being denoted with $x$) and the superscript $0$ denotes the neutrino flux before flavour conversion. 
The probabilities $\texttt{p}$ and $\bar{\texttt{p}}$ are defined according to the neutrino mass ordering as~\cite{Dighe:1999bi}%
\begin{align}
    \texttt{p} = \left \{
  \begin{aligned}
    &|U_{e3}|^2, && (\textrm{normal ordering, NO}) \\
    &|U_{e2}|^2, && (\textrm{inverted ordering, IO})
  \end{aligned} \right. \quad  \text{and} \quad   \bar{\texttt{p}} = \left \{
  \begin{aligned}
    &|U_{e1}|^2, && (\textrm{NO}) \\
    &|U_{e3}|^2, && (\textrm{IO})
  \end{aligned} \right. \, ,
\end{align}
and we calculated them using the best-fit values of the oscillation parameters from Ref.~\cite{deSalas:2020pgw}.
In the light of  uncertainties on the physics linked to  neutrino-neutrino interactions, we also consider other two mixing scenarios, one with maximum flavour conversion within the constraints of conservation of the neutrino lepton number per flavour~\cite{Just:2022flt}:
\begin{align}    \Phi_{\nu_e} &= \frac{1}{3}\text{min}(\Phi^0_{\nu_e}, \Phi^0_{\bar{\nu}_e}) + \text{max}(\Phi^0_{\nu_e} - \Phi^0_{\bar{\nu}_e}) + \frac{2}{3}\text{min}(\Phi^0_{\nu_x}, \Phi^0_{\bar{\nu}_x}) \, ,\\
    \Phi_{\bar{\nu}_e} &= \frac{1}{3}\text{min}(\Phi^0_{\nu_e}, \Phi^0_{\bar{\nu}_e}) - \text{min}(\Phi^0_{\nu_e} - \Phi^0_{\bar{\nu}_e}) + \frac{2}{3}\text{min}(\Phi^0_{\nu_x}, \Phi^0_{\bar{\nu}_x}) \, ,\\
    \Phi_{\nu_x} &= \Phi_{\bar{\nu}_x} = \frac{1}{3}\text{min}(\Phi^0_{\nu_e}, \Phi^0_{\bar{\nu}_e}) + \frac{2}{3}\text{min}(\Phi^0_{\nu_x}, \Phi^0_{\bar{\nu}_x})\, ;
\end{align}
and the case with  no  flavour conversion for comparison.

Similarly to the case of solar neutrinos, one can estimate the order of magnitude of the decay parameters testable from a core-collapse supernova explosion from the fact that
\begin{align}
\frac{\alpha_i d}{E_\nu} \simeq 0.16\left(\frac{\alpha_i}{10^{-22}\, \text{eV}^2}\frac{d}{10 \, \text{kpc}} \frac{1~\text{MeV}}{E_\nu}\right)\, .
\label{eq:solar-alpha-exp}
\end{align}

\subsection{Expected supernova neutrino event rate }
\begin{figure}
\includegraphics[width=\textwidth]{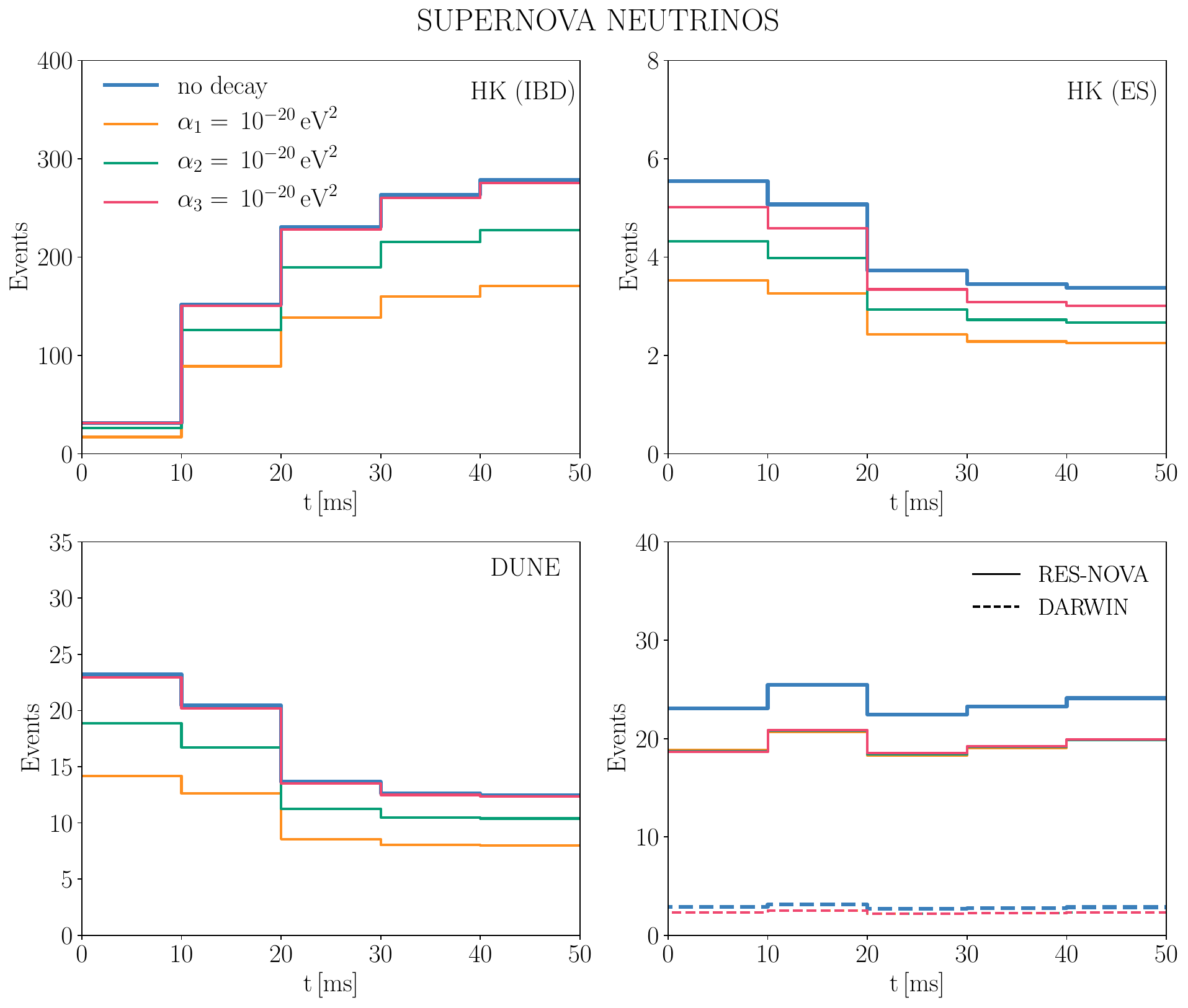}
\caption{Expected number of events  in time bins of $10$~ms for a $11.2 M_\odot$ stellar collapse occurring  at $10$~kpc from Earth. The top left and right panels correspond to the expected number of events for IBD and ES, respectively, in Hyper-Kamiokande. The bottom left panel displays the predicted number of events at DUNE and the bottom right shows the number of CE$\nu$NS events to be observed by RES-NOVA and DARWIN (solid and dashed lines, respectively). In all cases, inverted mass ordering is considered for illustrative purposes.}
\label{fig:sn-events}
\end{figure}

Figure~\ref{fig:sn-events} shows the expected sensitivity to  invisible decay of active neutrinos with  Hyper-Kamiokande, DUNE, DARWIN, and RES-NOVA for a CCSN with  mass of $11.2 M_\odot$ and  flavour composition resulting from the MSW effect for inverted ordering. Notice that, since the neutronisation burst lasts for about $50$~ms, we assume our analyses to be background free.
The main detection channel  in Hyper-Kamiokande is IBD, whose number of events for a  threshold of $3$~MeV in the positron energy~\cite{Hyper-Kamiokande:2018ofw} is displayed in the top left panel of Fig.~\ref{fig:sn-events}.
Hyper-Kamiokande is also sensitive to neutrinos and antineutrinos of all flavours via ES on electrons, as displayed in the top right panel of Fig.~\ref{fig:sn-events}. Its importance relies on the fact that it provides a measurement of the total flux.
For DUNE (bottom left panel of Fig.~\ref{fig:sn-events}), we consider a  threshold of $5$~MeV in the reconstructed neutrino energy~\cite{DUNE:2020zfm}. Being sensitive to electron neutrinos, DUNE is ideal to probe   the neutronisation burst.
The number of events expected in DARWIN and RES-NOVA is displayed in the bottom right panel of Fig.~\ref{fig:sn-events}. The employment of DARWIN and RES-NOVA is particularly relevant since these detectors are insensitive to the neutrino flavour and, therefore, allow to further constrain the hypothesis of  invisible neutrino decay.

We do not include JUNO in our forecast analysis  since its main detection channel is IBD, like in Hyper-Kamiokande, but with lower statistics due to the difference in their respective fiducial volume. Hence, our joint analysis would be dominated by the Hyper-Kamiokande constraints. However,  JUNO would be crucial  if the next galactic CCSN occurred before Hyper-Kamiokande was operational, as well as for real-time monitoring and to better discriminate the CCSN properties~\cite{JUNO:2023dnp}. 

Note that Ref.~\cite{Pompa:2023yzg} provided stronger projected sensitivity to $\alpha_1$ and $\alpha_2$ from Hyper-Kamiokande than the ones presented here. Such constraints rely on simplistic assumptions such as the accurate knowledge of the neutrino energy distribution and evolution of the neutrino signal  (including the accretion and cooling phases) as well as the flavour composition of the emitted flux.
In this work, we instead  account for uncertainties on the expected neutrino signal and  flavour composition focusing on the best understood phase of a CCSN: the neutronisation burst. 

\subsection{Constraints on invisible neutrino decay from supernova neutrinos}

\begin{figure}
\includegraphics[width = \textwidth]{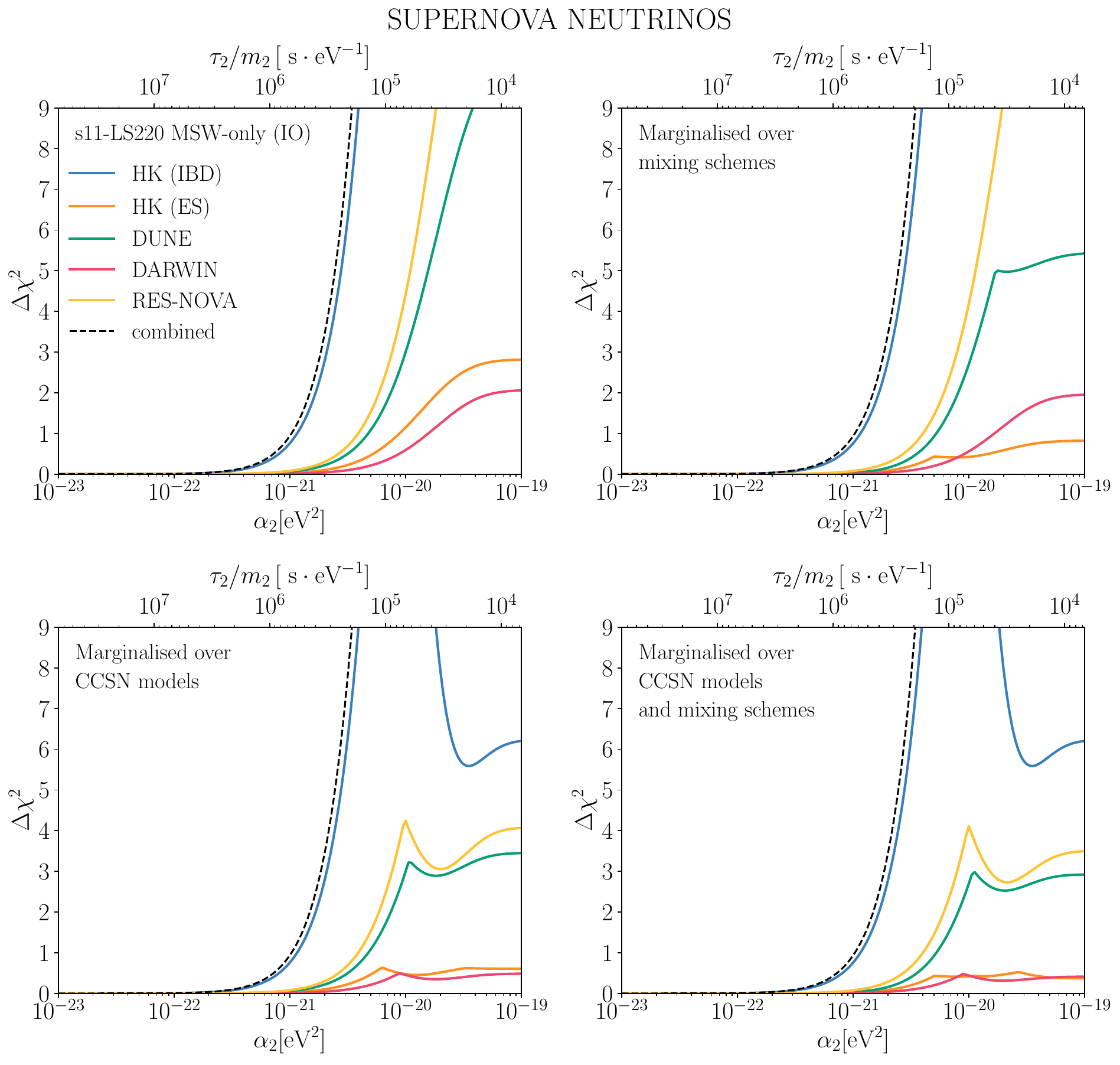}
\caption{Sensitivity to $\alpha_2$ for a stellar collapse occurring at $10$~kpc from Earth. We consider the following observatories:  Hyper-Kamiokande,  DUNE, RES-NOVA and DARWIN  corresponding to the blue, orange, green, red, and yellow curves, respectively. The outcome from the combined analysis is plotted with a black dashed line. The top left panel assumes that both the CCSN model and the flavour composition are known. The top right panel shows the result after marginalising over the mixing schemes,  whereas the bottom left panel corresponds to the results obtained  after marginalising over the CCSN models. The bottom right panel displays the sensitivity after marginalising over the CCSN models and mixing schemes. The uncertainties on the expected  neutrino signal  affect the  constraints  from different neutrino observatories. To this purpose it is crucial to exploit the complementary information provided by  the different detection channels to break the model  degeneracies and recover strong bounds on invisible neutrino decay. }
\label{fig:sn-uncertainties}
\end{figure}

A network of neutrino observatories sensitive to CCSN neutrinos  using different detection channels is necessary to derive meaningful constraints which account for the uncertainties intrinsic to the signal~\cite{Bendahman:2023hjj}. Focusing on the limits on $\alpha_2$ as an example, Fig.~\ref{fig:sn-uncertainties} illustrates the potential of a combined analysis of the data collected at different experiments for a CCSN with a mass of $11.2\ M_\odot$  occurring at $10$~kpc, and neutrino mixing resulting from the MSW effect for inverted ordering (cf.~Appendix~\ref{sec:stats} for  details on the statistical analysis).

The  top left panel of Fig.~\ref{fig:sn-uncertainties} shows  the sensitivity of each experiment to the decay parameter $\alpha_2$, if the emitted flux and  flavour composition were perfectly known. The top right panel displays the  expected sensitivity, if one knew the neutrino emission,   marginalising over the different mixing schemes introduced previously. Conversely, in the bottom left panel, we assume that the flavour composition is known and marginalise over our three  CCSN models. Finally, the bottom right panel shows how the  sensitivity of each  experiment worsens once one marginalises over the  mixing schemes and CCSN models. 

Figure~\ref{fig:sn-uncertainties}  highlights how the existing uncertainties on the CCSN neutrino signal  affect the  constraints inferred from different neutrino observatories. However, one can see that, in a combined analysis, strong bounds are recovered since the complementary information provided by the different detection channels breaks the existing degeneracies. 

Among the low-energy astrophysical signals  considered in this work, the neutrino signal from a galactic CCSN  is the only one  that in principle allows to set constraints on the decay of the three neutrino mass eigenstates individually. Figure~\ref{fig:sn-results}  shows the expected sensitivity for the three decay parameters  for a CCSN at $10$~kpc, for the MSW  scenario in both mass orderings. These limits take into account the uncertainties in the neutrino emission and flavour composition, following the marginalisation procedure previously outlined:
\begin{figure}
\centering
\includegraphics[width=0.6\textwidth]{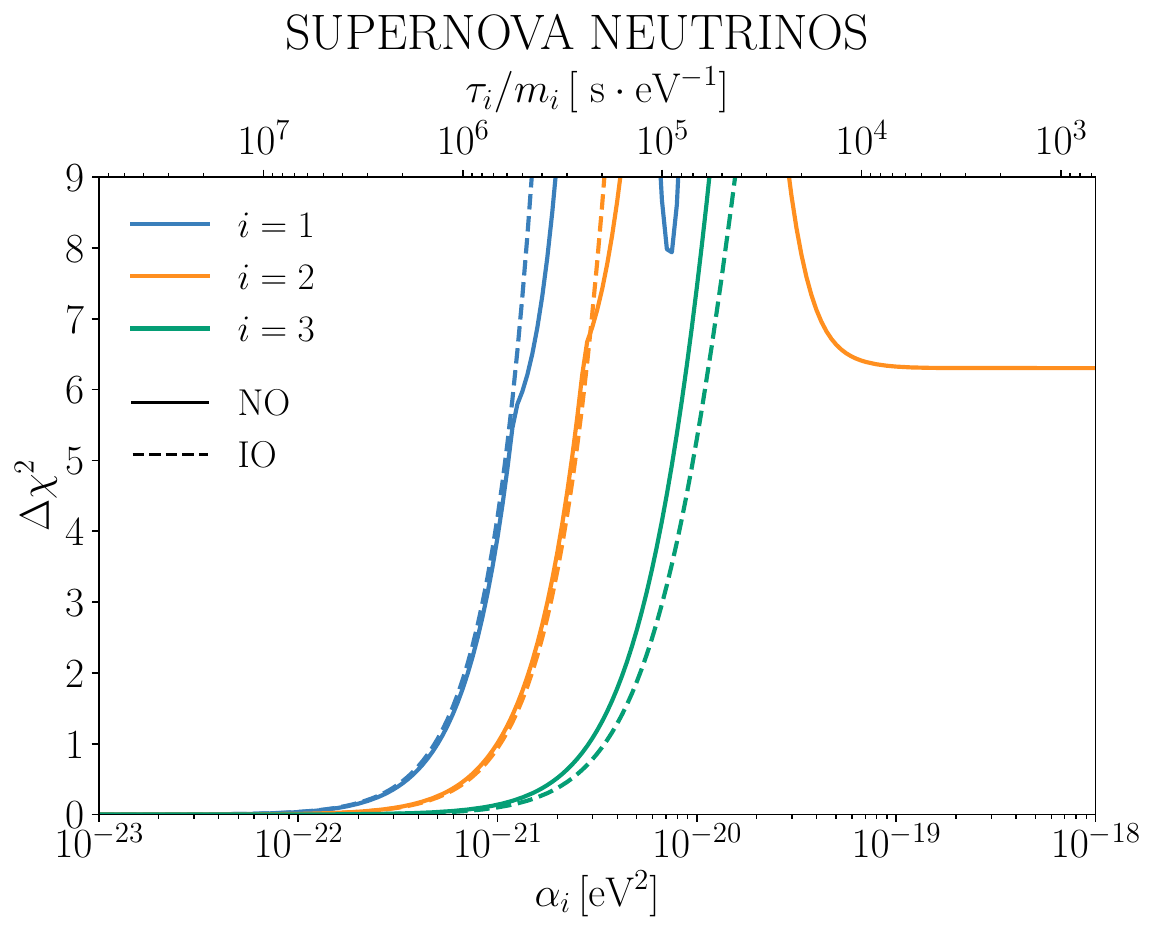}
\caption{Sensitivity to the decay parameters $\alpha_1$, $\alpha_2$ and $\alpha_3$ for the MSW  scenario in normal and inverted ordering for a CCSN at $10$~kpc. We rely on a combined analysis of the detection channels and neutrino observatories and marginalise with respect to the CCSN distance and model. }
\label{fig:sn-results}
\end{figure}
\begin{align}
    \alpha_1 &< 8.4 \,  (8.1) \times 10^{-22} \, \textrm{eV}^2 \, ,\\
    \alpha_2 &< 1.7 \, (1.8) \times 10^{-21} \, \textrm{eV}^2 \, ,\\
    \alpha_3 &< 5.2 \, (6.4) \times 10^{-21} \, \textrm{eV}^2\, ,
\end{align}
at $90\%$ C.L.~for the CCSN with a mass of $11.2 M_\odot$, for NO (IO) at a distance of $10$~kpc. Analogously, one can express these limits in terms of the lifetime-to-mass ratio
\begin{align}
    \tau_1/m_1 & >8.0 \,  (8.3) \times 10^{5} \, \textrm{s/eV} \, ,\\
    \tau_2/m_2 &> 3.9 \, (3.7) \times 10^{5} \, \textrm{s/eV} \, ,\\
    \tau_3/m_3 &> 1.3 \, (1.0) \times 10^{5} \, \textrm{s/eV}^2\, . 
\end{align}
Notice that, for the neutrino observatories considered in this work, the detection of IBD events in Hyper-Kamiokande drives the sensitivity to CCSN neutrinos. Then, the expected limits for $\alpha_1$ are stronger than for $\alpha_2$ and $\alpha_3$ because $\bar{\nu}_e$ has a larger fraction of $\bar{\nu}_1$ than $\bar{\nu}_2$ and $\bar{\nu}_3$.

\begin{figure}
    \centering
    \includegraphics[width = \textwidth]{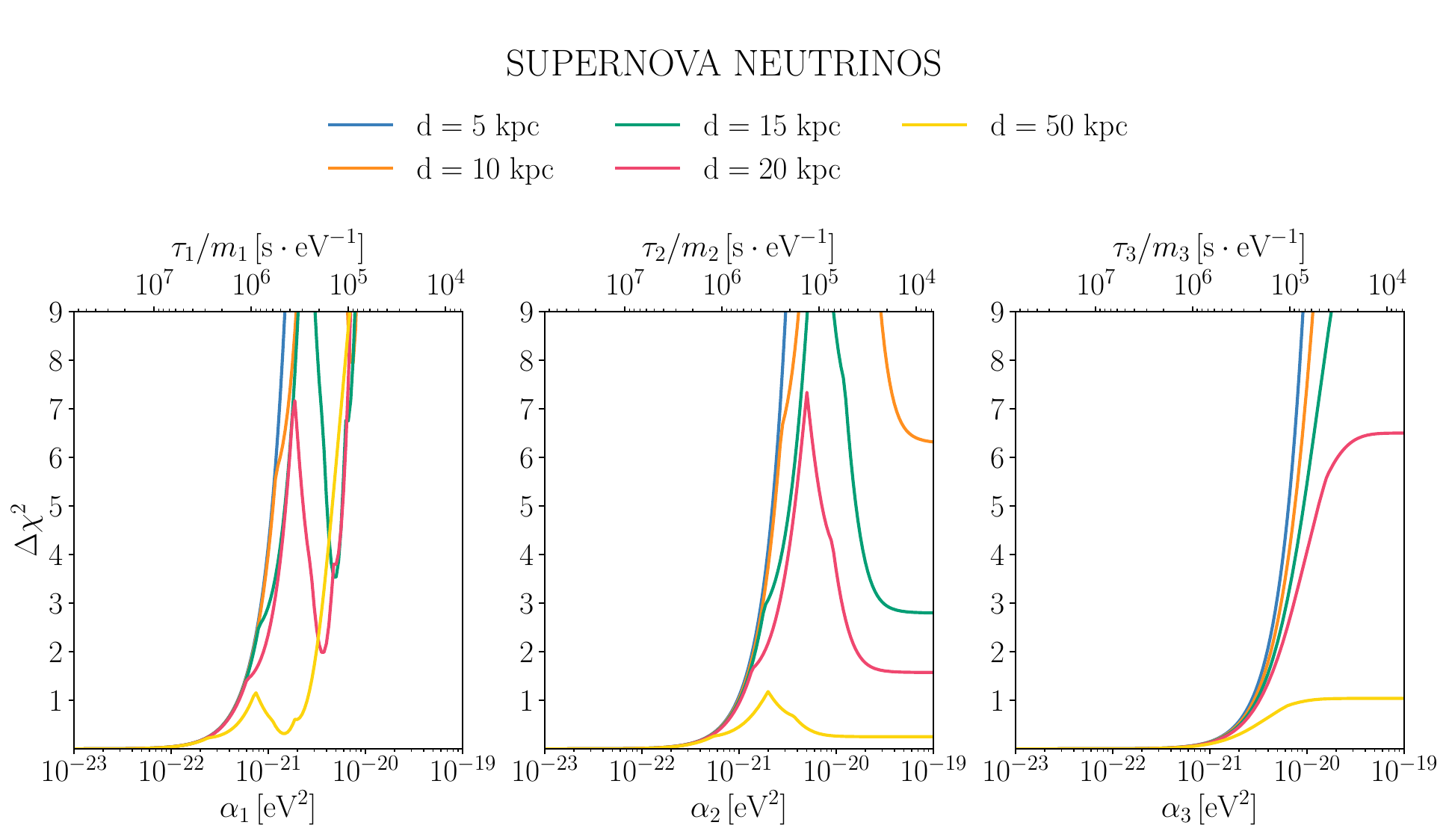}
    \caption{Dependence of the sensitivity on the three decay parameters $\alpha_1$ (left panel), $\alpha_2$ (middle panel), and $\alpha_3$ (right panel) on the CCSN  distance ($d = 5$, $10$, $15$, $20$, and $50$~kpc). The results assume a combined analysis of data collected at Hyper-Kamiokande, DUNE, RES-NOVA, and DARWIN after marginalising over the three SN models and the different mixing schemes. Note that the limits for $\alpha_2$ and $\alpha_3$ degrade significantly with the distance, while this is not the case for $\alpha_1$.}
    \label{fig:sn-distance}
\end{figure}

\begin{figure}
\includegraphics[width=\textwidth]{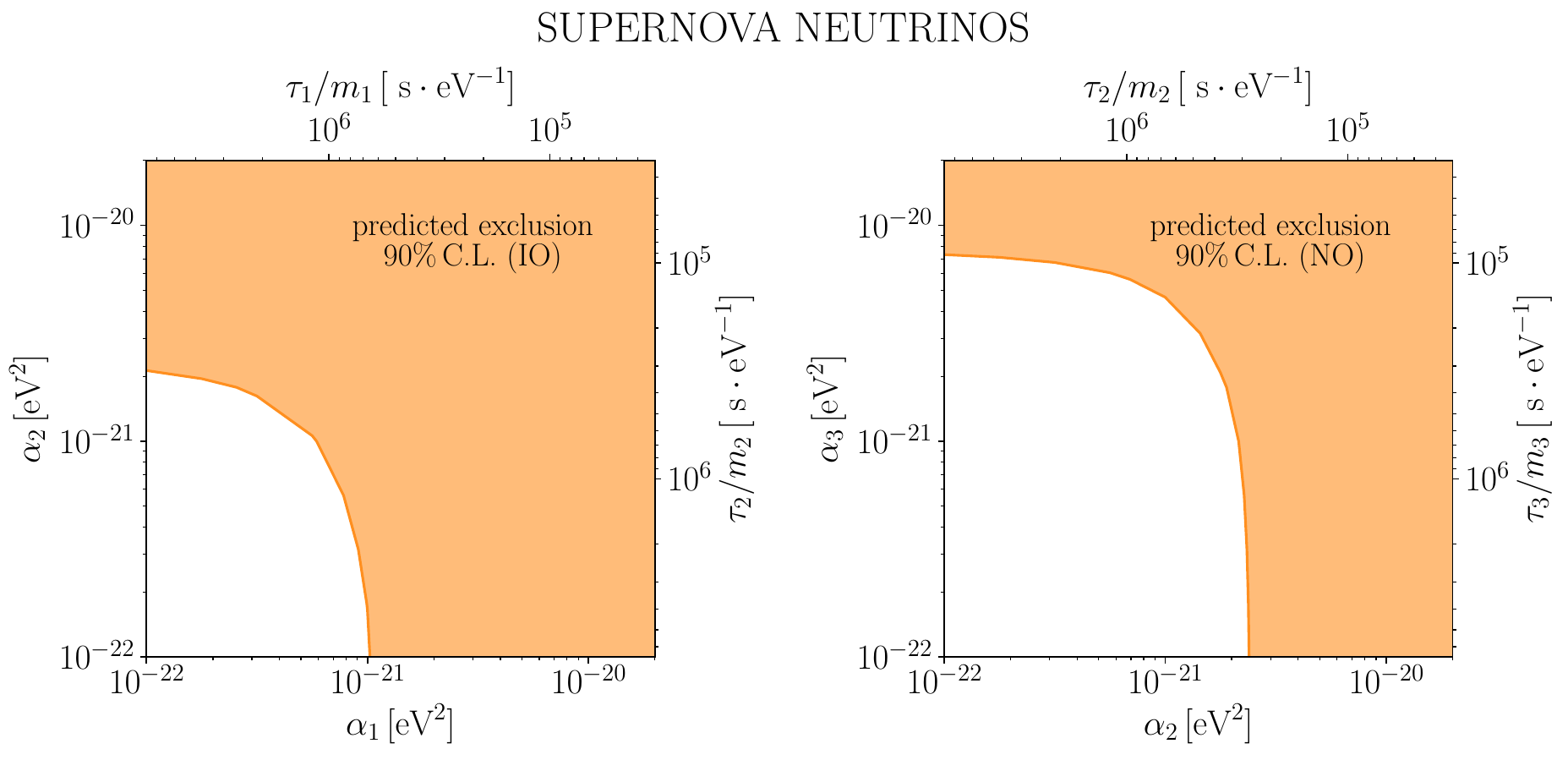}
\caption{Forecasted exclusion region at $90\%$~C.L.~from CCSN neutrinos in the plane spanned by  $\alpha_1$ and $\alpha_2$ (left panel) as well as $\alpha_2$ and $\alpha_3$ (right panel). Such predicted exclusion regions have been obtained after marginalising over the CCSN mass and mixing schemes.}
\label{fig:sn-2d}
\end{figure}

It is interesting to asses how the prospects of constraining invisible neutrino decay through CCSN neutrinos  would depend on the distance of the stellar collapse from Earth. To this end, Fig.~\ref{fig:sn-distance} shows the sensitivity to each of the three decay parameters from a combined analysis of Hyper-Kamiokande, DUNE, RES-NOVA, and DARWIN, assuming that the CCSN occurs at different distances ranging from $5$ to $50$~kpc. One can see that the limits for $\alpha_2$ and $\alpha_3$ degrade significantly with the distance. For $\alpha_2$, this is due to the challenges in  breaking the degeneracy between the mixing schemes with low statistics. For $\alpha_3$, the reason is that the limits are driven by RES-NOVA and DARWIN, which in general would collect fewer statistics than the IBD detectors.
Figure~\ref{fig:sn-2d} summarises the constraints on invisible neutrino decay in the planes ($\alpha_1$, $\alpha_2$) and ($\alpha_2$, $\alpha_3$). These limits assume a CCSN at $10$~kpc from Earth.

\subsection{Constraints on invisible neutrino decay from SN 1987A}
The detection of electron antineutrinos from SN 1987A has been employed  to constrain the neutrino lifetime~\cite{Hirata:1988ad}. In order to infer the lifetime for invisible neutrino decays, the observed time-integrated flux of electron antineutrinos from SN 1987A is approximately given by 
\begin{align}
    \Phi^{\rm SN1987A}_{\bar{\nu}_e} \simeq \Phi^0_{\bar{\nu}_e} \mathlarger{\mathlarger{\sum}}_{i=1,2,3} |U_{ei}|^2 \left[|U_{ei}|^2 + \frac{\Phi^0_{\bar{\nu}_x}}{\Phi^0_{\bar{\nu}_e}}(1 - |U_{ei}|^2)\right]e^{-\alpha_i D/\langle E\rangle} \, ,
    \label{eqn:sn1987A}
\end{align}
where $\Phi^0_{\bar{\nu}_e}$ and $\Phi^0_{\bar{\nu}_x}$ indicate the time-integrated fluxes of electron and non-electron antineutrinos emitted by SN 1987A (note that in this case we consider the whole time interval over which neutrinos have been observed, i.e.~$6$~s, and do not focus on the neutronisation burst only). The distance of  SN 1987A from Earth is estimated to be $D = 51.4$~kpc~\cite{1999IAUS..190..549P} and $\langle E \rangle = 12.5$ MeV is the average energy of the emitted antineutrinos from the events recorded up to $6$~s after bounce~\cite{Fiorillo:2023frv}.~\footnote{Notice that the measured distance to SN1987A depends on the method used. The choice of $D = 51.4$~kpc agrees well with other values in the literature~\cite{Cikota:2023ekm}.}

\begin{figure}
    \centering
    \includegraphics[width = 0.6\textwidth]{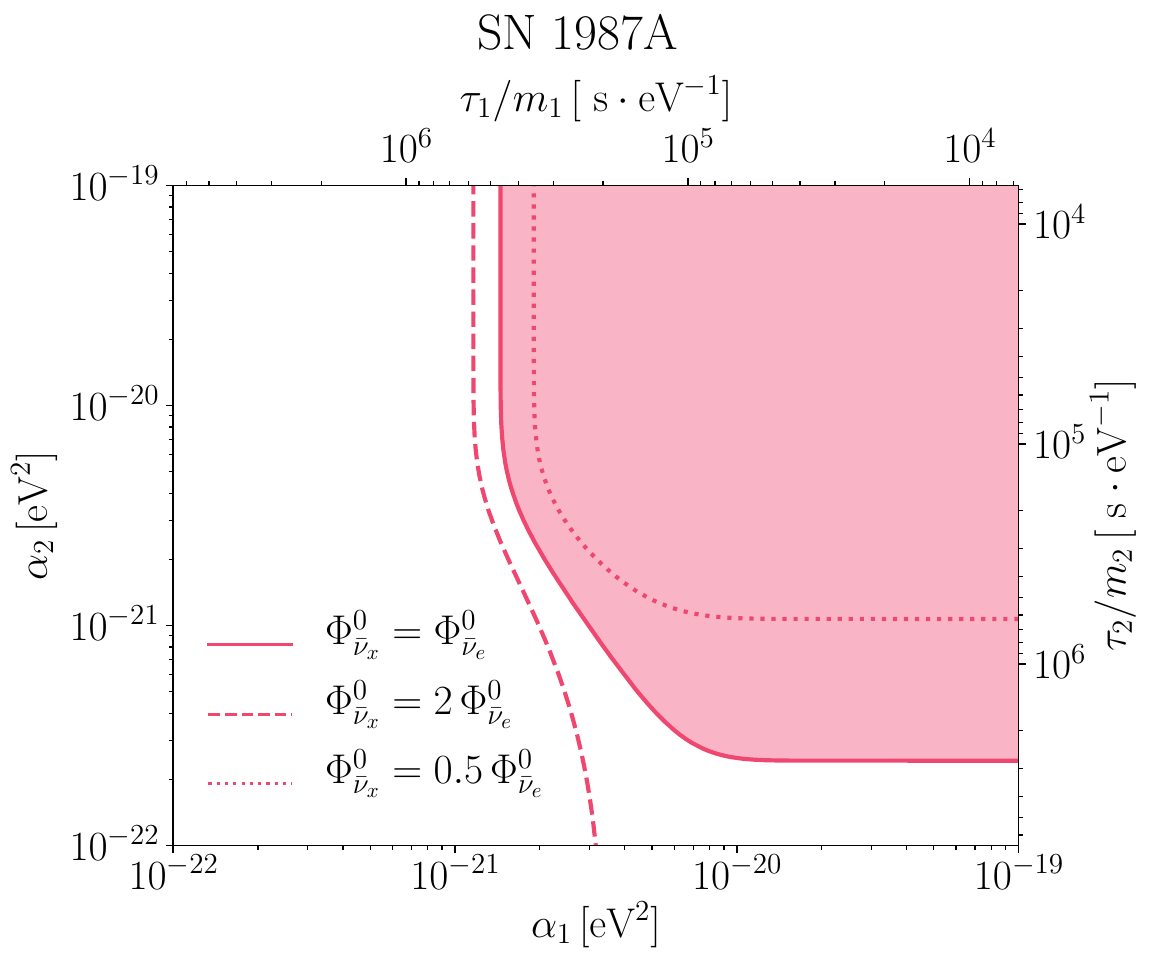}
    \caption{Exclusion region in the $\alpha_1$-$\alpha_2$ plane from the observation of SN 1987A electron antineutrinos. These bounds are shown for different choices of the ratio between the flux of electron and non-electron flavour antineutrinos emitted by the SN 1987A (cf.~solid, dashed and dotted contours). They  have been obtained  considering  the overall duration of the neutrino signal, instead of focusing on the neutronisation burst only and rely on the recent analysis of the SN 1987A data presented in Ref.~\cite{Fiorillo:2023frv} (see main text for details). }
    \label{fig:sn1987A}
\end{figure}

Figure~\ref{fig:sn1987A} shows our bounds on the invisible decay of $\bar{\nu}_1$ and $\bar{\nu}_2$. Such constraints have been obtained assuming that the total  energy emitted in $\bar\nu_e$'s from SN 1987A corresponds to the largest one obtained in the suite of simulations of Ref.~\cite{Fiorillo:2023frv},  i.e.~it is the total  energy of $\bar\nu_e$'s ($\mathcal{E}_{\bar{\nu}_e} \sim 7.5 \times 10^{52}$~erg) emitted from  the CCSN model with a $20 M_\odot$ progenitor, a proto-neutron star mass of $1.93 M_{\odot}$  and the SFHx version of Steiner, Fischer, and Hempel equation of state~\cite{Steiner:2012rk}. 
We contrast the total energy emitted in $\bar\nu_e$'s predicted from theory with the lowest value of the $\bar\nu_e$ energy allowed by the data in the analysis presented in Ref.~\cite{Fiorillo:2023frv}, $\mathcal{E}_{\bar{\nu}_e} \sim 2 \times 10^{52}$erg. Requiring that neutrino decay is responsible for the difference between the largest predicted $\bar\nu_e$ energy and the lowest bound on the observed one, we derive limits on the decay parameters following Eq.~\ref{eqn:sn1987A}. 

These limits on invisible neutrino decay depend on how large the emitted flux of non-electron antineutrinos is with respect to the one of electron antineutrinos. The impact of the assumptions of such ratio is shown in Fig.~\ref{fig:sn1987A}. We assume that the best-fit mean energy for $\bar\nu_e$'s  is the same as for the non-electron flavours, for simplicity. As expected, for a larger total neutrino emission (dashed line), the bounds become stronger.  
The asymmetry in the exclusion region in the plane $\alpha_1$--$\alpha_2$ displayed in Fig.~\ref{fig:sn1987A} results from the different fraction of $\bar{\nu}_1$ and $\bar{\nu}_2$ in $\bar{\nu}_e$. 

In Fig.~\ref{fig:summary}, we display the SN 1987A bound corresponding to equal luminosity for the three antineutrino flavours as representative case. Such exclusion regions are based on coarse assumptions and should be interpreted as indicative yet not strict limits. 

Finally,  since only electron antineutrinos were detected from SN 1987A, and those are mainly composed by $\bar{\nu}_1$ and $\bar{\nu}_2$, even if all $\bar{\nu}_3$  decayed  invisibly, the impact on the observed luminosity would be negligible. In the future, such limitation could be overcome through the inclusion of data from detection channels sensitive to non-electron flavours, i.e.~ES on electrons and CE$\nu$NS.

\section{Diffuse supernova neutrino background in the presence of invisible neutrino decay}
\label{sec:DSNB}
In this section, after modeling the DSNB signal expected at Hyper-Kamiokande, DUNE, and JUNO, we investigate the sensitivity of the DSNB to invisible neutrino decay  and its constraining power.

\subsection{Diffuse supernova neutrino background}
The DSNB is the  cumulative flux of neutrinos emitted from  all  CCSN explosions that  occur and have occurred in the Universe~\cite{Mirizzi:2015eza,Beacom:2010kk,Lunardini:2010ab,Vitagliano:2019yzm}. 
In order to model the DSNB, we rely on the CCSN models introduced in Sec.~\ref{sec:CCSN} and assume that flavour transformation is solely due to the MSW effect and  loss of coherence when travelling to Earth.

The DSNB  depends on the CCSN rate, $R_{\rm SN}$, which depends on the  redshift ($z$):
\begin{align}
    R_{\rm SN}(z, M)= \frac{\eta(M)}{\bigintss_{\, 0.5 M_\odot} ^{125M_\odot} \dd M \, M\eta(M)} \dot{\rho}_* (z) \, .
\end{align}
Here, we have introduced the initial mass function, $\eta(M)$, and the star formation rate, $\dot{\rho_*}$. For the former, we assume it follows the Salpeter law~\cite{Salpeter:1955it}, i.e.
\begin{align}
    \eta(M) \propto M^{-2.35}\, ,
\end{align}
whereas  the star-formation rate is defined as~\cite{Yuksel:2008cu}
\begin{align}
    \dot{\rho}_* \propto \left[(1 + z)^{-34} + \left(\frac{1+z}{5000}\right)^{3} + \left(\frac{1+z}{9}\right)^{35}\right]^{-0.1}\, .
\end{align}
In addition, we chose a normalisation of the supernova rate such that 
$\int_{8 M_\odot} ^{125 M_\odot} \dd M R_{\rm SN}(0,M) = (1.25 \pm 0.5)\times 10^{-4}\,\text{Mpc}^{-3} \text{yr}^{-1}$~\cite{Lien:2010yb}.

The DSNB  for each neutrino mass eigenstate is
\begin{align}
    \tilde{\Phi}_{\nu_i} (E) = &\frac{c}{H_0} \bigintss_{\, 8 M_\odot}^{125 M_\odot} \dd M \bigintss_{0}^{z_{\rm max}} \frac{R_{\rm SN}(z, M)}{\sqrt{\Omega_M (1+z)^3 + \Omega_\Lambda}} \Phi_{\nu_i} (E', M) \, ,
\end{align}
where $H_0 = 70\textrm{ km} s^{-1} \textrm{Mpc}^{-1}$~\cite{ParticleDataGroup:2022pth}, $c$ is the speed of light, and we adopt $z_{\rm max} = 5$. The dark matter and cosmological constant energy density are $\Omega_M = 0.3$ and $\Omega_\Lambda = 0.7$~\cite{Planck:2018vyg}. In addition, the redshifted neutrino energy at Earth is   $E' = E(1+z)$.
The DSNB  depends on the time-integrated neutrino flux $\Phi_{\nu_i} (E', M)$ for a given neutrino mass eigenstate.

In what follows, we adopt the following baseline DSNB model. We assume that CCSNe  with mass in the range $[8 M_\odot, 15 M_\odot]$ are well represented by our $11M_\odot$ model; for CCSNe whose mass is in the range $[15 M_\odot, 22 M_\odot]$ and $[25 
 M_\odot, 27 M_\odot]$, we take as a benchmark CCSN the model with  a mass of $27 M_\odot$. Finally, we assume that for CCSNe with mass in the ranges $[22 M_\odot, 25 M_\odot]$ and $[27 M_\odot, 125 M_\odot]$, the reference emission model is the  black hole forming collapse one with a mass of $40 M_\odot$. This parametrisation corresponds to   $21\%$  of all stellar collapses  leading to black hole formation, which is within theoretical and observational constraints~\cite{Kresse:2020nto,Adams:2016hit,Adams:2016ffj,Kochanek:2013yca}. 
 
\begin{figure}
    \centering
    \includegraphics[width = \textwidth]{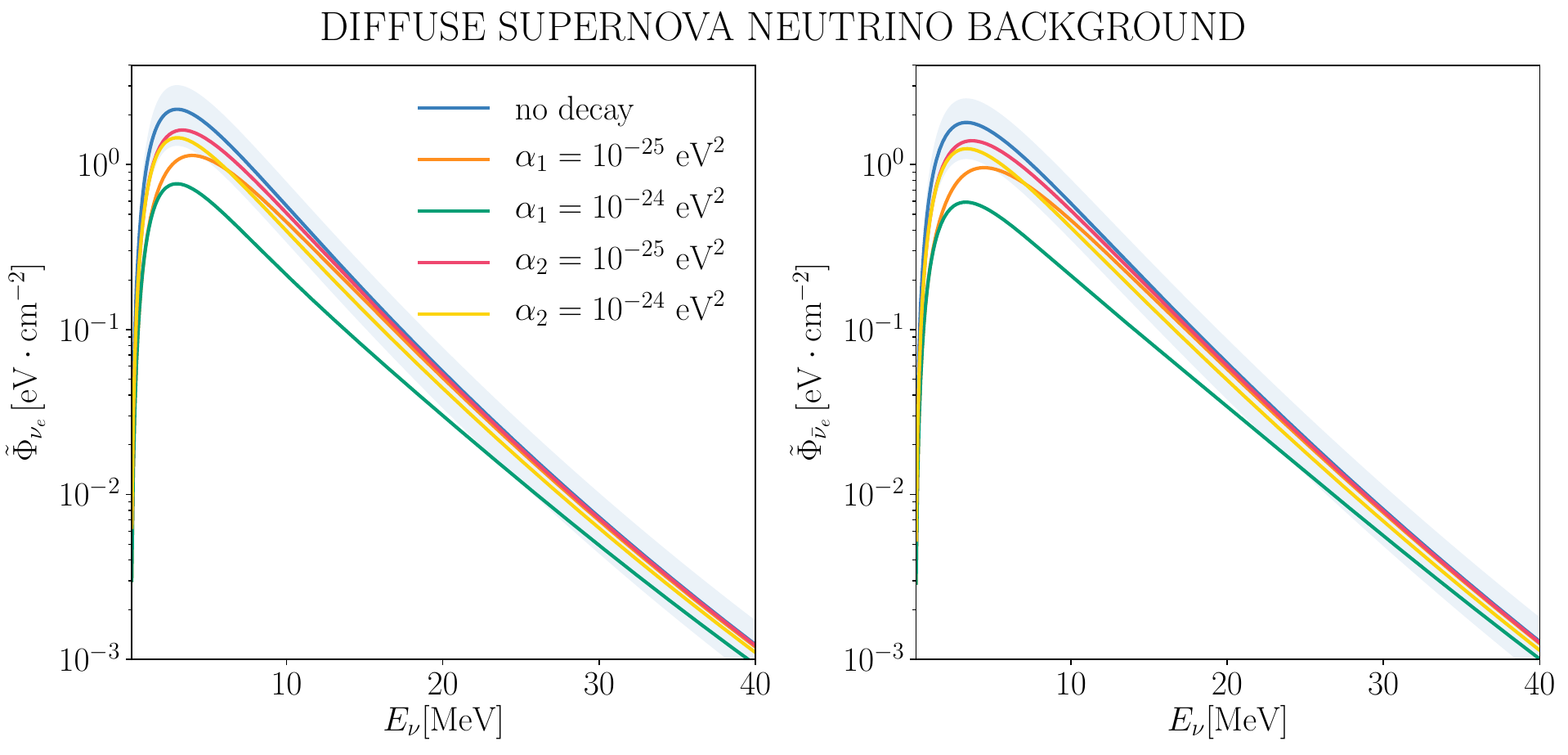}
    \caption{Electron neutrino (left panel) and antineutrino (right panel) DSNB fluxes as a function of the neutrino energy. The DSNB  in the absence of decay (blue) and for four values of the decay parameters $\alpha_1 = 10^{-25} \text{ eV}^2$ (orange), $\alpha_1 = 10^{-24} \text{ eV}^2$ (green), $\alpha_1 = 10^{-25} \text{ eV}^2$ (red), and $\alpha_1 = 10^{-25} \text{ eV}^2$ (yellow) is shown. The blue-shaded band corresponds to the uncertainty in the baseline prediction given the uncertainty in the CCSN rate. 
    Because of neutrino decay, the DSNB peak tends to decrease. The $\nu_e$  component of the DSNB flux is more sensitive to $\alpha_1$ than to $\alpha_2$. In addition,   even for large  $\alpha_2$,   the  DSNB flux falls within the uncertainty band of our baseline prediction; this implies that we cannot expect  stringent constraints on $\alpha_2$ from the DSNB.
    }
    \label{fig:dsnb-flux}
\end{figure}

 The DSNB electron neutrino and antineutrino signals (and their related uncertainty) in the absence of invisible neutrino decay are represented through the blue line (band) in Fig.~\ref{fig:dsnb-flux}.

Neutrino decay alters the expected DSNB spectrum~\cite{Ando:2003ie}. In particular, for  invisible decays, one expects a lower-than-nominal spectrally distorted flux. For each mass eigenstate, the expected flux is given by 
\begin{align}
    \tilde{\Phi}_{\nu_i} (E, \alpha_i) = &\frac{c}{H_0} \bigintss_{\, 8 M_\odot}^{125 M_\odot} \dd M \bigintss_{0}^{z_{\rm max}} \frac{R_{\rm SN}(z, M)}{\sqrt{\Omega_M (1+z)^3 + \Omega_\Lambda}} \Phi_{\nu_i} (E', M) e^{- \alpha_i \xi(z) /E} \, ,
\end{align}
where  the effective redshift-dependent parameter is~\cite{Fogli:2004gy}:
\begin{align}
    \xi(z) = \bigintss_{\,0} ^z \frac{\dd z'}{H_0\sqrt{\Omega_M (1+z')^3 + \Omega_\Lambda}} \frac{1}{(1+z')^2}\, .
\end{align}
Figure~\ref{fig:dsnb-flux} shows the  expected   DSNB spectra of $\nu_e$'s and $\bar\nu_e$'s at Earth for a few representative values of the decay parameters. 
Due to neutrino decay, the DSNB peak tends to decrease for increasing $\alpha_i$.
One can see that, since the fraction of $\nu_1$ in $\nu_e$ is larger than the one of $\nu_2$, the electron-neutrino component of the DSNB flux is more sensitive to $\alpha_1$ than  $\alpha_2$. Moreover,  despite the spectral difference, even for large values of $\alpha_2$, the  DSNB flux falls within the uncertainty band of our baseline prediction. Hence, we should not  expect stringent constraints  to be placed on $\alpha_2$ through the  DSNB, except for a  disfavoured region of the parameter space. Similar to what discussed earlier,  the decay of $\nu_3$ can not be constrained only from the  DSNB measurement of $\nu_e$ and $\bar\nu_e$. In fact, in this case,  one can not disentangle the predicted signatures from the existing astrophysical and experimental uncertainties.

\subsection{Expected diffuse supernova neutrino background event rate}

\begin{figure}
\includegraphics[width = \textwidth]{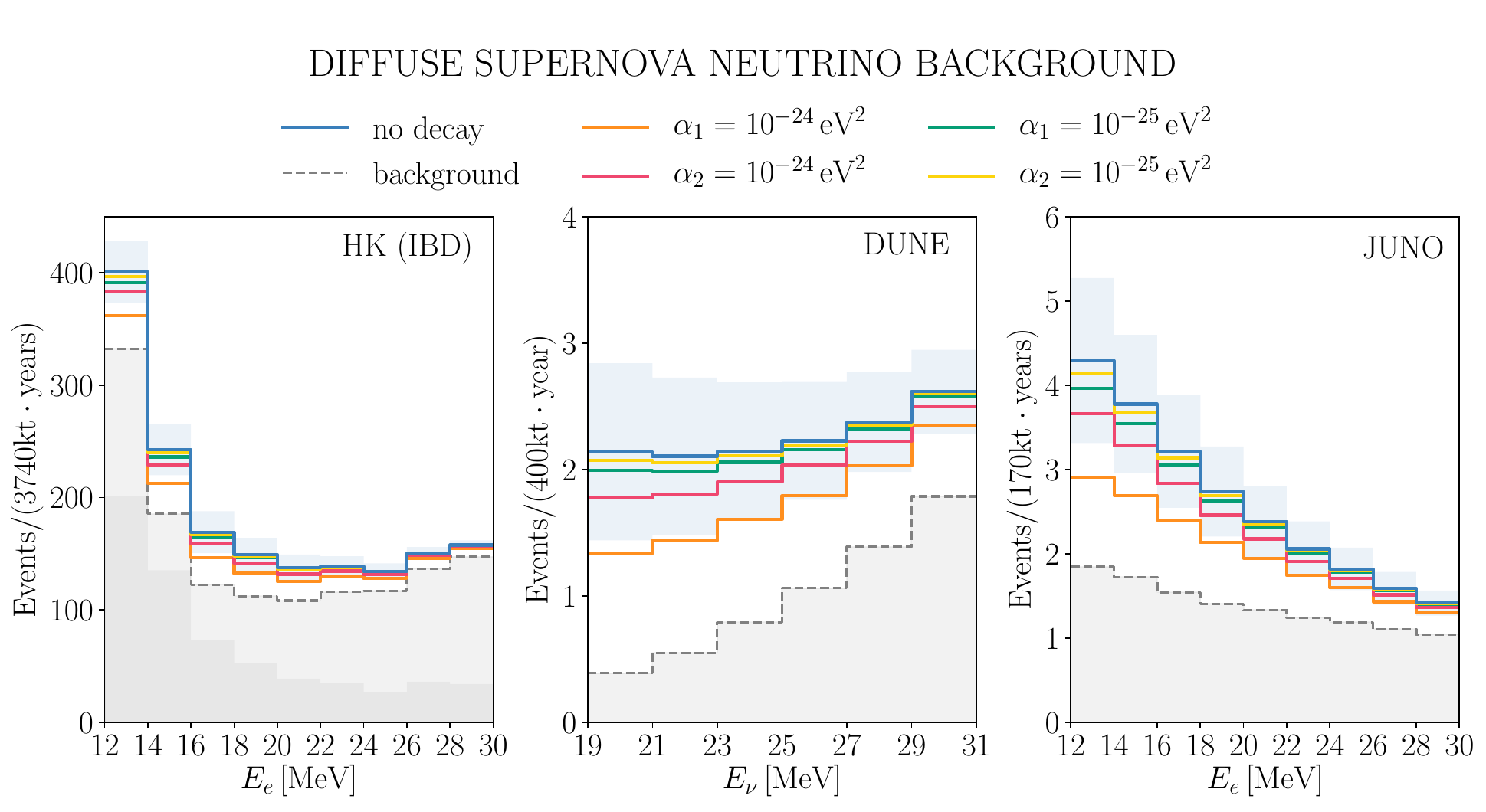}
\caption{Expected number of background and DSNB events in the energy region of interest at Hyper-Kamiokande, DUNE and JUNO (left, middle and right panels, respectively). The blue shaded band indicates the uncertainty in the number of signal events from the normalisation of the supernova rate. The fraction of background events is shaded in grey and marked with a grey dashed line, whereas the total number of events for different decay hypotheses is shown through the  solid colour lines. For comparison, we display the expected events for several values of the decay parameters $\alpha_1$ and $\alpha_2$. }
\label{fig:dsnb-events}
\end{figure}

Hyper-Kamiokande is  expected to be sensitive to electron antineutrinos from the DSNB via IBD. However, in contrast to a nearby CCSN  explosion, the backgrounds strongly limit the sensitivity to the DSNB. 
Atmospheric electron and muon neutrinos and antineutrinos are the main backgrounds for the DSNB detection~\cite{Hyper-Kamiokande:2018ofw}. Additionally, $^9$Li produced via spallation is another relevant background~\cite{Hyper-Kamiokande:2018ofw}. One should also take into account  neutral current atmospheric events, which are the dominant background for reconstructed positron energies in the range of $12$--$20$~MeV~\cite{Kunxian:2016joi,Ashida:2020erk}. It has been argued that neutral current atmospheric backgrounds could be almost completely removed with neutron tagging techniques~\cite{Maksimovic:2021dmz}. In the following, we report our findings with and without these backgrounds to illustrate their relevance for what concerns the bounds of invisible neutrino decay.  

At energies lower than $12$ MeV, the flux of reactor antineutrinos is much larger than the one expected for the DSNB. Therefore, we limit our analysis to the energy window between $12$  and $30$~MeV for the reconstructed positron energy. Note that, for Hyper-Kamiokande, atmospheric neutrinos can challenge  the DSNB  detection, hence dedicated investigations are required~\cite{Zhou:2023mou}. 

DUNE could observe the $\nu_e$ component of the DSNB. This measurement is of particular relevance to test the predicted flavour composition of the DSNB. In the case of DUNE, solar neutrinos dominate the signal up to energies close to $19$~MeV. The other main background that needs to be accounted for is due to atmospheric $\nu_e$'s~\cite{Cocco:2004ac}. These two backgrounds limit the region of interest in reconstructed neutrino energy to the range between $19$~MeV and $31$~MeV~\cite{DUNE:2020ypp}.

JUNO will  be sensitive to the DSNB, despite being significantly smaller than Hyper-Kamiokande. The charged current and neutral current  atmospheric backgrounds are not very prominent at lower energies after applying pulse-shape discrimination techniques~\cite{JUNO:2015zny,Mollenberg:2014pwa,Cheng:2023zds}. This is relevant to investigate neutrino decays, for which the lower the energy, the more prominent the signature of a decay is. The region of interest for the reconstructed positron energy is the same as for Hyper-Kamiokande. We conservatively assume a $50\%$ detection efficiency, although recent work suggests  the possibility of reaching an efficiency up to $80\%$~\cite{Cheng:2023zds}.

Figure~\ref{fig:dsnb-events} displays the  total number of events (both signal and background) for  the three detectors considered in this section. The blue shaded band indicates the uncertainty in the number of signal events from the normalisation of the supernova rate. In the left panel, which depicts the number of events expected in Hyper-Kamiokande, the darkest shading indicates the fraction of events corresponding to the neutral current atmospheric backgrounds, which might be reduced or even removed. In the same panel, the lightest shaded region indicates the fraction of events  corresponding to unavoidable backgrounds.  Similarly, the central and right panels show the expected number of events at DUNE and JUNO, respectively.

Other detection channels have been considered for the potential detection of the DSNB in all flavours, mainly CE$\nu$NS~\cite{Suliga:2021hek,Pattavina:2020cqc,Zhuang:2023dzd,Dutta:2019oaj}  and via neutrino interactions with  deuterium~\cite{SNO:2006dke}, carbon~\cite{Aglietta:1992yk}, oxygen~\cite{Lunardini:2008xd} or even ES  on electrons~\cite{Lunardini:2008xd}, but the prospects for constraining invisible neutrino decay are not encouraging. The reason is that the main signature of  invisible neutrino decay is an energy-dependent missing flux, yet only upper limits on the standard non-electron DSNB component could be extracted. 

\subsection{Constraints on invisible neutrino decay from the diffuse supernova neutrino background}

In order to explore the constraints on invisible neutrino decay, we consider $20$ years of exposure of Hyper-Kamiokande, JUNO, and DUNE. We adopt the inputs provided in  Sec.~\ref{sec:observatories}, except for JUNO, for which we consider a $50\%$ efficiency resulting from the cuts and background mitigation strategies optimised for the observation of the DSNB~\cite{JUNO:2015zny}.We account for the uncertainty in the DSNB flux prediction from the normalisation in the CCSN rate and a $20\%$ uncertainty in the normalisation of each of the backgrounds. Appendix~\ref{sec:stats} describes the statistical approach in more detail.

\begin{figure}
\includegraphics[width = \textwidth]{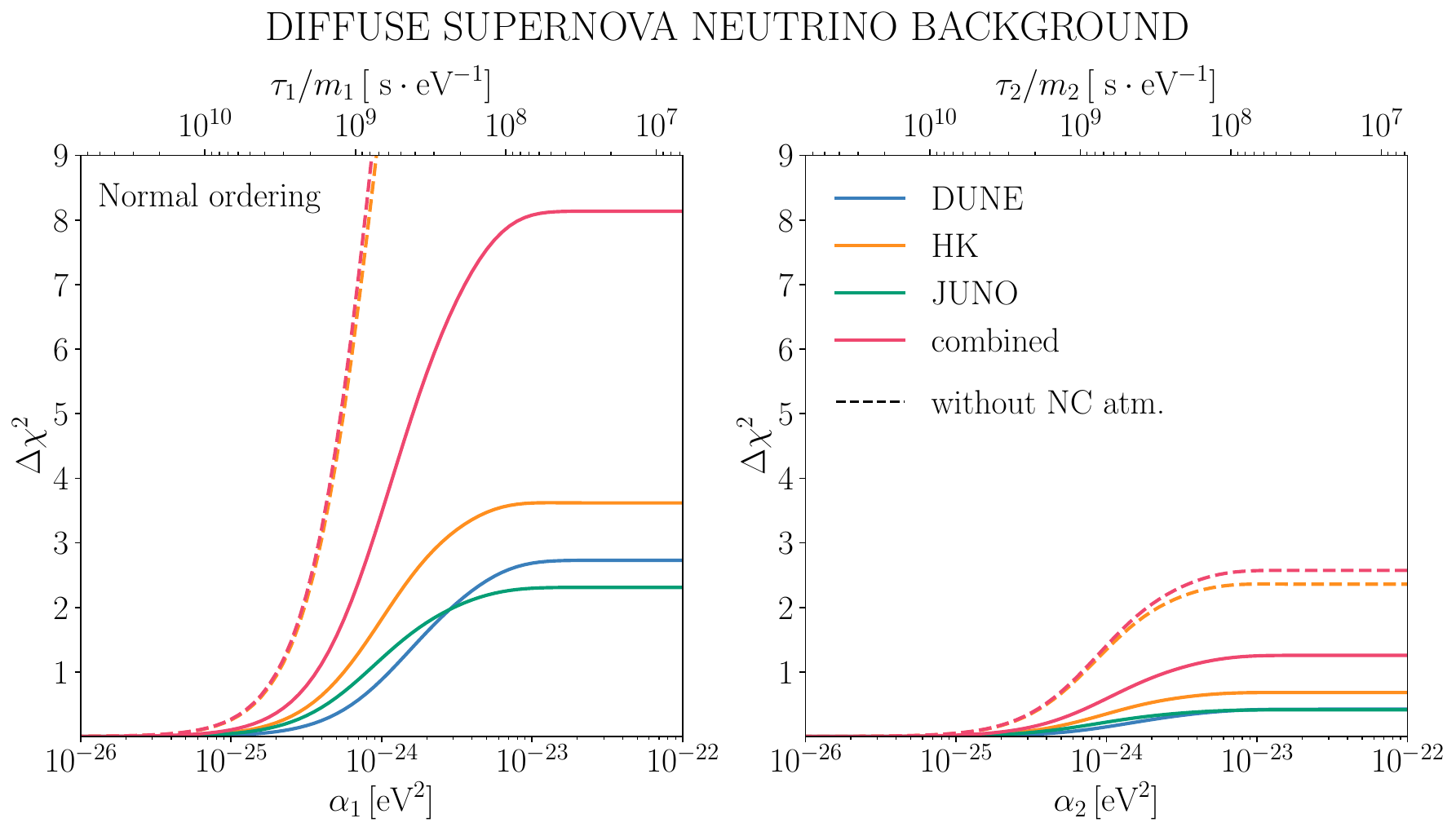}
\caption{Expected sensitivity of the DSNB to invisible neutrino decay parameters to $\alpha_1$ (left panel) and $\alpha_2$ (right panel), obtained relying on Hyper-Kamiokande (orange), DUNE (blue), JUNO (green) and all three detectors together (red). The solid lines correspond to the case  in which nominal backgrounds are considered for Hyper-Kamiokande. Dashed lines show the expected sensitivity assuming full discrimination of neutral-current atmospheric events was achieved in Hyper-Kamiokande. Hyper-Kamiokande is the most sensitive neutrino detector, and it would dominate the sensitivity to invisible neutrino decay, if neutral current atmospheric backgrounds were rejected. The DSNB cannot  robustly  constrain $\alpha_2$ because  the uncertainty in the supernova rate is larger than the fraction of $\nu_2$/$\bar{\nu}_2$ in electron neutrinos/antineutrinos.}
\label{fig:dsnb-detail}
\end{figure}

Figure~\ref{fig:dsnb-detail} shows the sensitivity of the expected DSNB signal in Hyper-Kamiokande, JUNO and DUNE to  $\alpha_1$ and $\alpha_2$. 
For Hyper-Kamiokande, the solid line corresponds to the analysis with nominal backgrounds and the dashed one is the expected sensitivity, if full discrimination of the neutral current atmospheric backgrounds was achieved. We also show the sensitivity for a joint analysis of Hyper-Kamiokande, JUNO and DUNE. As one can see from Fig.~\ref{fig:dsnb-detail}, Hyper-Kamiokande is the most sensitive of the three detectors to the point that it would completely dominate the sensitivity to invisible neutrino decay, if neutral current atmospheric backgrounds were rejected. This highlights the importance of further exploring background mitigation techniques in Hyper-Kamiokande. 

From Fig.~\ref{fig:dsnb-detail}, we also note that the DSNB would not be able to set meaningful constraints on $\alpha_2$  even in the most optimistic scenarios. The reason is that the uncertainty in the CCSN  rate is larger than the fraction of $\nu_2$ ($\bar{\nu}_2$) in electron neutrinos (antineutrinos). Therefore, it would not be possible to differentiate if a lower flux of neutrinos would be due to the decay of $\nu_2$ or if the reference DSNB flux prediction was computed using an optimistic CCSN rate. 
The projected limits from the DSNB at $90\%$~C.L. are
\begin{align}
    \alpha_1 < 8.0 \, (8.5) \, \times 10 ^{-25} \, \textrm{eV}^2\,  \quad \textrm{or} \quad \tau_1/m_1 > 8.4 \, (7.9) \times 10^{8}\, \textrm{s/eV}\, ,
\end{align}
for normal (inverted) ordering, assuming the nominal neutral current backgrounds in Hyper-Kamiokande. Conversely, for the decay parameter $\alpha_2$, non-zero values would be disfavoured at $\sim 1.1\sigma$ at most, both for normal and inverted ordering. If full discrimination of the neutral-current atmospheric backgrounds was achieved, the projected limits would read
\begin{align}
     \alpha_1 < 3.6 \, (3.7) \, \times 10 ^{-25} \, \textrm{eV}^2\,  \quad \textrm{or} \quad \tau_1/m_1 > 4.2 \, (4.4) \times 10^{8}\, \textrm{s/eV}\, ,
\end{align}
and a non-zero value of $\alpha_2$ could be discarded at $1.6~\sigma$.

\begin{figure}
\includegraphics[width = \textwidth]{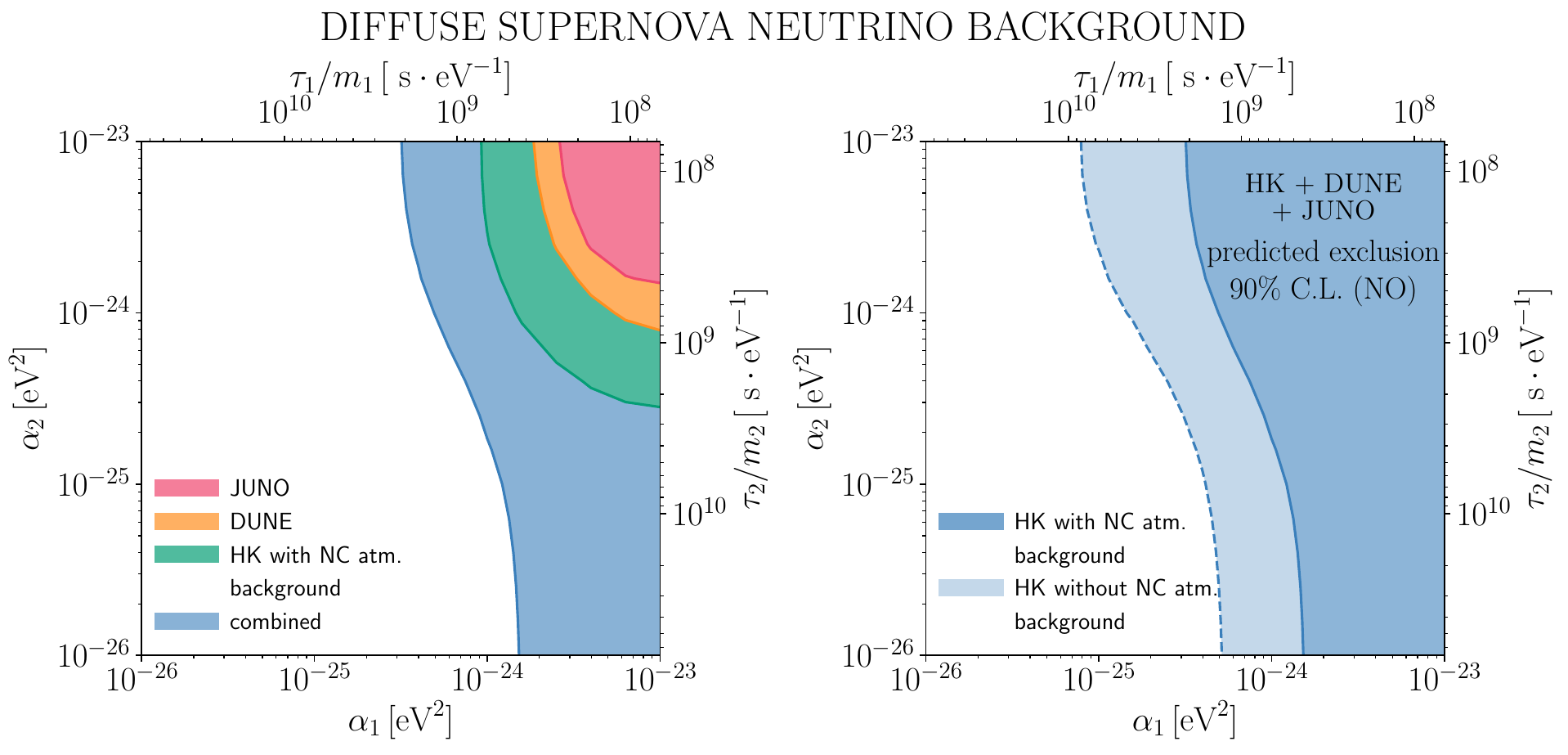}
\caption{Forecasted exclusion region at $90\%$ C.L.~from the DSNB in the plane spanned by $\alpha_1$ and $\alpha_2$. The left panel shows the regions of the parameter space excluded by  JUNO (red), DUNE (orange), Hyper-Kamiokande (green; obtained assuming  baseline backgrounds), and a combined analysis of the three experiments (blue).  The right panel displays the exclusion region  for a joint analysis of Hyper-Kamiokande, DUNE and JUNO with (without) neutral-current atmospheric backgrounds in dark (light) blue. Hyper-Kamiokande has the strongest potential to constrain invisible neutrino decay, especially if the neutral-current atmospheric background is fully tagged. }
\label{fig:dsnb-2d}
\end{figure}

In order to explore the potential interplay between a finite lifetime of two neutrino mass eigenstates, the left panel of Fig.~\ref{fig:dsnb-2d} shows the allowed regions of the parameter space for each neutrino observatory and for a joint analysis. The left panel shows how a joint analysis significantly boosts the sensitivity to non-zero neutrino decay parameters. The reason for the improvement in sensitivity is that a combination of the signals from different neutrino telescopes enhances the statistical significance with whom the decay hypothesis is rejected. Additionally, a combined analysis breaks existing degeneracies between the spectral shape of the signal and some of the backgrounds, which differ for the three detectors.  The right panel of Fig.~\ref{fig:dsnb-2d} shows  how the results of such combined analysis would change, if the neutral current atmospheric backgrounds in Hyper-Kamiokande were fully discriminated from the DSNB events.

\section{Conclusions}
\label{sec:summary}

Astrophysical neutrinos traveling over cosmic distances are ideal probes of  the stability of these elusive particles. In particular,  neutrinos from the Sun, CCSNe, and  the DSNB allow for an optimal combination between distance traveled and characteristic energy to test the possibility that neutrinos decay invisibly  (i.e.~the decay products evade detection). 

This work investigates the  future prospects for learning about  invisible neutrino decay while realistically addressing the astrophysical unknowns in the picture. Our main findings are based on a forecast of the signals from solar, CCSN neutrinos and the DSNB expected in next-generation neutrino observatories (such as Hyper-Kamiokande, DUNE, JUNO, DARWIN, and RES-NOVA) and are summarised in Fig.~\ref{fig:summary}.

When solar neutrinos detected by  DARWIN, DUNE and Hyper-Kamiokande are employed as probes of invisible neutrino decay, a priori knowledge on the magnitude of the solar flux from the SSM  allows to constrain the lifetime-mass ratios to be larger than $\sim 0.1$--$0.01$~s/eV for the  mass eigenstates $\nu_1$ and $\nu_2$. 

Intriguingly, we find that CCSN neutrinos can potentially provide constraints on the decay of all three mass eigenstates. 
The bounds on invisible neutrino decay derived relying on CCSN neutrinos  could be significantly more restrictive than the ones obtained  from solar neutrinos, if  a CCSN explosion occurs at a maximum distance of $10$--$20$~kpc. 
The upcoming detection of the DSNB could potentially set  the strongest limits on $\alpha_1$,  improving the ones from cosmology. Nonetheless, the current uncertainty on the CCSN rate limits the sensitivity to $\alpha_2$.

Our findings also suggest that  the employment of different detection channels at  next-generation neutrino observatories  would boost the sensitivity to invisible neutrino decay, allowing to break the degeneracy between the model uncertainties and the new physics parameters. 

\begin{figure}
    \centering
    \includegraphics[width = \textwidth]{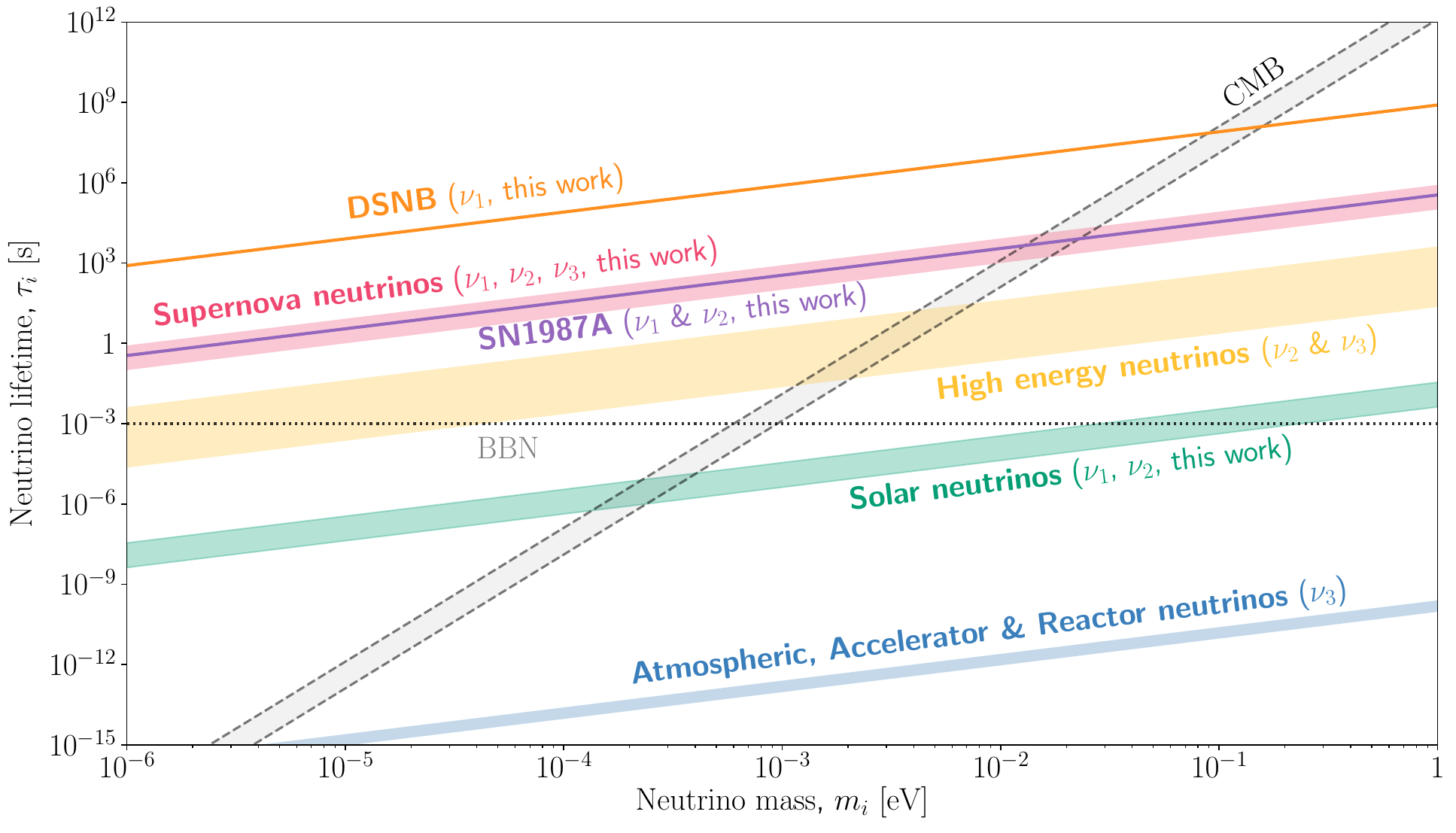}
    \caption{Sensitivity reach to finite neutrino lifetime from invisible neutrino decay as a function of the neutrino mass. Experimentally accessible regions are shown for atmospheric, accelerator and reactor neutrinos~\cite{Gonzalez-Garcia:2008mgl,Abrahao:2015rba, deSalas:2018kri,KM3NeT:2023ncz}, solar neutrinos (present limits~\cite{Berryman:2014qha,SNO:2018pvg} and forecast from this work), high energy neutrinos (existing limits and forecast~\cite{Song:2020nfh}), and supernova neutrinos for a collapse occurring at $10$~kpc (forecast from this work) and the DSNB (forecast from this work). The lifetime of the mass eigenstates that can be tested is indicated. Indirect constraints from Big Bang Nucleosynthesis (BBN) ~\cite{Escudero:2019gfk} and the cosmic microwave background (CMB) data~\cite{Barenboim:2020vrr} are also shown with thin lines  for comparison. The bands indicate the difference between the existing limits and the future sensitivity reach, except for CCSN  neutrinos for which the band is the difference between the sensitivity for $\alpha_1$ and $\alpha_3$. Note that the limits from high-energy neutrinos and SN1987A assume that the two mass eigenstates indicated decay at a similar rate. 
    }
    \label{fig:nulifetime}
\end{figure}

In summary,  after approximately $10$~years of data taking, solar data should provide stringent bounds on $\alpha_1$ and $\alpha_2$, namely
\begin{align}
    \alpha_1 < 7.8\times 10^{-15}\, \textrm{eV}^2 \quad \textrm{and} \quad \alpha_2 < 2.3 \times 10^{-14} \, \textrm{eV}^2\, ,
\end{align}
at  $2 \sigma$. This would correspond to a factor of $20$ improvement with respect to current bounds from solar data (cf.~Eq.~\ref{eqn:current-solar-1}).  After collecting approximately $20$~years of data, the DSNB measurements will improve the limits from solar neutrinos on $\alpha_1$ by ten orders of magnitude. 
If neutrinos from a nearby  CCSN should be detected, the limits on $\alpha_1$ and $\alpha_2$ would improve by about $6$  orders of magnitude with respect to the ones expected from future observation of solar neutrinos. Similar limits could also be set on $\alpha_3$, which would not be possible from solar data only. Actually, the limits on $\alpha_3$ would be more than $14$ orders of magnitude more stringent than those from accelerator and atmospheric data, which are $\alpha_3 \gtrsim 10^{-6} \textrm{eV}^2$~\cite{Gonzalez-Garcia:2008mgl}. Hence, although large exposures are  needed, this work shows that a network of next-generation neutrino observatories has the potential to probe finite neutrino lifetimes using keV--MeV astrophysical neutrinos while consistently accounting for the uncertainties in the source modelling and in the neutrino emission properties. This potential is highlighted in Fig.~\ref{fig:nulifetime}, where the  existing limits and the projected sensitivities derived in this work are contrasted.

In this work, we have focused on  invisible neutrino decay, which leads to a suppression of the expected neutrino flux  with respect to the standard theoretical prediction. Other decay channels may  also be realised in Nature. Neutrino decays involving the emission of photons are strongly constrained for instance from the non-observation of a gamma-ray burst in connection with  SN 1987A~\cite{Kolb:1988pe,Jaffe:1995sw}. For neutrino decays to lighter neutrinos and other daughter products, the main signature is not a reduction of the initial flux, but a spectral distortion and  modification of the detected flavour composition. Such visible neutrino decays have also been studied for instance using solar neutrinos~\cite{Funcke:2019grs,Picoreti:2021yct,deGouvea:2023jxn}, SN 1987A~\cite{Frieman:1987as,Funcke:2019grs,Ivanez-Ballesteros:2023lqa}, and could also be constrained with data from a future nearby CCSN  explosion~\cite{Ando:2004qe,deGouvea:2019goq} as well as the DSNB detection~\cite{DeGouvea:2020ang,Ivanez-Ballesteros:2022szu}, and the diffuse flux of high-energy astrophysical neutrinos detected by  the IceCube Neutrino Observatory~\cite{Bustamante:2016ciw,Beacom:2002vi, Baerwald:2012kc,Maltoni:2008jr}.
Yet, the ideal ratio between the large distances travelled and the energy range for solar and CCSN neutrinos  sets favourable conditions to investigate other forms of neutrino decay, extending  the strategies  presented here.

\acknowledgments
We acknowledge support from the the Danmarks Frie Forskningsfond (Project 
No.~8049-00038B), the Carlsberg Foundation (CF18-0183), and the Deutsche Forschungsgemeinschaft through Sonderforschungsbereich SFB 1258 ``Neutrinos and Dark Matter in Astro- and Particle Physics'' (NDM). This work was also supported by the Spanish grants PID2020-113775GBI00 (MCIN/AEI/10.13039/501100011033) and CIPROM/2021/054 (Generalitat Valenciana) and by Fundação para a Ciência e a Tecnologia (FCT, Portugal) through the project CERN/FIS-PAR/0019/2021. PMM thanks the European Consortium for Astroparticle Theory (EuCAPT) for partial support in the form of an Exchange Travel Grant during the early stages of this work.

\appendix
\section{Statistical analysis and $\mathbf{\chi^2}$-functions}
\label{sec:stats}

In this work, we have followed a frequentist approach for determining the sensitivity of solar and CCSN neutrinos to invisible neutrino decay. To do so, in the case of solar neutrinos at DARWIN, we have defined the following $\chi^2$-function:
\begin{align}
    \chi^2_{\rm \, DARWIN} = \sum_i\frac{\left(\mathcal{S}_i - (1 + \xi)N_i - \eta B_i \right)^2}{S^2 _i + B^2_i} + \left(\frac{\xi}{\sigma_{pp}}\right)^2 \, .
\end{align}
Here, $S_i$ and $B_i$ indicate the number of mock signal events and mock background events in the $i$-th energy bin. Similarly, $N_i$ denotes the number of events expected for a non-zero value of the decay parameters. We have included the pull parameters $\xi$ and $\eta$ to account for the uncertainties in the solar neutrino flux normalisation ($\sigma_{pp} = 0.5\%$) and on the background. Similarly, for the sensitivity of Hyper-Kamiokande and DUNE, we have followed the analysis from Ref.~\cite{Martinez-Mirave:2023fyb}.

For CCSN neutrinos, assuming poisson statistics, we define 
\begin{align}
    \chi^2_{\text{\, CCSN}, \,k} = 2 \sum_j \left[ N_{j,k}  - S_{j,k} + S_{jk} {\rm log}\left(\frac{S_{j,k}}{N_{j,k}}\right)\right]\ ,
\end{align}
as a function of the number of mock events, $S_{j,k}$, and the predicted one in the presence of neutrino decay, $N_{j,k}$, at the $j$-th time bin and at each experiment, $k$. Then, we compute the sum of the $\chi^2$-functions of all the experiments included in the analysis. We repeat this procedure for each CCSN model and mixing scheme. The sensitivity we report  corresponds to the minimum value of the total $\chi^2$ function after marginalising over CCSN  models and mixing scenarios.

Similar to  CCSN neutrinos, for our sensitivity study of neutrino decay from the DSNB, we take the following $\chi^2$ function for each experiment $k$:
\begin{align}
    \chi^2_{\textrm{DSNB},\,k} = \,& 2 \sum_i \left[(1 + \xi)N_{i,k} + \sum_n \eta_n B_{n,i,k}  - S_{i,k} \right. \nonumber \\
    &\left.+ (S_{i, k} + \sum_n B_{n,i,k}) {\rm log}\left(\frac{S_{i,k} + \sum_n B_{n,i,k}}{(1+ \xi)N_{i,k} + \sum_n (1 + \eta_n) B_{n,i,k}}\right)\right] \\ \nonumber &+ \left(\frac{\xi}{\sigma_{R_{\rm SN}}}\right)^2 + \sum_n \left(\frac{\eta_n}{\sigma_n}\right)^2\, .
\end{align}
As in the previous cases, $S_{i,k}$ and $B_{i,k}$ are the mock number of signal and background events in the $i$-th energy bin of the $k$-th experiment, and $N_{i,k}$ is the number of events expected in the presence of invisible decay. The pull parameter $\xi$ accounts for the uncertainty from the normalisation of the CCSN rate, $\sigma_{R_{\rm SN}} = 0.2$, and $\eta_n$ parameterises the uncertainties in the background ($\sigma_n = 0.2$). Note that the index $n$ denotes the different sources of backgrounds in each experiment.

\bibliographystyle{JHEP}
\bibliography{bibliography}

\providecommand{\href}[2]{#2}\begingroup\raggedright\begin{thebibliography}{100}

\bibitem{Giunti:2014ixa}
C.~Giunti and A.~Studenikin, \emph{{Neutrino electromagnetic interactions: a
  window to new physics}},
  \href{https://doi.org/10.1103/RevModPhys.87.531}{\emph{Rev. Mod. Phys.}
  {\bfseries 87} (2015) 531} [\href{https://arxiv.org/abs/1403.6344}{{\ttfamily
  1403.6344}}].

\bibitem{Bahcall:1972my}
J.N.~Bahcall, N.~Cabibbo and A.~Yahil, \emph{{Are neutrinos stable
  particles?}}, \href{https://doi.org/10.1103/PhysRevLett.28.316}{\emph{Phys.
  Rev. Lett.} {\bfseries 28} (1972) 316}.

\bibitem{Shrock:1974nd}
R.~Shrock, \emph{{Decay l0 ---\ensuremath{>} nu(lepton) gamma in gauge theories
  of weak and electromagnetic interactions}},
  \href{https://doi.org/10.1103/PhysRevD.9.743}{\emph{Phys. Rev. D} {\bfseries
  9} (1974) 743}.

\bibitem{Petcov:1976ff}
S.T.~Petcov, \emph{{The Processes $\mu \rightarrow e + \gamma, \mu \rightarrow
  e + \overline{e}, \nu' \rightarrow \nu + \gamma$ in the Weinberg-Salam Model
  with Neutrino Mixing}}, {\emph{Sov. J. Nucl. Phys.} {\bfseries 25} (1977)
  340}.

\bibitem{Marciano:1977wx}
W.J.~Marciano and A.I.~Sanda, \emph{{Exotic Decays of the Muon and Heavy
  Leptons in Gauge Theories}},
  \href{https://doi.org/10.1016/0370-2693(77)90377-X}{\emph{Phys. Lett. B}
  {\bfseries 67} (1977) 303}.

\bibitem{Zatsepin:1978iy}
G.T.~Zatsepin and A.Y.~Smirnov, \emph{{Neutrino Decay in Gauge Theories}},
  {\emph{Yad. Fiz.} {\bfseries 28} (1978) 1569}.

\bibitem{Chikashige:1980qk}
Y.~Chikashige, R.N.~Mohapatra and R.D.~Peccei, \emph{{Spontaneously Broken
  Lepton Number and Cosmological Constraints on the Neutrino Mass Spectrum}},
  \href{https://doi.org/10.1103/PhysRevLett.45.1926}{\emph{Phys. Rev. Lett.}
  {\bfseries 45} (1980) 1926}.

\bibitem{Gelmini:1980re}
G.B.~Gelmini and M.~Roncadelli, \emph{{Left-Handed Neutrino Mass Scale and
  Spontaneously Broken Lepton Number}},
  \href{https://doi.org/10.1016/0370-2693(81)90559-1}{\emph{Phys. Lett. B}
  {\bfseries 99} (1981) 411}.

\bibitem{Pal:1981rm}
P.B.~Pal and L.~Wolfenstein, \emph{{Radiative Decays of Massive Neutrinos}},
  \href{https://doi.org/10.1103/PhysRevD.25.766}{\emph{Phys. Rev. D} {\bfseries
  25} (1982) 766}.

\bibitem{Schechter:1981cv}
J.~Schechter and J.W.F.~Valle, \emph{{Neutrino Decay and Spontaneous Violation
  of Lepton Number}},
  \href{https://doi.org/10.1103/PhysRevD.25.774}{\emph{Phys. Rev. D} {\bfseries
  25} (1982) 774}.

\bibitem{Shrock:1982sc}
R.E.~Shrock, \emph{{Electromagnetic Properties and Decays of Dirac and Majorana
  Neutrinos in a General Class of Gauge Theories}},
  \href{https://doi.org/10.1016/0550-3213(82)90273-5}{\emph{Nucl. Phys. B}
  {\bfseries 206} (1982) 359}.

\bibitem{Gelmini:1983ea}
G.B.~Gelmini and J.W.F.~Valle, \emph{{Fast Invisible Neutrino Decays}},
  \href{https://doi.org/10.1016/0370-2693(84)91258-9}{\emph{Phys. Lett. B}
  {\bfseries 142} (1984) 181}.

\bibitem{Bahcall:1986gq}
J.N.~Bahcall, S.T.~Petcov, S.~Toshev and J.W.F.~Valle, \emph{{Tests of Neutrino
  Stability}}, \href{https://doi.org/10.1016/0370-2693(86)90065-1}{\emph{Phys.
  Lett. B} {\bfseries 181} (1986) 369}.

\bibitem{Nussinov:1987pc}
S.~Nussinov, \emph{{Some Comments on Decaying Neutrinos and the Triplet Majoron
  Model}}, \href{https://doi.org/10.1016/0370-2693(87)91548-6}{\emph{Phys.
  Lett. B} {\bfseries 185} (1987) 171}.

\bibitem{Frieman:1987as}
J.A.~Frieman, H.E.~Haber and K.~Freese, \emph{{Neutrino Mixing, Decays and
  Supernova Sn1987a}},
  \href{https://doi.org/10.1016/0370-2693(88)91120-3}{\emph{Phys. Lett. B}
  {\bfseries 200} (1988) 115}.

\bibitem{Kim:1990km}
C.W.~Kim and W.P.~Lam, \emph{{Some remarks on neutrino decay via a
  Nambu-Goldstone boson}},
  \href{https://doi.org/10.1142/S0217732390000354}{\emph{Mod. Phys. Lett. A}
  {\bfseries 5} (1990) 297}.

\bibitem{Abdullahi:2020rge}
A.~Abdullahi and P.B.~Denton, \emph{{Visible Decay of Astrophysical Neutrinos
  at IceCube}}, \href{https://doi.org/10.1103/PhysRevD.102.023018}{\emph{Phys.
  Rev. D} {\bfseries 102} (2020) 023018}
  [\href{https://arxiv.org/abs/2005.07200}{{\ttfamily 2005.07200}}].

\bibitem{Ackermann:2022rqc}
M.~Ackermann et~al., \emph{{High-energy and ultra-high-energy neutrinos: A
  Snowmass white paper}},
  \href{https://doi.org/10.1016/j.jheap.2022.08.001}{\emph{JHEAp} {\bfseries
  36} (2022) 55} [\href{https://arxiv.org/abs/2203.08096}{{\ttfamily
  2203.08096}}].

\bibitem{Abrahao:2015rba}
T.~Abrah\~ao, H.~Minakata, H.~Nunokawa and A.A.~Quiroga, \emph{{Constraint on
  Neutrino Decay with Medium-Baseline Reactor Neutrino Oscillation
  Experiments}}, \href{https://doi.org/10.1007/JHEP11(2015)001}{\emph{JHEP}
  {\bfseries 11} (2015) 001}
  [\href{https://arxiv.org/abs/1506.02314}{{\ttfamily 1506.02314}}].

\bibitem{Gonzalez-Garcia:2008mgl}
M.C.~Gonzalez-Garcia and M.~Maltoni, \emph{{Status of Oscillation plus Decay of
  Atmospheric and Long-Baseline Neutrinos}},
  \href{https://doi.org/10.1016/j.physletb.2008.04.041}{\emph{Phys. Lett. B}
  {\bfseries 663} (2008) 405}
  [\href{https://arxiv.org/abs/0802.3699}{{\ttfamily 0802.3699}}].

\bibitem{Gomes:2014yua}
R.A.~Gomes, A.L.G.~Gomes and O.L.G.~Peres, \emph{{Constraints on neutrino decay
  lifetime using long-baseline charged and neutral current data}},
  \href{https://doi.org/10.1016/j.physletb.2014.12.014}{\emph{Phys. Lett. B}
  {\bfseries 740} (2015) 345}
  [\href{https://arxiv.org/abs/1407.5640}{{\ttfamily 1407.5640}}].

\bibitem{Choubey:2017dyu}
S.~Choubey, S.~Goswami and D.~Pramanik, \emph{{A study of invisible neutrino
  decay at DUNE and its effects on $\theta_{23}$ measurement}},
  \href{https://doi.org/10.1007/JHEP02(2018)055}{\emph{JHEP} {\bfseries 02}
  (2018) 055} [\href{https://arxiv.org/abs/1705.05820}{{\ttfamily
  1705.05820}}].

\bibitem{Choubey_2018}
S.~Choubey, D.~Dutta and D.~Pramanik, \emph{Invisible neutrino decay in the
  light of nova and t2k data},
  \href{https://doi.org/10.1007/jhep08(2018)141}{\emph{Journal of High Energy
  Physics} {\bfseries 2018} (2018) }.

\bibitem{deSalas:2018kri}
P.F.~de~Salas, S.~Pastor, C.A.~Ternes, T.~Thakore and M.~T\'ortola,
  \emph{{Constraining the invisible neutrino decay with KM3NeT-ORCA}},
  \href{https://doi.org/10.1016/j.physletb.2018.12.066}{\emph{Phys. Lett. B}
  {\bfseries 789} (2019) 472}
  [\href{https://arxiv.org/abs/1810.10916}{{\ttfamily 1810.10916}}].

\bibitem{Ternes:2024qui}
C.A.~Ternes and G.~Pagliaroli, \emph{{Invisible neutrino decay at long-baseline
  neutrino oscillation experiments}},
  \href{https://arxiv.org/abs/2401.14316}{{\ttfamily 2401.14316}}.

\bibitem{Super-Kamiokande:2006jvq}
{\scshape Super-Kamiokande} collaboration, \emph{{Three flavor neutrino
  oscillation analysis of atmospheric neutrinos in Super-Kamiokande}},
  \href{https://doi.org/10.1103/PhysRevD.74.032002}{\emph{Phys. Rev. D}
  {\bfseries 74} (2006) 032002}
  [\href{https://arxiv.org/abs/hep-ex/0604011}{{\ttfamily hep-ex/0604011}}].

\bibitem{MINOS:2006foh}
{\scshape MINOS} collaboration, \emph{{Observation of muon neutrino
  disappearance with the MINOS detectors and the NuMI neutrino beam}},
  \href{https://doi.org/10.1103/PhysRevLett.97.191801}{\emph{Phys. Rev. Lett.}
  {\bfseries 97} (2006) 191801}
  [\href{https://arxiv.org/abs/hep-ex/0607088}{{\ttfamily hep-ex/0607088}}].

\bibitem{K2K:2006yov}
{\scshape K2K} collaboration, \emph{{Measurement of Neutrino Oscillation by the
  K2K Experiment}},
  \href{https://doi.org/10.1103/PhysRevD.74.072003}{\emph{Phys. Rev. D}
  {\bfseries 74} (2006) 072003}
  [\href{https://arxiv.org/abs/hep-ex/0606032}{{\ttfamily hep-ex/0606032}}].

\bibitem{Escudero:2019gfk}
M.~Escudero and M.~Fairbairn, \emph{{Cosmological Constraints on Invisible
  Neutrino Decays Revisited}},
  \href{https://doi.org/10.1103/PhysRevD.100.103531}{\emph{Phys. Rev. D}
  {\bfseries 100} (2019) 103531}
  [\href{https://arxiv.org/abs/1907.05425}{{\ttfamily 1907.05425}}].

\bibitem{Basboll:2008fx}
A.~Basboll, O.E.~Bjaelde, S.~Hannestad and G.G.~Raffelt, \emph{{Are
  cosmological neutrinos free-streaming?}},
  \href{https://doi.org/10.1103/PhysRevD.79.043512}{\emph{Phys. Rev. D}
  {\bfseries 79} (2009) 043512}
  [\href{https://arxiv.org/abs/0806.1735}{{\ttfamily 0806.1735}}].

\bibitem{Archidiacono:2013dua}
M.~Archidiacono and S.~Hannestad, \emph{{Updated constraints on non-standard
  neutrino interactions from Planck}},
  \href{https://doi.org/10.1088/1475-7516/2014/07/046}{\emph{JCAP} {\bfseries
  07} (2014) 046} [\href{https://arxiv.org/abs/1311.3873}{{\ttfamily
  1311.3873}}].

\bibitem{Hannestad:2004qu}
S.~Hannestad, \emph{{Structure formation with strongly interacting neutrinos -
  Implications for the cosmological neutrino mass bound}},
  \href{https://doi.org/10.1088/1475-7516/2005/02/011}{\emph{JCAP} {\bfseries
  02} (2005) 011} [\href{https://arxiv.org/abs/astro-ph/0411475}{{\ttfamily
  astro-ph/0411475}}].

\bibitem{Hannestad:2005ex}
S.~Hannestad and G.G.~Raffelt, \emph{{Constraining invisible neutrino decays
  with the cosmic microwave background}},
  \href{https://doi.org/10.1103/PhysRevD.72.103514}{\emph{Phys. Rev. D}
  {\bfseries 72} (2005) 103514}
  [\href{https://arxiv.org/abs/hep-ph/0509278}{{\ttfamily hep-ph/0509278}}].

\bibitem{Barenboim:2020vrr}
G.~Barenboim, J.Z.~Chen, S.~Hannestad, I.M.~Oldengott, T.~Tram and Y.Y.Y.~Wong,
  \emph{{Invisible neutrino decay in precision cosmology}},
  \href{https://doi.org/10.1088/1475-7516/2021/03/087}{\emph{JCAP} {\bfseries
  03} (2021) 087} [\href{https://arxiv.org/abs/2011.01502}{{\ttfamily
  2011.01502}}].

\bibitem{Chen:2022idm}
J.Z.~Chen, I.M.~Oldengott, G.~Pierobon and Y.Y.Y.~Wong, \emph{{Weaker yet
  again: mass spectrum-consistent cosmological constraints on the neutrino
  lifetime}}, \href{https://doi.org/10.1140/epjc/s10052-022-10518-3}{\emph{Eur.
  Phys. J. C} {\bfseries 82} (2022) 640}
  [\href{https://arxiv.org/abs/2203.09075}{{\ttfamily 2203.09075}}].

\bibitem{Chacko:2019nej}
Z.~Chacko, A.~Dev, P.~Du, V.~Poulin and Y.~Tsai, \emph{{Cosmological Limits on
  the Neutrino Mass and Lifetime}},
  \href{https://doi.org/10.1007/JHEP04(2020)020}{\emph{JHEP} {\bfseries 04}
  (2020) 020} [\href{https://arxiv.org/abs/1909.05275}{{\ttfamily
  1909.05275}}].

\bibitem{FrancoAbellan:2021hdb}
G.~Franco~Abell\'an, Z.~Chacko, A.~Dev, P.~Du, V.~Poulin and Y.~Tsai,
  \emph{{Improved cosmological constraints on the neutrino mass and lifetime}},
  \href{https://doi.org/10.1007/JHEP08(2022)076}{\emph{JHEP} {\bfseries 08}
  (2022) 076} [\href{https://arxiv.org/abs/2112.13862}{{\ttfamily
  2112.13862}}].

\bibitem{Chacko:2020hmh}
Z.~Chacko, A.~Dev, P.~Du, V.~Poulin and Y.~Tsai, \emph{{Determining the
  Neutrino Lifetime from Cosmology}},
  \href{https://doi.org/10.1103/PhysRevD.103.043519}{\emph{Phys. Rev. D}
  {\bfseries 103} (2021) 043519}
  [\href{https://arxiv.org/abs/2002.08401}{{\ttfamily 2002.08401}}].

\bibitem{Valera:2023bud}
V.B.~Valera, D.~Fiorillo, I.~Esteban and M.~Bustamante, \emph{{Probing Neutrino
  Decay Using the First Steady-State Source of High-Energy Astrophysical
  Neutrinos, NGC 1068}}, \href{https://doi.org/10.22323/1.444.1066}{\emph{PoS}
  {\bfseries ICRC2023} (2023) 1066}.

\bibitem{Denton:2018aml}
P.B.~Denton and I.~Tamborra, \emph{{Invisible Neutrino Decay Could Resolve
  IceCube\textquoteright{}s Track and Cascade Tension}},
  \href{https://doi.org/10.1103/PhysRevLett.121.121802}{\emph{Phys. Rev. Lett.}
  {\bfseries 121} (2018) 121802}
  [\href{https://arxiv.org/abs/1805.05950}{{\ttfamily 1805.05950}}].

\bibitem{Song:2020nfh}
N.~Song, S.W.~Li, C.A.~Arg\"uelles, M.~Bustamante and A.C.~Vincent, \emph{{The
  Future of High-Energy Astrophysical Neutrino Flavor Measurements}},
  \href{https://doi.org/10.1088/1475-7516/2021/04/054}{\emph{JCAP} {\bfseries
  04} (2021) 054} [\href{https://arxiv.org/abs/2012.12893}{{\ttfamily
  2012.12893}}].

\bibitem{Bandyopadhyay:2002qg}
A.~Bandyopadhyay, S.~Choubey and S.~Goswami, \emph{{Neutrino decay confronts
  the SNO data}},
  \href{https://doi.org/10.1016/S0370-2693(03)00044-3}{\emph{Phys. Lett. B}
  {\bfseries 555} (2003) 33}
  [\href{https://arxiv.org/abs/hep-ph/0204173}{{\ttfamily hep-ph/0204173}}].

\bibitem{Picoreti:2015ika}
R.~Picoreti, M.M.~Guzzo, P.C.~de~Holanda and O.L.G.~Peres, \emph{{Neutrino
  Decay and Solar Neutrino Seasonal Effect}},
  \href{https://doi.org/10.1016/j.physletb.2016.08.007}{\emph{Phys. Lett. B}
  {\bfseries 761} (2016) 70}
  [\href{https://arxiv.org/abs/1506.08158}{{\ttfamily 1506.08158}}].

\bibitem{Choubey:2000an}
S.~Choubey, S.~Goswami and D.~Majumdar, \emph{{Status of the neutrino decay
  solution to the solar neutrino problem}},
  \href{https://doi.org/10.1016/S0370-2693(00)00608-0}{\emph{Phys. Lett. B}
  {\bfseries 484} (2000) 73}
  [\href{https://arxiv.org/abs/hep-ph/0004193}{{\ttfamily hep-ph/0004193}}].

\bibitem{Bellini:2011rx}
G.~Bellini et~al., \emph{{Precision measurement of the 7Be solar neutrino
  interaction rate in Borexino}},
  \href{https://doi.org/10.1103/PhysRevLett.107.141302}{\emph{Phys. Rev. Lett.}
  {\bfseries 107} (2011) 141302}
  [\href{https://arxiv.org/abs/1104.1816}{{\ttfamily 1104.1816}}].

\bibitem{KamLAND:2014gul}
{\scshape KamLAND} collaboration, \emph{{$^7$Be Solar Neutrino Measurement with
  KamLAND}}, \href{https://doi.org/10.1103/PhysRevC.92.055808}{\emph{Phys. Rev.
  C} {\bfseries 92} (2015) 055808}
  [\href{https://arxiv.org/abs/1405.6190}{{\ttfamily 1405.6190}}].

\bibitem{SNO:2011hxd}
{\scshape SNO} collaboration, \emph{{Combined Analysis of all Three Phases of
  Solar Neutrino Data from the Sudbury Neutrino Observatory}},
  \href{https://doi.org/10.1103/PhysRevC.88.025501}{\emph{Phys. Rev. C}
  {\bfseries 88} (2013) 025501}
  [\href{https://arxiv.org/abs/1109.0763}{{\ttfamily 1109.0763}}].

\bibitem{Super-Kamiokande:2016yck}
{\scshape Super-Kamiokande} collaboration, \emph{{Solar Neutrino Measurements
  in Super-Kamiokande-IV}},
  \href{https://doi.org/10.1103/PhysRevD.94.052010}{\emph{Phys. Rev. D}
  {\bfseries 94} (2016) 052010}
  [\href{https://arxiv.org/abs/1606.07538}{{\ttfamily 1606.07538}}].

\bibitem{KamLAND:2011fld}
{\scshape KamLAND} collaboration, \emph{{Measurement of the 8B Solar Neutrino
  Flux with the KamLAND Liquid Scintillator Detector}},
  \href{https://doi.org/10.1103/PhysRevC.84.035804}{\emph{Phys. Rev. C}
  {\bfseries 84} (2011) 035804}
  [\href{https://arxiv.org/abs/1106.0861}{{\ttfamily 1106.0861}}].

\bibitem{Borexino:2017uhp}
{\scshape Borexino} collaboration, \emph{{Improved measurement of $^8$B solar
  neutrinos with $1.5 kt·y$ of Borexino exposure}},
  \href{https://doi.org/10.1103/PhysRevD.101.062001}{\emph{Phys. Rev. D}
  {\bfseries 101} (2020) 062001}
  [\href{https://arxiv.org/abs/1709.00756}{{\ttfamily 1709.00756}}].

\bibitem{Cleveland:1998nv}
B.T.~Cleveland, T.~Daily, R.~Davis, Jr., J.R.~Distel, K.~Lande, C.K.~Lee
  et~al., \emph{{Measurement of the solar electron neutrino flux with the
  Homestake chlorine detector}},
  \href{https://doi.org/10.1086/305343}{\emph{Astrophys. J.} {\bfseries 496}
  (1998) 505}.

\bibitem{SAGE:2009eeu}
{\scshape SAGE} collaboration, \emph{{Measurement of the solar neutrino capture
  rate with gallium metal. III: Results for the 2002--2007 data-taking
  period}}, \href{https://doi.org/10.1103/PhysRevC.80.015807}{\emph{Phys. Rev.
  C} {\bfseries 80} (2009) 015807}
  [\href{https://arxiv.org/abs/0901.2200}{{\ttfamily 0901.2200}}].

\bibitem{Berryman:2014qha}
J.M.~Berryman, A.~de~Gouv\^ea and D.~Hernandez, \emph{{Solar Neutrinos and the
  Decaying Neutrino Hypothesis}},
  \href{https://doi.org/10.1103/PhysRevD.92.073003}{\emph{Phys. Rev. D}
  {\bfseries 92} (2015) 073003}
  [\href{https://arxiv.org/abs/1411.0308}{{\ttfamily 1411.0308}}].

\bibitem{SNO:2018pvg}
{\scshape SNO} collaboration, \emph{{Constraints on Neutrino Lifetime from the
  Sudbury Neutrino Observatory}},
  \href{https://doi.org/10.1103/PhysRevD.99.032013}{\emph{Phys. Rev. D}
  {\bfseries 99} (2019) 032013}
  [\href{https://arxiv.org/abs/1812.01088}{{\ttfamily 1812.01088}}].

\bibitem{Hirata:1988ad}
K.S.~Hirata et~al., \emph{{Observation in the Kamiokande-II Detector of the
  Neutrino Burst from Supernova SN 1987a}},
  \href{https://doi.org/10.1103/PhysRevD.38.448}{\emph{Phys. Rev. D} {\bfseries
  38} (1988) 448}.

\bibitem{Super-Kamiokande:2021jaq}
{\scshape Super-Kamiokande} collaboration, \emph{{Diffuse supernova neutrino
  background search at Super-Kamiokande}},
  \href{https://doi.org/10.1103/PhysRevD.104.122002}{\emph{Phys. Rev. D}
  {\bfseries 104} (2021) 122002}
  [\href{https://arxiv.org/abs/2109.11174}{{\ttfamily 2109.11174}}].

\bibitem{Super-Kamiokande:2023xup}
{\scshape Super-Kamiokande} collaboration, \emph{{Search for Astrophysical
  Electron Antineutrinos in Super-Kamiokande with 0.01\% Gadolinium-loaded
  Water}}, \href{https://doi.org/10.3847/2041-8213/acdc9e}{\emph{Astrophys. J.
  Lett.} {\bfseries 951} (2023) L27}
  [\href{https://arxiv.org/abs/2305.05135}{{\ttfamily 2305.05135}}].

\bibitem{Hyper-Kamiokande:2018ofw}
{\scshape Hyper-Kamiokande} collaboration, \emph{{Hyper-Kamiokande Design
  Report}},  \href{https://arxiv.org/abs/1805.04163}{{\ttfamily 1805.04163}}.

\bibitem{Strumia:2003zx}
A.~Strumia and F.~Vissani, \emph{{Precise quasielastic neutrino/nucleon
  cross-section}},
  \href{https://doi.org/10.1016/S0370-2693(03)00616-6}{\emph{Phys. Lett. B}
  {\bfseries 564} (2003) 42}
  [\href{https://arxiv.org/abs/astro-ph/0302055}{{\ttfamily
  astro-ph/0302055}}].

\bibitem{Super-Kamiokande:2020frs}
{\scshape Super-Kamiokande} collaboration, \emph{{Search for solar electron
  anti-neutrinos due to spin-flavor precession in the Sun with
  Super-Kamiokande-IV}},
  \href{https://doi.org/10.1016/j.astropartphys.2022.102702}{\emph{Astropart.
  Phys.} {\bfseries 139} (2022) 102702}
  [\href{https://arxiv.org/abs/2012.03807}{{\ttfamily 2012.03807}}].

\bibitem{Super-Kamiokande:2010tar}
{\scshape Super-Kamiokande} collaboration, \emph{{Solar neutrino results in
  Super-Kamiokande-III}},
  \href{https://doi.org/10.1103/PhysRevD.83.052010}{\emph{Phys. Rev. D}
  {\bfseries 83} (2011) 052010}
  [\href{https://arxiv.org/abs/1010.0118}{{\ttfamily 1010.0118}}].

\bibitem{Laha:2013hva}
R.~Laha and J.F.~Beacom, \emph{{Gadolinium in water Cherenkov detectors
  improves detection of supernova $\nu_e$}},
  \href{https://doi.org/10.1103/PhysRevD.89.063007}{\emph{Phys. Rev. D}
  {\bfseries 89} (2014) 063007}
  [\href{https://arxiv.org/abs/1311.6407}{{\ttfamily 1311.6407}}].

\bibitem{Kolbe:2002gk}
E.~Kolbe, K.~Langanke and P.~Vogel, \emph{{Estimates of weak and
  electromagnetic nuclear decay signatures for neutrino reactions in
  Super-Kamiokande}},
  \href{https://doi.org/10.1103/PhysRevD.66.013007}{\emph{Phys. Rev. D}
  {\bfseries 66} (2002) 013007}.

\bibitem{JUNO:2015zny}
{\scshape JUNO} collaboration, \emph{{Neutrino Physics with JUNO}},
  \href{https://doi.org/10.1088/0954-3899/43/3/030401}{\emph{J. Phys. G}
  {\bfseries 43} (2016) 030401}
  [\href{https://arxiv.org/abs/1507.05613}{{\ttfamily 1507.05613}}].

\bibitem{DUNE:2020lwj}
{\scshape DUNE} collaboration, \emph{{Deep Underground Neutrino Experiment
  (DUNE), Far Detector Technical Design Report, Volume I Introduction to
  DUNE}}, \href{https://doi.org/10.1088/1748-0221/15/08/T08008}{\emph{JINST}
  {\bfseries 15} (2020) T08008}
  [\href{https://arxiv.org/abs/2002.02967}{{\ttfamily 2002.02967}}].

\bibitem{Capozzi:2018dat}
F.~Capozzi, S.W.~Li, G.~Zhu and J.F.~Beacom, \emph{{DUNE as the Next-Generation
  Solar Neutrino Experiment}},
  \href{https://doi.org/10.1103/PhysRevLett.123.131803}{\emph{Phys. Rev. Lett.}
  {\bfseries 123} (2019) 131803}
  [\href{https://arxiv.org/abs/1808.08232}{{\ttfamily 1808.08232}}].

\bibitem{DUNE:2020ypp}
{\scshape DUNE} collaboration, \emph{{Deep Underground Neutrino Experiment
  (DUNE), Far Detector Technical Design Report, Volume II: DUNE Physics}},
  \href{https://arxiv.org/abs/2002.03005}{{\ttfamily 2002.03005}}.

\bibitem{DUNE:2020zfm}
{\scshape DUNE} collaboration, \emph{{Supernova neutrino burst detection with
  the Deep Underground Neutrino Experiment}},
  \href{https://doi.org/10.1140/epjc/s10052-021-09166-w}{\emph{Eur. Phys. J. C}
  {\bfseries 81} (2021) 423}
  [\href{https://arxiv.org/abs/2008.06647}{{\ttfamily 2008.06647}}].

\bibitem{snowglobes}
K.~{Scholberg}, J.B.~{Albert} and J.~{Vasel}, ``{SNOwGLoBES: SuperNova
  Observatories with GLoBES}.'' Astrophysics Source Code Library, record
  ascl:2109.019, Sept., 2021.

\bibitem{Castiglioni:2020tsu}
W.~Castiglioni, W.~Foreman, I.~Lepetic, B.R.~Littlejohn, M.~Malaker and
  A.~Mastbaum, \emph{{Benefits of MeV-scale reconstruction capabilities in
  large liquid argon time projection chambers}},
  \href{https://doi.org/10.1103/PhysRevD.102.092010}{\emph{Phys. Rev. D}
  {\bfseries 102} (2020) 092010}
  [\href{https://arxiv.org/abs/2006.14675}{{\ttfamily 2006.14675}}].

\bibitem{Barenboim:2023krl}
G.~Barenboim, P.~Mart\'\i{}nez-Mirav\'e, C.A.~Ternes and M.~T\'ortola,
  \emph{{Neutrino CPT violation in the solar sector}},
  \href{https://doi.org/10.1103/PhysRevD.108.035039}{\emph{Phys. Rev. D}
  {\bfseries 108} (2023) 035039}
  [\href{https://arxiv.org/abs/2305.06384}{{\ttfamily 2305.06384}}].

\bibitem{Aalbers:2022dzr}
J.~Aalbers et~al., \emph{{A next-generation liquid xenon observatory for dark
  matter and neutrino physics}},
  \href{https://doi.org/10.1088/1361-6471/ac841a}{\emph{J. Phys. G} {\bfseries
  50} (2023) 013001} [\href{https://arxiv.org/abs/2203.02309}{{\ttfamily
  2203.02309}}].

\bibitem{DARWIN:2016hyl}
{\scshape DARWIN} collaboration, \emph{{DARWIN: towards the ultimate dark
  matter detector}},
  \href{https://doi.org/10.1088/1475-7516/2016/11/017}{\emph{JCAP} {\bfseries
  11} (2016) 017} [\href{https://arxiv.org/abs/1606.07001}{{\ttfamily
  1606.07001}}].

\bibitem{Lang:2016zhv}
R.F.~Lang, C.~McCabe, S.~Reichard, M.~Selvi and I.~Tamborra, \emph{{Supernova
  neutrino physics with xenon dark matter detectors: A timely perspective}},
  \href{https://doi.org/10.1103/PhysRevD.94.103009}{\emph{Phys. Rev. D}
  {\bfseries 94} (2016) 103009}
  [\href{https://arxiv.org/abs/1606.09243}{{\ttfamily 1606.09243}}].

\bibitem{Schumann:2015cpa}
M.~Schumann, L.~Baudis, L.~B\"utikofer, A.~Kish and M.~Selvi, \emph{{Dark
  matter sensitivity of multi-ton liquid xenon detectors}},
  \href{https://doi.org/10.1088/1475-7516/2015/10/016}{\emph{JCAP} {\bfseries
  10} (2015) 016} [\href{https://arxiv.org/abs/1506.08309}{{\ttfamily
  1506.08309}}].

\bibitem{DARWIN:2020bnc}
{\scshape DARWIN} collaboration, \emph{{Solar neutrino detection sensitivity in
  DARWIN via electron scattering}},
  \href{https://doi.org/10.1140/epjc/s10052-020-08602-7}{\emph{Eur. Phys. J. C}
  {\bfseries 80} (2020) 1133}
  [\href{https://arxiv.org/abs/2006.03114}{{\ttfamily 2006.03114}}].

\bibitem{Pattavina:2020cqc}
L.~Pattavina, N.~Ferreiro~Iachellini and I.~Tamborra, \emph{{Neutrino
  observatory based on archaeological lead}},
  \href{https://doi.org/10.1103/PhysRevD.102.063001}{\emph{Phys. Rev. D}
  {\bfseries 102} (2020) 063001}
  [\href{https://arxiv.org/abs/2004.06936}{{\ttfamily 2004.06936}}].

\bibitem{RES-NOVAGroupofInterest:2022glt}
{\scshape RES-NOVA Group of Interest} collaboration, \emph{{Radiopurity of a
  kg-scale PbWO$_4$ cryogenic detector produced from archaeological Pb for the
  RES-NOVA experiment}},
  \href{https://doi.org/10.1140/epjc/s10052-022-10656-8}{\emph{Eur. Phys. J. C}
  {\bfseries 82} (2022) 692}
  [\href{https://arxiv.org/abs/2203.07441}{{\ttfamily 2203.07441}}].

\bibitem{RES-NOVA:2021gqp}
{\scshape RES-NOVA} collaboration, \emph{{RES-NOVA sensitivity to core-collapse
  and failed core-collapse supernova neutrinos}},
  \href{https://doi.org/10.1088/1475-7516/2021/10/064}{\emph{JCAP} {\bfseries
  10} (2021) 064} [\href{https://arxiv.org/abs/2103.08672}{{\ttfamily
  2103.08672}}].

\bibitem{Gann:2021ndb}
G.D.O.~Gann, K.~Zuber, D.~Bemmerer and A.~Serenelli, \emph{{The Future of Solar
  Neutrinos}},
  \href{https://doi.org/10.1146/annurev-nucl-011921-061243}{\emph{Ann. Rev.
  Nucl. Part. Sci.} {\bfseries 71} (2021) 491}
  [\href{https://arxiv.org/abs/2107.08613}{{\ttfamily 2107.08613}}].

\bibitem{BOREXINO:2020aww}
{\scshape BOREXINO} collaboration, \emph{{Experimental evidence of neutrinos
  produced in the CNO fusion cycle in the Sun}},
  \href{https://doi.org/10.1038/s41586-020-2934-0}{\emph{Nature} {\bfseries
  587} (2020) 577} [\href{https://arxiv.org/abs/2006.15115}{{\ttfamily
  2006.15115}}].

\bibitem{BOREXINO:2023ygs}
{\scshape BOREXINO} collaboration, \emph{{Final results of Borexino on CNO
  solar neutrinos}},
  \href{https://doi.org/10.1103/PhysRevD.108.102005}{\emph{Phys. Rev. D}
  {\bfseries 108} (2023) 102005}
  [\href{https://arxiv.org/abs/2307.14636}{{\ttfamily 2307.14636}}].

\bibitem{Vinyoles:2016djt}
N.~Vinyoles, A.M.~Serenelli, F.L.~Villante, S.~Basu, J.~Bergstr\"om,
  M.C.~Gonzalez-Garcia et~al., \emph{{A new Generation of Standard Solar
  Models}}, \href{https://doi.org/10.3847/1538-4357/835/2/202}{\emph{Astrophys.
  J.} {\bfseries 835} (2017) 202}
  [\href{https://arxiv.org/abs/1611.09867}{{\ttfamily 1611.09867}}].

\bibitem{Wolfenstein:1977ue}
L.~Wolfenstein, \emph{{Neutrino Oscillations in Matter}},
  \href{https://doi.org/10.1103/PhysRevD.17.2369}{\emph{Phys. Rev. D}
  {\bfseries 17} (1978) 2369}.

\bibitem{Mikheyev:1985zog}
S.P.~Mikheyev and A.Y.~Smirnov, \emph{{Resonance Amplification of Oscillations
  in Matter and Spectroscopy of Solar Neutrinos}}, {\emph{Sov. J. Nucl. Phys.}
  {\bfseries 42} (1985) 913}.

\bibitem{Maltoni:2015kca}
M.~Maltoni and A.Y.~Smirnov, \emph{{Solar neutrinos and neutrino physics}},
  \href{https://doi.org/10.1140/epja/i2016-16087-0}{\emph{Eur. Phys. J. A}
  {\bfseries 52} (2016) 87} [\href{https://arxiv.org/abs/1507.05287}{{\ttfamily
  1507.05287}}].

\bibitem{deSalas:2020pgw}
P.F.~de~Salas, D.V.~Forero, S.~Gariazzo, P.~Mart\'\i{}nez-Mirav\'e, O.~Mena,
  C.A.~Ternes et~al., \emph{{2020 global reassessment of the neutrino
  oscillation picture}},
  \href{https://doi.org/10.1007/JHEP02(2021)071}{\emph{JHEP} {\bfseries 02}
  (2021) 071} [\href{https://arxiv.org/abs/2006.11237}{{\ttfamily
  2006.11237}}].

\bibitem{Bouchez:1986kb}
J.~Bouchez, M.~Cribier, J.~Rich, M.~Spiro, D.~Vignaud and W.~Hampel,
  \emph{{Matter Effects for Solar Neutrino Oscillations}},
  \href{https://doi.org/10.1007/BF01550771}{\emph{Z. Phys. C} {\bfseries 32}
  (1986) 499}.

\bibitem{Cribier:1986ak}
M.~Cribier, W.~Hampel, J.~Rich and D.~Vignaud, \emph{{Msw Regeneration of Solar
  $\nu_e$ in the Earth}},
  \href{https://doi.org/10.1016/0370-2693(86)91083-X}{\emph{Phys. Lett. B}
  {\bfseries 182} (1986) 89}.

\bibitem{Akhmedov:2004rq}
E.K.~Akhmedov, M.A.~Tortola and J.W.F.~Valle, \emph{{A Simple analytic three
  flavor description of the day night effect in the solar neutrino flux}},
  \href{https://doi.org/10.1088/1126-6708/2004/05/057}{\emph{JHEP} {\bfseries
  05} (2004) 057} [\href{https://arxiv.org/abs/hep-ph/0404083}{{\ttfamily
  hep-ph/0404083}}].

\bibitem{Super-Kamiokande:2023jbt}
{\scshape Super-Kamiokande} collaboration, \emph{{Solar neutrino measurements
  using the full data period of Super-Kamiokande-IV}},
  \href{https://arxiv.org/abs/2312.12907}{{\ttfamily 2312.12907}}.

\bibitem{Funcke:2019grs}
L.~Funcke, G.~Raffelt and E.~Vitagliano, \emph{{Distinguishing Dirac and
  Majorana neutrinos by their decays via Nambu-Goldstone bosons in the
  gravitational-anomaly model of neutrino masses}},
  \href{https://doi.org/10.1103/PhysRevD.101.015025}{\emph{Phys. Rev. D}
  {\bfseries 101} (2020) 015025}
  [\href{https://arxiv.org/abs/1905.01264}{{\ttfamily 1905.01264}}].

\bibitem{Yano:2021usb}
{\scshape Hyper-Kamiokande Proto} collaboration, \emph{{Sensitivity Study for
  Astrophysical Neutrinos at Hyper-Kamiokande}},
  \href{https://doi.org/10.22323/1.390.0191}{\emph{PoS} {\bfseries ICHEP2020}
  (2021) 191}.

\bibitem{Martinez-Mirave:2021cvh}
P.~Mart\'\i{}nez-Mirav\'e, S.M.~Sedgwick and M.~T\'ortola, \emph{{Nonstandard
  interactions from the future neutrino solar sector}},
  \href{https://doi.org/10.1103/PhysRevD.105.035004}{\emph{Phys. Rev. D}
  {\bfseries 105} (2022) 035004}
  [\href{https://arxiv.org/abs/2111.03031}{{\ttfamily 2111.03031}}].

\bibitem{Nakano:2016uws}
Y.~Nakano, \emph{{$^8$B solar neutrino spectrum measurement using
  Super-Kamiokande IV}}, Ph.D. thesis, Tokyo U., 2016.
\newblock 10.15083/00073298.

\bibitem{deGouvea:2021ymm}
A.~de~Gouv\^ea, E.~McGinness, I.~Martinez-Soler and Y.F.~Perez-Gonzalez,
  \emph{{pp solar neutrinos at DARWIN}},
  \href{https://doi.org/10.1103/PhysRevD.106.096017}{\emph{Phys. Rev. D}
  {\bfseries 106} (2022) 096017}
  [\href{https://arxiv.org/abs/2111.02421}{{\ttfamily 2111.02421}}].

\bibitem{DARWIN:2020jme}
{\scshape DARWIN} collaboration, \emph{{Sensitivity of the DARWIN observatory
  to the neutrinoless double beta decay of $^{136}$Xe}},
  \href{https://doi.org/10.1140/epjc/s10052-020-8196-z}{\emph{Eur. Phys. J. C}
  {\bfseries 80} (2020) 808}
  [\href{https://arxiv.org/abs/2003.13407}{{\ttfamily 2003.13407}}].

\bibitem{Suliga:2020jfa}
A.M.~Suliga and I.~Tamborra, \emph{{Astrophysical constraints on nonstandard
  coherent neutrino-nucleus scattering}},
  \href{https://doi.org/10.1103/PhysRevD.103.083002}{\emph{Phys. Rev. D}
  {\bfseries 103} (2021) 083002}
  [\href{https://arxiv.org/abs/2010.14545}{{\ttfamily 2010.14545}}].

\bibitem{JUNO:2022jkf}
{\scshape JUNO} collaboration, \emph{{Model Independent Approach of the JUNO
  $^8$B Solar Neutrino Program}},
  \href{https://arxiv.org/abs/2210.08437}{{\ttfamily 2210.08437}}.

\bibitem{JUNO:2023zty}
{\scshape JUNO} collaboration, \emph{{JUNO sensitivity to $^{7}$Be, pep, and
  CNO solar neutrinos}},
  \href{https://doi.org/10.1088/1475-7516/2023/10/022}{\emph{JCAP} {\bfseries
  10} (2023) 022} [\href{https://arxiv.org/abs/2303.03910}{{\ttfamily
  2303.03910}}].

\bibitem{Beacom:2002cb}
J.F.~Beacom and N.F.~Bell, \emph{{Do Solar Neutrinos Decay?}},
  \href{https://doi.org/10.1103/PhysRevD.65.113009}{\emph{Phys. Rev. D}
  {\bfseries 65} (2002) 113009}
  [\href{https://arxiv.org/abs/hep-ph/0204111}{{\ttfamily hep-ph/0204111}}].

\bibitem{Martinez-Mirave:2023fyb}
P.~Mart\'\i{}nez-Mirav\'e, \emph{{Neutrino properties from the laboratory and
  the cosmos}}, Ph.D. thesis, Valencia U., 5, 2023.
\newblock \href{https://arxiv.org/abs/2309.15446}{{\ttfamily 2309.15446}}.

\bibitem{Huang:2018nxj}
G.-Y.~Huang and S.~Zhou, \emph{{Constraining Neutrino Lifetimes and Magnetic
  Moments via Solar Neutrinos in the Large Xenon Detectors}},
  \href{https://doi.org/10.1088/1475-7516/2019/02/024}{\emph{JCAP} {\bfseries
  02} (2019) 024} [\href{https://arxiv.org/abs/1810.03877}{{\ttfamily
  1810.03877}}].

\bibitem{Mirizzi:2015eza}
A.~Mirizzi, I.~Tamborra, H.-T.~Janka, N.~Saviano, K.~Scholberg, R.~Bollig
  et~al., \emph{{Supernova Neutrinos: Production, Oscillations and Detection}},
  \href{https://doi.org/10.1393/ncr/i2016-10120-8}{\emph{Riv. Nuovo Cim.}
  {\bfseries 39} (2016) 1} [\href{https://arxiv.org/abs/1508.00785}{{\ttfamily
  1508.00785}}].

\bibitem{Tamborra:2020cul}
I.~Tamborra and S.~Shalgar, \emph{{New Developments in Flavor Evolution of a
  Dense Neutrino Gas}},
  \href{https://doi.org/10.1146/annurev-nucl-102920-050505}{\emph{Ann. Rev.
  Nucl. Part. Sci.} {\bfseries 71} (2021) 165}
  [\href{https://arxiv.org/abs/2011.01948}{{\ttfamily 2011.01948}}].

\bibitem{Richers:2022zug}
S.~Richers and M.~Sen, \emph{{Fast Flavor Transformations}},  in
  \emph{{Handbook of Nuclear Physics}}, I.~Tanihata, H.~Toki and T.~Kajino,
  eds., pp.~1--17 (2022),
  \href{https://doi.org/10.1007/978-981-15-8818-1_125-1}{DOI}
  [\href{https://arxiv.org/abs/2207.03561}{{\ttfamily 2207.03561}}].

\bibitem{Mezzacappa:2022hmk}
A.~Mezzacappa, \emph{{Toward Realistic Models of Core Collapse Supernovae: A
  Brief Review}},  5, 2022,
  \href{https://doi.org/10.1017/S1743921322001831}{DOI}
  [\href{https://arxiv.org/abs/2205.13438}{{\ttfamily 2205.13438}}].

\bibitem{Mezzacappa:2020oyq}
A.~Mezzacappa, E.~Endeve, O.E.~Bronson~Messer and S.W.~Bruenn, \emph{{Physical,
  numerical, and computational challenges of modeling neutrino transport in
  core-collapse supernovae}},
  \href{https://doi.org/10.1007/s41115-020-00010-8}{\emph{Liv. Rev. Comput.
  Astrophys.} {\bfseries 6} (2020) 4}
  [\href{https://arxiv.org/abs/2010.09013}{{\ttfamily 2010.09013}}].

\bibitem{Janka:2017vlw}
H.-T.~Janka, \emph{{Neutrino Emission from Supernovae}},
  \href{https://arxiv.org/abs/1702.08713}{{\ttfamily 1702.08713}}.

\bibitem{OConnor:2018sti}
E.~O'Connor et~al., \emph{{Global Comparison of Core-Collapse Supernova
  Simulations in Spherical Symmetry}},
  \href{https://doi.org/10.1088/1361-6471/aadeae}{\emph{J. Phys. G} {\bfseries
  45} (2018) 104001} [\href{https://arxiv.org/abs/1806.04175}{{\ttfamily
  1806.04175}}].

\bibitem{Kachelriess:2004ds}
M.~Kachelriess, R.~Tomas, R.~Buras, H.-T.~Janka, A.~Marek and M.~Rampp,
  \emph{{Exploiting the neutronization burst of a galactic supernova}},
  \href{https://doi.org/10.1103/PhysRevD.71.063003}{\emph{Phys. Rev. D}
  {\bfseries 71} (2005) 063003}
  [\href{https://arxiv.org/abs/astro-ph/0412082}{{\ttfamily
  astro-ph/0412082}}].

\bibitem{garching}
``{The Garching Core-Collapse Supernova Archive}.''

\bibitem{Lattimer:1991nc}
J.M.~Lattimer and F.D.~Swesty, \emph{{A Generalized equation of state for hot,
  dense matter}},
  \href{https://doi.org/10.1016/0375-9474(91)90452-C}{\emph{Nucl. Phys. A}
  {\bfseries 535} (1991) 331}.

\bibitem{Dighe:1999bi}
A.S.~Dighe and A.Y.~Smirnov, \emph{{Identifying the neutrino mass spectrum from
  the neutrino burst from a supernova}},
  \href{https://doi.org/10.1103/PhysRevD.62.033007}{\emph{Phys. Rev. D}
  {\bfseries 62} (2000) 033007}
  [\href{https://arxiv.org/abs/hep-ph/9907423}{{\ttfamily hep-ph/9907423}}].

\bibitem{Just:2022flt}
O.~Just, S.~Abbar, M.-R.~Wu, I.~Tamborra, H.-T.~Janka and F.~Capozzi,
  \emph{{Fast neutrino conversion in hydrodynamic simulations of
  neutrino-cooled accretion disks}},
  \href{https://doi.org/10.1103/PhysRevD.105.083024}{\emph{Phys. Rev. D}
  {\bfseries 105} (2022) 083024}
  [\href{https://arxiv.org/abs/2203.16559}{{\ttfamily 2203.16559}}].

\bibitem{JUNO:2023dnp}
{\scshape JUNO} collaboration, \emph{{Real-time Monitoring for the Next
  Core-Collapse Supernova in JUNO}},
  \href{https://arxiv.org/abs/2309.07109}{{\ttfamily 2309.07109}}.

\bibitem{Pompa:2023yzg}
F.~Pompa and O.~Mena, \emph{{How much do neutrinos live and weigh?}},
  \href{https://arxiv.org/abs/2310.05474}{{\ttfamily 2310.05474}}.

\bibitem{Bendahman:2023hjj}
M.~Bendahman et~al., \emph{{Prospects for realtime characterization of
  core-collapse supernova and neutrino properties}},
  \href{https://arxiv.org/abs/2311.06216}{{\ttfamily 2311.06216}}.

\bibitem{Fiorillo:2023frv}
D.F.G.~Fiorillo, M.~Heinlein, H.-T.~Janka, G.G.~Raffelt, E.~Vitagliano and
  R.~Bollig, \emph{{Supernova simulations confront SN 1987A neutrinos}},
  \href{https://doi.org/10.1103/PhysRevD.108.083040}{\emph{Phys. Rev. D}
  {\bfseries 108} (2023) 083040}
  [\href{https://arxiv.org/abs/2308.01403}{{\ttfamily 2308.01403}}].

\bibitem{Steiner:2012rk}
A.W.~Steiner, M.~Hempel and T.~Fischer, \emph{{Core-collapse supernova
  equations of state based on neutron star observations}},
  \href{https://doi.org/10.1088/0004-637X/774/1/17}{\emph{Astrophys. J.}
  {\bfseries 774} (2013) 17} [\href{https://arxiv.org/abs/1207.2184}{{\ttfamily
  1207.2184}}].

\bibitem{Beacom:2010kk}
J.F.~Beacom, \emph{{The Diffuse Supernova Neutrino Background}},
  \href{https://doi.org/10.1146/annurev.nucl.010909.083331}{\emph{Ann. Rev.
  Nucl. Part. Sci.} {\bfseries 60} (2010) 439}
  [\href{https://arxiv.org/abs/1004.3311}{{\ttfamily 1004.3311}}].

\bibitem{Lunardini:2010ab}
C.~Lunardini, \emph{{Diffuse supernova neutrinos at underground laboratories}},
  \href{https://doi.org/10.1016/j.astropartphys.2016.02.005}{\emph{Astropart.
  Phys.} {\bfseries 79} (2016) 49}
  [\href{https://arxiv.org/abs/1007.3252}{{\ttfamily 1007.3252}}].

\bibitem{Vitagliano:2019yzm}
E.~Vitagliano, I.~Tamborra and G.G.~Raffelt, \emph{{Grand Unified Neutrino
  Spectrum at Earth: Sources and Spectral Components}},
  \href{https://doi.org/10.1103/RevModPhys.92.045006}{\emph{Rev. Mod. Phys.}
  {\bfseries 92} (2020) 45006}
  [\href{https://arxiv.org/abs/1910.11878}{{\ttfamily 1910.11878}}].

\bibitem{Salpeter:1955it}
E.E.~Salpeter, \emph{{The Luminosity function and stellar evolution}},
  \href{https://doi.org/10.1086/145971}{\emph{Astrophys. J.} {\bfseries 121}
  (1955) 161}.

\bibitem{Yuksel:2008cu}
H.~Yuksel, M.D.~Kistler, J.F.~Beacom and A.M.~Hopkins, \emph{{Revealing the
  High-Redshift Star Formation Rate with Gamma-Ray Bursts}},
  \href{https://doi.org/10.1086/591449}{\emph{Astrophys. J. Lett.} {\bfseries
  683} (2008) L5} [\href{https://arxiv.org/abs/0804.4008}{{\ttfamily
  0804.4008}}].

\bibitem{Lien:2010yb}
A.~Lien, B.D.~Fields and J.F.~Beacom, \emph{{Synoptic Sky Surveys and the
  Diffuse Supernova Neutrino Background: Removing Astrophysical Uncertainties
  and Revealing Invisible Supernovae}},
  \href{https://doi.org/10.1103/PhysRevD.81.083001}{\emph{Phys. Rev. D}
  {\bfseries 81} (2010) 083001}
  [\href{https://arxiv.org/abs/1001.3678}{{\ttfamily 1001.3678}}].

\bibitem{ParticleDataGroup:2022pth}
{\scshape Particle Data Group} collaboration, \emph{{Review of Particle
  Physics}}, \href{https://doi.org/10.1093/ptep/ptac097}{\emph{PTEP} {\bfseries
  2022} (2022) 083C01}.

\bibitem{Planck:2018vyg}
{\scshape Planck} collaboration, \emph{{Planck 2018 results. VI. Cosmological
  parameters}},
  \href{https://doi.org/10.1051/0004-6361/201833910}{\emph{Astron. Astrophys.}
  {\bfseries 641} (2020) A6}
  [\href{https://arxiv.org/abs/1807.06209}{{\ttfamily 1807.06209}}].

\bibitem{Kresse:2020nto}
D.~Kresse, T.~Ertl and H.-T.~Janka, \emph{{Stellar Collapse Diversity and the
  Diffuse Supernova Neutrino Background}},
  \href{https://doi.org/10.3847/1538-4357/abd54e}{\emph{Astrophys. J.}
  {\bfseries 909} (2021) 169}
  [\href{https://arxiv.org/abs/2010.04728}{{\ttfamily 2010.04728}}].

\bibitem{Adams:2016hit}
S.M.~Adams, C.S.~Kochanek, J.R.~Gerke and K.Z.~Stanek, \emph{{The search for
  failed supernovae with the Large Binocular Telescope: constraints from 7 yr
  of data}}, \href{https://doi.org/10.1093/mnras/stx898}{\emph{Mon. Not. Roy.
  Astron. Soc.} {\bfseries 469} (2017) 1445}
  [\href{https://arxiv.org/abs/1610.02402}{{\ttfamily 1610.02402}}].

\bibitem{Adams:2016ffj}
S.M.~Adams, C.S.~Kochanek, J.R.~Gerke, K.Z.~Stanek and X.~Dai, \emph{{The
  search for failed supernovae with the Large Binocular Telescope: confirmation
  of a disappearing star}},
  \href{https://doi.org/10.1093/mnras/stx816}{\emph{Mon. Not. Roy. Astron.
  Soc.} {\bfseries 468} (2017) 4968}
  [\href{https://arxiv.org/abs/1609.01283}{{\ttfamily 1609.01283}}].

\bibitem{Kochanek:2013yca}
C.S.~Kochanek, \emph{{Failed Supernovae Explain the Compact Remnant Mass
  Function}},
  \href{https://doi.org/10.1088/0004-637X/785/1/28}{\emph{Astrophys. J.}
  {\bfseries 785} (2014) 28} [\href{https://arxiv.org/abs/1308.0013}{{\ttfamily
  1308.0013}}].

\bibitem{Ando:2003ie}
S.~Ando, \emph{{Decaying neutrinos and implications from the supernova relic
  neutrino observation}},
  \href{https://doi.org/10.1016/j.physletb.2003.07.009}{\emph{Phys. Lett. B}
  {\bfseries 570} (2003) 11}
  [\href{https://arxiv.org/abs/hep-ph/0307169}{{\ttfamily hep-ph/0307169}}].

\bibitem{Fogli:2004gy}
G.L.~Fogli, E.~Lisi, A.~Mirizzi and D.~Montanino, \emph{{Three generation
  flavor transitions and decays of supernova relic neutrinos}},
  \href{https://doi.org/10.1103/PhysRevD.70.013001}{\emph{Phys. Rev. D}
  {\bfseries 70} (2004) 013001}
  [\href{https://arxiv.org/abs/hep-ph/0401227}{{\ttfamily hep-ph/0401227}}].

\bibitem{Kunxian:2016joi}
H.~Kunxian, \emph{{Measurement of the Neutrino-Oxygen Neutral Current
  Quasi-elastic Interaction Cross-section by Observing Nuclear De-excitation
  $\gamma$-rays in the T2K Experiment}}, Ph.D. thesis, Kyoto U., 2016.
\newblock 10.14989/doctor.k19501.

\bibitem{Ashida:2020erk}
Y.~Ashida, \emph{{Measurement of Neutrino and Antineutrino Neutral-Current
  Quasielastic-like Interactions and Applications to Supernova Relic Neutrino
  Searches}}, Ph.D. thesis, Kyoto U., 2020.

\bibitem{Maksimovic:2021dmz}
D.~Maksimovi\'c, M.~Nieslony and M.~Wurm, \emph{{CNNs for enhanced background
  discrimination in DSNB searches in large-scale water-Gd detectors}},
  \href{https://doi.org/10.1088/1475-7516/2021/11/051}{\emph{JCAP} {\bfseries
  11} (2021) 051} [\href{https://arxiv.org/abs/2104.13426}{{\ttfamily
  2104.13426}}].

\bibitem{Zhou:2023mou}
B.~Zhou and J.F.~Beacom, \emph{{First Detailed Calculation of Atmospheric
  Neutrino Foregrounds to the Diffuse Supernova Neutrino Background in
  Super-Kamiokande}},  \href{https://arxiv.org/abs/2311.05675}{{\ttfamily
  2311.05675}}.

\bibitem{Cocco:2004ac}
A.G.~Cocco, A.~Ereditato, G.~Fiorillo, G.~Mangano and V.~Pettorino,
  \emph{{Supernova relic neutrinos in liquid argon detectors}},
  \href{https://doi.org/10.1088/1475-7516/2004/12/002}{\emph{JCAP} {\bfseries
  12} (2004) 002} [\href{https://arxiv.org/abs/hep-ph/0408031}{{\ttfamily
  hep-ph/0408031}}].

\bibitem{Mollenberg:2014pwa}
R.~M\"ollenberg, F.~von Feilitzsch, D.~Hellgartner, L.~Oberauer, M.~Tippmann,
  V.~Zimmer et~al., \emph{{Detecting the Diffuse Supernova Neutrino Background
  with LENA}}, \href{https://doi.org/10.1103/PhysRevD.91.032005}{\emph{Phys.
  Rev. D} {\bfseries 91} (2015) 032005}
  [\href{https://arxiv.org/abs/1409.2240}{{\ttfamily 1409.2240}}].

\bibitem{Cheng:2023zds}
J.~Cheng, X.-J.~Luo, G.-S.~Li, Y.-F.~Li, Z.-P.~Li, H.-Q.~Lu et~al.,
  \emph{{Pulse shape discrimination technique for diffuse supernova neutrino
  background search with JUNO}},
  \href{https://arxiv.org/abs/2311.16550}{{\ttfamily 2311.16550}}.

\bibitem{Suliga:2021hek}
A.M.~Suliga, J.F.~Beacom and I.~Tamborra, \emph{{Towards probing the diffuse
  supernova neutrino background in all flavors}},
  \href{https://doi.org/10.1103/PhysRevD.105.043008}{\emph{Phys. Rev. D}
  {\bfseries 105} (2022) 043008}
  [\href{https://arxiv.org/abs/2112.09168}{{\ttfamily 2112.09168}}].

\bibitem{Zhuang:2023dzd}
Y.~Zhuang, L.E.~Strigari, L.~Jin and S.~Sinha, \emph{{Detection of
  astrophysical neutrinos at prospective locations of dark matter detectors}},
  \href{https://arxiv.org/abs/2307.13792}{{\ttfamily 2307.13792}}.

\bibitem{Dutta:2019oaj}
B.~Dutta and L.E.~Strigari, \emph{{Neutrino physics with dark matter
  detectors}},
  \href{https://doi.org/10.1146/annurev-nucl-101918-023450}{\emph{Ann. Rev.
  Nucl. Part. Sci.} {\bfseries 69} (2019) 137}
  [\href{https://arxiv.org/abs/1901.08876}{{\ttfamily 1901.08876}}].

\bibitem{SNO:2006dke}
{\scshape SNO} collaboration, \emph{{A Search for Neutrinos from the Solar hep
  Reaction and the Diffuse Supernova Neutrino Background with the Sudbury
  Neutrino Observatory}},
  \href{https://doi.org/10.1086/508768}{\emph{Astrophys. J.} {\bfseries 653}
  (2006) 1545} [\href{https://arxiv.org/abs/hep-ex/0607010}{{\ttfamily
  hep-ex/0607010}}].

\bibitem{Aglietta:1992yk}
M.~Aglietta et~al., \emph{{Limits on low-energy neutrino fluxes with the Mont
  Blanc liquid scintillator detector}},
  \href{https://doi.org/10.1016/0927-6505(92)90004-J}{\emph{Astropart. Phys.}
  {\bfseries 1} (1992) 1}.

\bibitem{Lunardini:2008xd}
C.~Lunardini and O.L.G.~Peres, \emph{{Upper limits on the diffuse supernova
  neutrino flux from the SuperKamiokande data}},
  \href{https://doi.org/10.1088/1475-7516/2008/08/033}{\emph{JCAP} {\bfseries
  08} (2008) 033} [\href{https://arxiv.org/abs/0805.4225}{{\ttfamily
  0805.4225}}].

\bibitem{KM3NeT:2023ncz}
{\scshape KM3NeT} collaboration, \emph{{Probing invisible neutrino decay with
  KM3NeT/ORCA}}, \href{https://doi.org/10.1007/JHEP04(2023)090}{\emph{JHEP}
  {\bfseries 04} (2023) 090}
  [\href{https://arxiv.org/abs/2302.02717}{{\ttfamily 2302.02717}}].

\bibitem{Kolb:1988pe}
E.W.~Kolb and M.S.~Turner, \emph{{Limits to the Radiative Decays of Neutrinos
  and Axions from Gamma-Ray Observations of SN 1987a}},
  \href{https://doi.org/10.1103/PhysRevLett.62.509}{\emph{Phys. Rev. Lett.}
  {\bfseries 62} (1989) 509}.

\bibitem{Jaffe:1995sw}
A.H.~Jaffe and M.S.~Turner, \emph{{Gamma-rays and the decay of neutrinos from
  SN1987A}}, \href{https://doi.org/10.1103/PhysRevD.55.7951}{\emph{Phys. Rev.
  D} {\bfseries 55} (1997) 7951}
  [\href{https://arxiv.org/abs/astro-ph/9601104}{{\ttfamily
  astro-ph/9601104}}].

\bibitem{Picoreti:2021yct}
R.~Picoreti, D.~Pramanik, P.C.~de~Holanda and O.L.G.~Peres, \emph{{Updating
  \ensuremath{\nu}3 lifetime from solar antineutrino spectra}},
  \href{https://doi.org/10.1103/PhysRevD.106.015025}{\emph{Phys. Rev. D}
  {\bfseries 106} (2022) 015025}
  [\href{https://arxiv.org/abs/2109.13272}{{\ttfamily 2109.13272}}].

\bibitem{deGouvea:2023jxn}
A.~de~Gouv\^ea, J.~Weill and M.~Sen, \emph{{Solar neutrinos and $\nu_2$ visible
  decays to $\nu_1$}},  \href{https://arxiv.org/abs/2308.03838}{{\ttfamily
  2308.03838}}.

\bibitem{Ivanez-Ballesteros:2023lqa}
P.~Iv\'a\~nez Ballesteros and M.C.~Volpe, \emph{{SN1987A and neutrino
  non-radiative decay}},
  \href{https://doi.org/10.1016/j.physletb.2023.138252}{\emph{Phys. Lett. B}
  {\bfseries 847} (2023) 138252}
  [\href{https://arxiv.org/abs/2307.03549}{{\ttfamily 2307.03549}}].

\bibitem{Ando:2004qe}
S.~Ando, \emph{{Appearance of neutronization peak and decaying supernova
  neutrinos}}, \href{https://doi.org/10.1103/PhysRevD.70.033004}{\emph{Phys.
  Rev. D} {\bfseries 70} (2004) 033004}
  [\href{https://arxiv.org/abs/hep-ph/0405200}{{\ttfamily hep-ph/0405200}}].

\bibitem{deGouvea:2019goq}
A.~de~Gouv\^ea, I.~Martinez-Soler and M.~Sen, \emph{{Impact of neutrino decays
  on the supernova neutronization-burst flux}},
  \href{https://doi.org/10.1103/PhysRevD.101.043013}{\emph{Phys. Rev. D}
  {\bfseries 101} (2020) 043013}
  [\href{https://arxiv.org/abs/1910.01127}{{\ttfamily 1910.01127}}].

\bibitem{DeGouvea:2020ang}
A.~De~Gouv\^ea, I.~Martinez-Soler, Y.F.~Perez-Gonzalez and M.~Sen,
  \emph{{Fundamental physics with the diffuse supernova background neutrinos}},
  \href{https://doi.org/10.1103/PhysRevD.102.123012}{\emph{Phys. Rev. D}
  {\bfseries 102} (2020) 123012}
  [\href{https://arxiv.org/abs/2007.13748}{{\ttfamily 2007.13748}}].

\bibitem{Ivanez-Ballesteros:2022szu}
P.~Ivanez-Ballesteros and M.C.~Volpe, \emph{{Neutrino nonradiative decay and
  the diffuse supernova neutrino background}},
  \href{https://doi.org/10.1103/PhysRevD.107.023017}{\emph{Phys. Rev. D}
  {\bfseries 107} (2023) 023017}
  [\href{https://arxiv.org/abs/2209.12465}{{\ttfamily 2209.12465}}].

\bibitem{Bustamante:2016ciw}
M.~Bustamante, J.F.~Beacom and K.~Murase, \emph{{Testing decay of astrophysical
  neutrinos with incomplete information}},
  \href{https://doi.org/10.1103/PhysRevD.95.063013}{\emph{Phys. Rev. D}
  {\bfseries 95} (2017) 063013}
  [\href{https://arxiv.org/abs/1610.02096}{{\ttfamily 1610.02096}}].

\bibitem{Beacom:2002vi}
J.F.~Beacom, N.F.~Bell, D.~Hooper, S.~Pakvasa and T.J.~Weiler, \emph{{Decay of
  High-Energy Astrophysical Neutrinos}},
  \href{https://doi.org/10.1103/PhysRevLett.90.181301}{\emph{Phys. Rev. Lett.}
  {\bfseries 90} (2003) 181301}
  [\href{https://arxiv.org/abs/hep-ph/0211305}{{\ttfamily hep-ph/0211305}}].

\bibitem{Baerwald:2012kc}
P.~Baerwald, M.~Bustamante and W.~Winter, \emph{{Neutrino Decays over
  Cosmological Distances and the Implications for Neutrino Telescopes}},
  \href{https://doi.org/10.1088/1475-7516/2012/10/020}{\emph{JCAP} {\bfseries
  10} (2012) 020} [\href{https://arxiv.org/abs/1208.4600}{{\ttfamily
  1208.4600}}].

\bibitem{Maltoni:2008jr}
M.~Maltoni and W.~Winter, \emph{{Testing neutrino oscillations plus decay with
  neutrino telescopes}},
  \href{https://doi.org/10.1088/1126-6708/2008/07/064}{\emph{JHEP} {\bfseries
  07} (2008) 064} [\href{https://arxiv.org/abs/0803.2050}{{\ttfamily
  0803.2050}}].

\end{thebibliography}\endgroup

\end{document}